\documentclass[final,3p,12pt,times]{elsarticle}

\journal{Prog. Part. Nucl. Phys.}

\usepackage{graphicx,color}
\usepackage{amsmath,amssymb}
\usepackage{afterpage}

\newcommand{\eq}[1]{\begin{equation}#1\end{equation}}
\newcommand{\eqmulti}[1]{\begin{equation}\begin{split}#1\end{split}\end{equation}}

\newcommand{\eqmultline}[1]{\begin{multline}#1\end{multline}}

\newcommand{\bra}[1]{\ensuremath{\langle{#1}|\,}}

\newcommand{\ket}[1]{\ensuremath{\,|{#1}\rangle}}

\newcommand{\braket}[2]{\ensuremath{\langle{#1}|{#2}\rangle}}

\newcommand{\matrixe}[3]{\ensuremath{\langle{#1}|\,{#2}\,|{#3}\rangle}}

\newcommand{\expect}[1]{\ensuremath{\langle{#1}\rangle}}

\newcommand{\op}[1]{\ensuremath{\mathrm{#1}}}
\newcommand{\conj}[1]{\ensuremath{{{#1}}^{\star}}}
\newcommand{\adj}[1]{\ensuremath{{{#1}}^{\dag}}}
\newcommand{\comm}[2]{\ensuremath{[{#1},{#2}]}}

\newcommand{\corr}[1]{\ensuremath{\widetilde{#1}}}

\newcommand{\ii}{\ensuremath{\mathrm{i}}}
\newcommand{\dd}{\ensuremath{\mathrm{d}}}

\newcommand{\clebschgordan}[6]{\ensuremath{\braket{#1 #4, #2 #5}{#3 #6}}}

\newcommand{\half}{\ensuremath{\tfrac{1}{2}}}
\newcommand{\UCOM}{\ensuremath{\mathrm{UCOM}}}
\newcommand{\intr}{\ensuremath{\mathrm{int}}}

\newcommand{\cm}{\ensuremath{\mathrm{cm}}}

\newcommand{\fm}{\ensuremath{\,\text{fm}}}

\newcommand{\MeV}{\ensuremath{\,\text{MeV}}}

\newcommand{\elem}[2]{\ensuremath{{}^{#2}\text{#1}}}

\newcommand{\gO}{\ensuremath{\op{g}}}

\newcommand{\qO}{\ensuremath{\op{q}}}
\newcommand{\rO}{\ensuremath{\op{r}}}

\newcommand{\AO}{\ensuremath{\op{A}}}
\newcommand{\CO}{\ensuremath{\op{C}}}
\newcommand{\CCO}{\ensuremath{\adj{\op{C}}}}

\newcommand{\HO}{\ensuremath{\op{H}}}

\newcommand{\OO}{\ensuremath{\op{O}}}

\newcommand{\RO}{\ensuremath{\op{R}}}

\newcommand{\TO}{\ensuremath{\op{T}}}
\newcommand{\VO}{\ensuremath{\op{V}}}

\newcommand{\PiO}{\ensuremath{\op{\Pi}}}

\newcommand{\qV}{\ensuremath{\vec{q}}}

\newcommand{\aOV}{\ensuremath{\vec{\op{a}}}}
\newcommand{\bOV}{\ensuremath{\vec{\op{b}}}}

\newcommand{\pOV}{\ensuremath{\vec{\op{p}}}}
\newcommand{\qOV}{\ensuremath{\vec{\op{q}}}}
\newcommand{\rOV}{\ensuremath{\vec{\op{r}}}}

\newcommand{\xOV}{\ensuremath{\vec{\op{x}}}}

\newcommand{\LOV}{\ensuremath{\vec{\op{L}}}}

\newcommand{\SOV}{\ensuremath{\vec{\op{S}}}}
\newcommand{\XOV}{\ensuremath{\vec{\op{X}}}}

\newcommand{\sigmaOV}{\ensuremath{\vec{\op{\sigma}}}}

\newcommand{\tautauO}{\ensuremath{(\vec{\op{\tau}}_{\!1}\!\cdot\!\vec{\op{\tau}}_{\!2})}}
\newcommand{\sigmasigmaO}{\ensuremath{(\vec{\op{\sigma}}_{\!1}\!\cdot\!\vec{\op{\sigma}}_{\!2})}}
\newcommand{\tensorO}{\ensuremath{\op{S}_{12}}}
\newcommand{\tensorRRO}{\ensuremath{\op{S}_{12}(\tfrac{\rOV}{\rO},\tfrac{\rOV}{\rO})}}
\newcommand{\tensorLLO}{\ensuremath{\op{S}_{12}(\LOV,\LOV)}}
\newcommand{\tensorQQO}{\ensuremath{\op{S}_{12}(\qOV_{\Omega},\qOV_{\Omega})}}
\newcommand{\tensorRQO}{\ensuremath{\op{S}_{12}(\rOV,\qOV_{\Omega})}}
\newcommand{\tensorbarQQO}{\ensuremath{\bar{\op{S}}_{12}(\qOV_{\Omega},\qOV_{\Omega})}}
\newcommand{\spinorbitO}{\ensuremath{(\vec{\op{L}}\cdot\vec{\op{S}})}}
\newcommand{\orbitsqrO}{\ensuremath{\vec{\op{L}}^2}}

\newcommand{\pmS}{\ensuremath{\!\pm\!}}
\newcommand{\mpS}{\ensuremath{\!\mp\!}}

\newcommand{\Rm}{\ensuremath{R_-}}
\newcommand{\DRm}{\ensuremath{R'_-}}

\newcommand{\Rp}{\ensuremath{R_+}}
\newcommand{\DRp}{\ensuremath{R'_+}}
\newcommand{\DDRp}{\ensuremath{R''_+}}
\newcommand{\DDDRp}{\ensuremath{R'''_+}}

\newcommand{\Rpm}{\ensuremath{R_{\pm}}}
\newcommand{\Rmp}{\ensuremath{R_{\mp}}}

\definecolor{FGViolet}{rgb}{0.61,0.32,0.61}
\definecolor{FGDarkBlue}{rgb}{0,0,0.6}
\definecolor{FGBlue}{rgb}{0,0,0.8}
\definecolor{FGLightBlue}{rgb}{0.2, 0.6, 0.8}
\definecolor{FGGreen}{rgb}{0.2,0.7,0.2}
\definecolor{FGLightGreen}{rgb}{0.4,1,0.4}
\definecolor{FGYellow}{rgb}{1,0.95,0}
\definecolor{FGOrange}{rgb}{0.95,0.5,0.1}
\definecolor{FGRed}{rgb}{0.8,0,0}
\definecolor{FGWhite}{rgb}{1,1,1}
\definecolor{FGLightGray}{rgb}{0.8,0.8,0.8}
\definecolor{FGGray}{rgb}{0.5,0.5,0.5}
\definecolor{FGDarkGray}{rgb}{0.3,0.3,0.3}
\definecolor{FGBlack}{rgb}{0,0,0}

\newcommand{\symboldiamond}[1][black]{{\color{#1}$\blacklozenge$}}

\newcommand{\symbolbox}[1][black]{{\color{#1}$\blacksquare$}}
\newcommand{\symbolcircle}[1][black]{{\color{#1}$\medbullet$}}

\newcommand{\linemediumsolid}[1][black]{\unitlength 1ex
  {\color{#1}
  \begin{picture}(6,1)
  \linethickness{0.4mm}
  \put(0,0.5){\line(1,0){6.0}}
  \end{picture}}\nolinebreak
}

\newcommand{\linemediumdashed}[1][black]{\unitlength 1ex
  {\color{#1}
  \begin{picture}(6,1)
  \linethickness{0.4mm}
  \put(0,0.5){\line(1,0){1.5}}
  \put(2.2,0.5){\line(1,0){1.5}}
  \put(4.4,0.5){\line(1,0){1.5}}
  \end{picture}}\nolinebreak
}

\newcommand{\linemediumdotted}[1][black]{\unitlength 1ex
  {\color{#1}
  \begin{picture}(6,1)
  \linethickness{0.4mm}
  \put(0,0.5){\line(1,0){0.8}}
  \put(1.2,0.5){\line(1,0){0.8}}
  \put(2.4,0.5){\line(1,0){0.8}}
  \put(3.6,0.5){\line(1,0){0.8}}
  \put(4.8,0.5){\line(1,0){0.8}}
  \end{picture}}\nolinebreak
}

\newcommand{\linemediumdashdot}[1][black]{\unitlength 1ex
  {\color{#1}
  \begin{picture}(6,1)
  \linethickness{0.4mm}
  \put(0,0.5){\line(1,0){0.4}}
  \put(0.9,0.5){\line(1,0){1.5}}
  \put(2.9,0.5){\line(1,0){0.4}}
  \put(3.8,0.5){\line(1,0){1.5}}
  \put(5.8,0.5){\line(1,0){0.4}}
  \end{picture}}\nolinebreak
}

\begin{document}

\begin{frontmatter}

\title{Nuclear Structure in the Framework of the\\ Unitary Correlation Operator Method}

\author[tud]{Robert Roth}
\ead{robert.roth@physik.tu-darmstadt.de}

\author[gsi]{Thomas Neff}
\ead{t.neff@gsi.de}

\author[gsi]{Hans Feldmeier}
\ead{h.feldmeier@gsi.de}

\address[tud]{Institut f\"ur Kernphysik, Technische Universit\"at Darmstadt, Schlossgartenstra{\ss}e 9, 64289 Darmstadt, Germany}
\address[gsi]{GSI Helmholtzzentrum f{\"u}r Schwerionenforschung GmbH, Planckstra{\ss}e 1, 64291 Darmstadt, Germany}

\begin{abstract}

Correlations play a crucial role in the nuclear many-body problem. We give an overview of recent developments in nuclear structure theory aiming at the description of these interaction-induced correlations by unitary transformations. We focus on the Unitary Correlation Operator Method (UCOM), which offers a very intuitive, universal and robust approach for the treatment of short-range correlations. We discuss the UCOM formalism in detail and highlight the connections to other methods for the description of short-range correlations and the construction of effective interactions. In particular, we juxtapose UCOM with the Similarity Renormalization Group (SRG) approach, which implements the unitary transformation of the Hamiltonian through a very flexible flow-equation formulation. The UCOM- and SRG-transformed interactions are compared on the level of matrix elements and in many-body calculations within the no-core shell model and with Hartree-Fock plus perturbation theory for a variety of nuclei and observables. These calculations provide a detailed picture of the similarities and differences as well as the advantages and limitations of unitary transformation methods.

\end{abstract}


\end{frontmatter}
\clearpage

%%%%%%%%%%%%%%%%%%%%%%%%%%%%%%%%%%%%%%%%%%%%%%%%%%%%%%%%%%%%%%%%%%%%%%
%%%%%%%%%%%%%%%%%%%%%%%%%%%%%%%%%%%%%%%%%%%%%%%%%%%%%%%%%%%%%%%%%%%%%%
%%%%%%%%%%%%%%%%%%%%%%%%%%%%%%%%%%%%%%%%%%%%%%%%%%%%%%%%%%%%%%%%%%%%%%

\tableofcontents

%%%%%%%%%%%%%%%%%%%%%%%%%%%%%%%%%%%%%%%%%%%%%%%%%%%%%%%%%%%%%%%%%%%%%%
%%%%%%%%%%%%%%%%%%%%%%%%%%%%%%%%%%%%%%%%%%%%%%%%%%%%%%%%%%%%%%%%%%%%%%
%%%%%%%%%%%%%%%%%%%%%%%%%%%%%%%%%%%%%%%%%%%%%%%%%%%%%%%%%%%%%%%%%%%%%%
\clearpage
\section{Introduction}
\label{sec:intro}

\newcommand{\Heff}{\op{H}_\mathrm{eff}}

Recent years have seen substantial progress in theoretical methods describing the 
many-body problem of low energy nuclear structure in an {\it ab initio} sense.
{\it Ab initio} means from the beginning, without further assumptions or uncontrolled
approximations. At present for a system of particles interacting by the strong interaction,
the most elementary degrees of freedom  are considered to be the point-like quarks 
and gluons, whose dynamics is derived from the Lagrangian of Quantum Chromo Dynamics (QCD). 
However, in the low-energy regime QCD cannot be treated by perturbation
theory because of the confinement phenomenon.
The lowest bound states of QCD are baryons and mesons, which are the
natural degrees of freedom at low energies. 
In a system at very low energy, below about 50 MeV per baryon,
only the proton and neutron are left.
Other baryons like the $\Delta$-resonance appear only as intermediate
virtual excitations. Also the mesons, among which the pions are the lightest ones,
do not occur as real particles on their mass shell, but may be regarded as
bosons mediating the interaction among the baryons.

Recent QCD lattice simulations \cite{IsAo07} give hope that in the near future the 
baryon-baryon interaction might be ``measured'' on the lattice with sufficient
precision. This would complement the scattering data, which can give only 
indirect information on the interaction in form of
the phase shifts measured at large distances, where the particles do not
interact anymore.

For proton and neutron precise scattering data exist that allow to
fit the parameters of nucleon-nucleon potential models.  
Using global symmetries the form of the potential can be written as a sum 
of central, spin-orbit, and tensor interactions.
Precise fits to scattering phase shifts 
reveal that the nucleon-nucleon potential cannot be assumed to be local
in coordinate space, 
which means the potential depends not only on the relative distance and the 
spin orientations but also on the relative momentum.
The minimal momentum dependence besides the spin-orbit terms is
of $\vec{\op{L}}^2$ type (as in the Argonne V18 potential \cite{WiSt95}) 
or terms like $\vec{\op{q}}^2 V(\op{r})+ V(\op{r})\vec{\op{q}}^2$ (as in the Bonn A/B potentials
\cite{Mach89}).
Recent developments use the approximate chiral symmetry 
in the light quark sector of QCD to establish for the nucleon-nucleon
interaction a perturbation scheme in terms of diagrams contributing to the so 
called chiral potentials \cite{EnMa03,EpNo02,Epelbaum06}.
By adding higher-order terms to the expansion they can be improved systematically.
However, with increasing number of 
contributions more and more parameters have to be fixed by
experimental data. One great advantage of the chiral potentials is that
two- and three-body forces are treated on equal footing.

The nuclear structure problem consists in solving the stationary
many-body Schr\"odinger equation for $A$ nucleons
%-------------------------------------------------------------
\begin{equation}\label{eq:Schroedinger}
\op{H}\ \ket{\corr{\Psi}_n}=E_n\, \ket{\corr{\Psi}_n}
\end{equation}
%--------------------------------------------------------------
and in investigating the properties of the $A$-body eigenstates $\ket{\corr{\Psi}_n}$
by computing observable quantities as expectation values or transition
matrix elements. The subset of the discrete eigenvalues $E_n$ 
represents the excitation spectrum of a nucleus and can be compared to
data directly. The continuous part of the spectrum corresponds to
scattering states which usually have a rich resonance structure.

As the Schr\"odinger equations already indicates, solving the many-body
problem requires two main ingredients, the Hamiltonian $\op{H}$ and a 
representation for the many-body eigenstates $\ket{\corr{\Psi}_n}$
or a basis in the $A$-body Hilbert space. 

For more than about 6 particles Slater determinants $\ket{\Phi_{[\nu]}}$ provide 
the most convenient basis for treating the many-body problem numerically.
They are usually constructed as eigenstates of a one-body mean-field Hamiltonian $\op{H}_0$. 
Matrix elements of one-, two-, or three-body operators in $A$-body space 
can be calculated easily using an occupation number representation.
Modern computers and numerical techniques can tackle $A$-body Hilbert spaces 
dimensions up to about $10^{10}$.
These numbers are impressive but by far not sufficient. The dilemma is 
that realistic nuclear potentials, which fit the scattering
data at low energy and induce the high momentum components observed
in bound states of nuclei,
possess a strong repulsion at short distances and a strong tensor force.
Figure \ref{fig:AV18-projected} illustrates that
for the Argonne V18 potential.
These properties of the nuclear Hamiltonian induce short- and long-range
correlations in the many-body eigenstate $\ket{\corr{\Psi}_n}$. 
Especially the short-range repulsive and tensor correlations
cannot be properly represented by Slater determinants. The reason is
that a Slater determinant is an antisymmetrized product of 
single-particle states ($\op{\cal A}$ antisymmetrization operator): 

%----------------------------------------------------------------------- 
\begin{equation}
\braket{\xi_1,\xi_2,\dots,\xi_A}{\Phi_{[\nu]}}=\op{\cal A}\ 
\phi_{\nu_1}\!(\xi_1)\,\phi_{\nu_2}\!(\xi_2)\dots\phi_{\nu_{\!A}}\!(\xi_A),
\ \ \  \xi_i=(\vec{x}_i,{\sigma}_i,\tau_i)
\end{equation}
%--------------------------------------------------------------------------   
describing the motion of independent particles and as such cannot describe correlations 
in the relative distances $(\vec{x}_i-\vec{x}_j)$ between particles. 
Of course, Slater determinants form a complete basis and thus can in principle
represent any state. However, as we show in Sec.~\ref{sec:ncsm} 
the number of Slater determinants needed to describe the ground state 
that includes the short-range correlations exceeds 
soon any numerically tractable number for $A > 4$. 
The Hamiltonian represented in Slater determinants results in
matrices with large matrix elements even far off the diagonal.
In physical terms, in this basis the Hamiltonian scatters to
very high lying $\op{H}_0$ eigenstates.
Therefore, one introduces an effective Hamiltonian based on the following
concepts.

Common to many conventional derivations of effective Hamiltonians is that one 
first divides the many-body Hilbert space into a so called P-space 
or model space and a Q-space. The projection operators $\op{P}$ and 
$\op{Q}=\op{1}-\op{P}$ are defined with the eigenstates of a Hamiltonian $\op{H}_0$,
which is usually a one-body operator like kinetic energy or the harmonic oscillator
Hamiltonian. 
One is then looking for an effective Hamiltonian $\Heff$ that should have the same 
eigenvalues $E_n$ as the full Hamiltonian $\op{H}$ 
%----------------------------------------------------------------------
\begin{equation}\label{eq:Schroedinger-eff}
\Heff\ \ket{\Phi_n}=E_n\, \ket{\Phi_n}
\end{equation}
%----------------------------------------------------------------------
and its eigenstates should be contained in the model-space, i.e. they can be represented
as a finite sum of $\op{H}_0$ eigenstates that span the P-space,
%----------------------------------------------------------------------------------------
\begin{equation}\label{eq:Proj}
\ket{\Phi_n}=\op{P}\,\ket{\Phi_n}
      =\sum_{[\nu]}^\mathrm{P-space}\ket{\Phi_{[\nu]}} \braket{\Phi_{[\nu]}}{\Phi_n}\ .
\end{equation}
%----------------------------------------------------------------------------------------
Therefore, $\Heff$ should not connect P- and Q-space, 
$\op{P}\Heff\op{Q}=0$.
Usually one is satisfied with a few low lying eigenvalues and eigenstates of
Eq.~(\ref{eq:Schroedinger-eff}) and regards the higher eigenstates as
orthogonal rest that carries less and less physics. 
Therefore, in practice only the decoupling of the lowest eigenstates is needed, 
i.e. $\bra{\Phi_n}\Heff\op{Q}=0$ for few low lying $E_n$. A sharp separation 
between P- and Q-space, as given in the mathematical definition of Eq.~(\ref{eq:Proj}), 
is difficult to identify on physical arguments anyhow.

As the P-space consists of many-body product states that cannot describe short-range 
correlations or high relative momenta $q$, it is regarded as a low-momentum Hilbert space.
Sometimes one also speaks of a low-energy Hilbert space, but here energy refers to 
$\op{H}_0$ and its eigenvalues and not to the true Hamiltonian $\op{H}$, which is the 
physical energy operator. To avoid confusion we prefer to speak of low or high momentum states.

As a unitary transformation leaves eigenvalues invariant it is quite natural
to perform a suitable unitary transformation of the 
uncorrelated many-body states $\ket{\Phi_{[\nu]}}$ to
another basis $\ket{\corr{\Phi}_{[\nu]}}$, which already includes 
correlations to a certain extent so that the Hamiltonian matrix becomes 
more band-diagonal and is not scattering to high momentum states 
anymore. Whenever many-body states include short-range correlations we mark them by $\corr{\phantom{m}}$ like the fully correlated eigenstates
$\ket{\corr{\Psi}_n}$ of the Hamiltonian $\op{H}$ which includes 
short-range repulsive and tensor potentials.

In this contribution we discuss in depth the 
Unitary Correlation Operator Method (UCOM) which tries to achieve
this goal by an explicitly given unitary operator that transforms the Hamiltonian and 
all other observables to effective operators that include 
the effects of the short-range correlations.

Before discussing effective interactions in general and
UCOM in particular, we would like to
address a few more general issues related to the nuclear potential and
effective interactions.

%ssssssssssssssssssssssssssssssssssssssssssssssssssssssssssssssssssssssssssssssssssssssssss
\subsection{Many-body potentials}
%ssssssssssssssssssssssssssssssssssssssssssssssssssssssssssssssssssssssssssssssssssssssssss
In recent years it has become clear that the nuclear many-body problem 
needs at least three-body forces to reproduce data with sufficient precision
%--------------------------------------------------------------------------------
\begin{equation}\label{eq:3body-H}
\op{H}=\op{T}^{[1]}+\op{V}_\mathrm{ NN}^{[2]} +\op{V}_\mathrm{NNN}^{[3]} + \dots
   =\sum_{i=1}^A \op{T}_i+\sum_{j>i=1}^A \op{V}_{ij}+
             \sum_{k>j>i=1}^A \op{V}_{ijk}+\dots \ .
\end{equation}
%--------------------------------------------------------------------------------- 
An irreducible three-body potential $\op{V}_\mathrm{NNN}^{[3]}$ is the remaining 
part of the Hamiltonian acting in three-body space that cannot be described by a 
sum over one-body kinetic energies and two-body potentials.
Likewise, irreducible four-body potentials are the remaining parts of the Hamiltonian acting 
in four-body space that cannot be described by a sum over two-body and three-body potentials
and so on.

\begin{figure}
\begin{center}
\includegraphics[width=7cm]{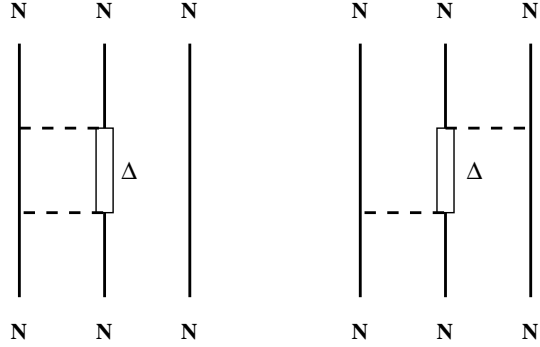}
\end{center}
\caption{$\Delta$-resonance as intermediate off-shell excitation in a three-body system.}
\label{fig:FujMiya}
\end{figure}

A typical example of an irreducible or genuine three-body potential is the
Fujita-Miyazawa interaction \cite{fujita57}, which arises from the restriction to proton and neutron
as elementary degrees of freedom. Due to the strong coupling with the pion, a nucleon can
be converted into a $\Delta$-resonance, which then has to
decay back into a nucleon, because the energy of the total system 
is not sufficient to produce a $\Delta$ on its mass shell. This intermediate
virtual excitation is not contained in the nucleonic Hilbert space, but 
is effectively already included in a two-body potential that is fitted 
to data. The left-hand graph displayed in Fig.~\ref{fig:FujMiya}, where the 
third nucleon is only a spectator, therefore, is a contribution that
makes the two-body potential an effective one with respect to excluded 
$\Delta$-degrees of freedom from the nucleonic Hilbert space. 
In the right-hand graph the pion is coupling to a third nucleon. Thus
this interaction cannot be written as an effective two-body potential.
It has to be seen as a genuine three-body potential.
%---------------------------------------------------------------------------------
%--------------------------------------------------------------------------------

Already from Fig.~\ref{fig:FujMiya} it is obvious that two-body and three-body 
forces cannot be treated independently as they may originate from the same physical process. 
Furthermore, this simple example shows that omitting parts of the many-body Hilbert space
that are in principle reachable with the strong interaction leads to effective
operators and induces many-body forces. This effect will also occur,
and is of major concern, in the following sections 
when the nucleonic Hilbert space itself is truncated 
to low-energy states.

%ssssssssssssssssssssssssssssssssssssssssssssssssssssssssssssssssssssssssssssssssssssssssss
\subsection{Off-shell properties}
%ssssssssssssssssssssssssssssssssssssssssssssssssssssssssssssssssssssssssssssssssssssssssss
The elastic nucleon-nucleon scattering cross section is represented in terms
of phase shifts and mixing angles as a function of energy. As cross sections are 
measured far away from the interaction region, these quantities give
only indirect information about the nuclear forces and cannot determine
the potential in a unique way. In a stationary scattering state the two
nucleons have a sharp energy $E_q$ and the modulus of the asymptotic relative momentum 
$q$ of the ingoing wave has to come back on-shell to the same momentum $q$ in the outgoing 
wave, both related to the eigenenergy by $E_q=q^2/(2\mu)$.

In an interacting many-body system the total energy $E_n$ corresponding 
to an eigenstate of the Hamiltonian is sharp but a pair of particles within
the system has neither a sharp relative energy nor a sharp momentum. While they feel
their mutual interaction they are also interacting with other particles
exchanging energy and momentum. Therefore, the interaction is tested
for different in- and outgoing momenta $q$ and $q'$ that are not connected
by any on-shell relation. Studying many-body systems will, therefore, give 
information on the off-shell behavior of the nuclear force. 

As we will see in Sec.~\ref{sec:ucom} a unitary transformation that
acts only at short distances and does not affect the asymptotic behavior,
like the UCOM correlator, leaves the phase shifts untouched 
but creates a new potential that is not distinguishable from the original one by 
measurements of the elastic cross section. As one can devise an infinite number of such
transformations, there are infinitely many phase-shift equivalent potentials.
These, however, differ in their off-shell behavior.
Therefore, scattering data will never be able to determine the nuclear force uniquely.
In Sec.~\ref{sec:ncsm} and \ref{sec:hf} we show that phase-shift equivalent 
potentials with different off-shell properties can produce very different 
results when applied in many-body systems. One of them produces with increasing
mass number a dramatic overbinding in a variational calculation, where 
the calculated binding energy is, anyhow, only an upper limit. 
In order to be sure that the limited Hilbert space used is not the reason 
that the other potentials are not overbinding as well one needs  
many-body approaches models that can solve
the many-body problem exactly for a given Hamiltonian.
 
One goal of modern nuclear structure theory is to disentangle
as much as possible the off-shell nature of the potential from 
effects originating in the use of restricted many-body Hilbert spaces. 
However, one should always keep in mind that in any theory, not only
in nuclear structure, any interaction is an effective low-momentum interaction
in the sense that it is constructed for the degrees of freedom and Hilbert space 
one is using. With increasing momentum transfer one always opens up
new degrees of freedom not contained in the original model space.

Exact benchmark 
calculations for the three- and four-body system \cite{deltuva07,KaNo01,nogga00,NaKa00,GlKa93}
were very helpful in this respect. They also showed that three-body forces
are needed, though different ones for different phase-shift equivalent 
two-body potentials. For example the contribution 
to the binding energy of $^4$He coming from three-body terms complementing 
the Argonne V18 potential are about 50\% larger than those accompanying the 
CD Bonn potential \cite{Mach01}, see Fig.~\ref{fig:tjon-line}. The CD Bonn potential 
has a softer short-range repulsion than the AV18 but contains a radial momentum dependence
which is absent in Argonne V18. Gl\"ockle and collaborators have shown in a quite general way that 
off-shell properties of two-body interactions can be traded against three-body forces \cite{polyzou90}.
We will encounter this effect again when we perform the UCOM and SRG similarity
transformations of the Hamiltonian.

%ssssssssssssssssssssssssssssssssssssssssssssssssssssssssssssssssssssssssssssssssssssssssss
\subsection{Effective potentials}
%ssssssssssssssssssssssssssssssssssssssssssssssssssssssssssssssssssssssssssssssssssssssssss
Besides the conceptual problems of deriving and defining the nuclear interaction
there is also the already mentioned technical problem that the solution of 
the many-body Schr\"odinger equation~(\ref{eq:Schroedinger}) requires a representation
of the many-body eigenstates. Since realistic nuclear interactions
induce various kinds of correlations, in particular short-range correlations,
one possibility is to work with many-body states that can represent the correlations.
In that case the main numerical effort goes into calculating the matrix elements of the 
Hamiltonian. Examples are 
the exact Faddeev and Faddeev-Yakubovsky equations for the 3- and 4-body system
\cite{Gloe83}, 
the hyperspherical harmonics basis \cite{BaLe99},
or the Green's Function Monte Carlo method (GFMC) \cite{PiWi01,PiWi02,PiVa02,PiWi04,Piep05}.
The other possibility is to represent the eigenstates of the Hamiltonian 
with many-body basis states $\ket{\Phi_{[\nu]}}$ which are chosen such that the 
numerical effort for calculating the matrix elements 
$\matrixe{\Phi_{[\nu]}}{\op{T}^{[1]}}{\Phi_{[\mu]}}$,
$\matrixe{\Phi_{[\nu]}}{\op{V}_\mathrm{NN}^{[2]}}{\Phi_{[\mu]}}$,
$\matrixe{\Phi_{[\nu]}}{\op{V}_\mathrm{NNN}^{[3]}}{\Phi_{[\mu]}}$
of one-, two-, and three-body operators
is minimized. 
In that respect the best choice are antisymmetrized product states (Slater determinants).
However, the strong short-range correlations induced by realistic nuclear forces 
cannot be represented in a reasonable way by a product state basis. 
For example the four-body system \elem{He}{4}, when represented in harmonic oscillator states,
needs a three-body Hilbert space of dimension more than $10^8$ to get a converged
ground-state energy with the Argonne V18 potential, see Sec.~\ref{sec:ncsm}. 
Therefore, so called effective interactions for truncated Hilbert spaces are introduced.
At present there is no other possibility for mass numbers larger than about 12.  

%ssssssssssssssssssssssssssssssssssssssssssssssssssssssssssssssssssssssssssssssssssssssssss
\subsection{Unitary approaches}
%ssssssssssssssssssssssssssssssssssssssssssssssssssssssssssssssssssssssssssssssssssssssssss
In a product basis representation the nucleon-nucleon interaction scatters
to energetically very high-lying basis states. Or in other words, 
the short-range repulsive and tensor correlations imply 
components in the many-body state with large relative momenta,
which necessitate very large many-body Hilbert 
spaces in order to accommodate the correlations in this basis.
To still work with the numerically convenient product basis one uses 
so called effective interactions. The effective interaction should decouple
the Hilbert space containing high momenta from the one with low momenta, 
which can be represented more easily by product states.

The most straightforward and intuitive way is to consider a unitary transformation
of the product basis $\ket{\Phi_{[\nu]}}$
%-------------------------------------------------------------
\begin{equation}\label{eq:unitary-trafo}
\ket{\corr{\Phi}_{[\nu]}}=\op{U}\, \ket{\Phi_{[\nu]}}
\end{equation}
%-------------------------------------------------------------
to render the Hamiltonian matrix into a more diagonal form
not connecting low and high momenta.
The new basis $\ket{\corr{\Phi}_{[\nu]}}$ 
should contain already the main
properties of the short-range correlations like a depletion
of the many-body wave function
%------------------------------------------------------------------------
\begin{equation} 
\braket{\xi_1,\xi_2,\dots,\xi_A}{\corr{\Phi}_{[\nu]}}\ \ \mathrm{with}
                                \ \ \xi_i=(\vec{x}_i,{\sigma}_i,\tau_i)
\end{equation}
%-----------------------------------------------------------------------
whenever the distance $|\vec{x}_i-\vec{x}_j|$ 
between two nucleons of the many-body system is within the 
range of the repulsive core of the interaction.  
Similarly their spins $\vec{\sigma}_i$, $\vec{\sigma}_j$
should be aligned with $\vec{r}_{ij}=\vec{x}_i-\vec{x}_j$
if their isospin is in the $T=0$ configuration,
because this correlation produces binding energy, see Fig.~\ref{fig:AV18-projected}.

This unitary transformation can be used to define an effective Hamiltonian
through the similarity transformation
%---------------------------------------------------------
\begin{equation}
\Heff = \op{U}^{-1} \op{H}\, \op{U} =\op{U}^{\dagger} \op{H}\, \op{U} \ .
\end{equation}
%----------------------------------------------------------
The unitary transformation fulfills the principal requirement
formulated in Eq.~(\ref{eq:Schroedinger-eff}) that the eigenvalues
for the energy are invariant. Later approximations
should be checked regarding this aspect.
Another advantage is that orthonormality relations between eigenstates
of $\op{H}$ or between basis states are not changed
%----------------------------------------------------------------------------
\begin{eqnarray}
\braket{\corr{\Psi}_n}{\corr{\Psi}_m}=
  \matrixe{\Psi_n}{\op{U}^{-1}\op{U}}{\Psi_m}=\braket{\Psi_n}{\Psi_m}=\delta_{n,m}\nonumber\\
\braket{\corr{\Phi}_{[\nu]}}{\corr{\Phi}_{[\mu]}}=
  \matrixe{\Phi_{[\nu]}}{\op{U}^{-1}\op{U}}{\Phi_{[\mu]}}=
  \braket{\Phi_{[\nu]}}{\Phi_{[\mu]}}=\delta_{[\nu],[\mu]} \;.  
\end{eqnarray}
%--------------------------------------------------------------------------

Any unitary transformation can be written in terms of a hermitian generator
$\op{G}$ as 
%-------------------------------------------------------------------------------
\begin{equation}
\op{U}=e^{-\ii\,\op{G}} \ .
\end{equation}
%-------------------------------------------------------------------------------
If $\op{G}=\sum_{j=1}^A \op{g}_j$ is a one-body operator no correlations
are induced because in that case $\op{U}=\prod_{j=1}^A e^{-\ii\,\op{g}_j}$ just transforms
each single-particle state in $\ket{\Phi_{[\nu]}}$ independently.
Therefore, the generator $\op{G}$ has to be at least
a two-body operator.
This implies that the unitary transformation of a one-body operator
$\op{T}^{[1]}$ yields \cite{ucom98,ucom03,ucom04}
%----------------------------------------------------------------------------
\begin{equation}\label{eq:one-body-eff}
\op{U}^{\dagger}\ \op{T}^{[1]}\, \op{U}= \op{T}^{[1]}
                      + \corr{\op{T}}^{[2]}
                      + \corr{\op{T}}^{[3]} + \dots \;.
\end{equation}
%--------------------------------------------------------------------------
The transformation of a two-body operator $\op{V}_\mathrm{NN}^{[2]}$
%----------------------------------------------------------------------------
\begin{equation}
\op{U}^{\dagger}\ \op{V}_\mathrm{NN}^{[2]}\, \op{U}= \corr{\op{V}}_\mathrm{NN}^{[2]}
                      + \corr{\op{V}}_\mathrm{NN}^{[3]}
                      + \corr{\op{V}}_\mathrm{NN}^{[4]} + \dots
\end{equation}
%--------------------------------------------------------------------------
yields a new two-body operator $\corr{\op{V}}_\mathrm{NN}^{[2]}$ and additional operators of higher-order. Likewise a transformed three-body operator results in a new irreducible
three-body operator with its higher-body companions. 
The $\corr{\phantom{m}}$ above the operators indicates that effects from
short-range correlations are now moved to the operators in the sense that
the potential $\corr{\op{V}}_\mathrm{NN}^{[2]}$ is softer or less repulsive at 
short distances than  the original $\op{V}_\mathrm{NN}^{[2]}$. This ``taming'' of the
potential comes at the expense of introducing two-, three- and more-body terms
originating from the kinetic energy as well as from the potential.
The different contributions will be discussed and shown in Sec.~\ref{sec:ucom}. 

An important message that holds for any method of deriving effective
interactions is that the effective Hamiltonian contains irreducible
$n$-body interactions, where $n$ goes in principle from $1$ to $A$. 
For example, if we start from a Hamiltonian with two- and three-body forces
like in Eq.~(\ref{eq:3body-H}), we obtain
%----------------------------------------------------------------------------
\begin{equation}\label{eq:Heff-many-body}
\op{U}^{-1}\ \op{H} \op{U}= \op{T}^{[1]}
            + \left(\corr{\op{T}}^{[2]}+\corr{\op{V}}_\mathrm{NN}^{[2]}\right)
            + \left(\corr{\op{T}}^{[3]}+\corr{\op{V}}_\mathrm{NN}^{[3]}
                     +\corr{\op{V}}_\mathrm{NNN}^{[3]}\right) +\dots \;.
\end{equation}
%--------------------------------------------------------------------------
The hope is to keep the $n>2$ body terms small in order to reduce the numerical effort.

One important advantage of formulating the effective Hamiltonian through 
a unitary similarity transformation is that any other observable $\op{B}$
can and should be transformed the same way. 
An arbitrary matrix element of $\op{B}$ between two eigenstates of the
Hamiltonian can be written as
%----------------------------------------------------------------------------
\begin{equation}
\matrixe{\corr{\Psi}_n}{\op{B}}{\corr{\Psi}_m}=
\matrixe{\Psi_n}{\op{U}^{\dagger}\op{B}\,\op{U}}{\Psi_m}=
\matrixe{\Psi_n}{\op{B}_\mathrm{eff}}{\Psi_m}
\end{equation}
%--------------------------------------------------------------------------
which implies the definition of $\op{B}_\mathrm{eff}$ as
%----------------------------------------------------------------------------
\begin{equation}
\op{B}_\mathrm{eff}=\op{U}^{\dagger}\op{B}\,\op{U} \;.
\end{equation}
%--------------------------------------------------------------------------
Again $\op{B}_\mathrm{eff}$ becomes a many-body operator even if 
$\op{B}$ is an one-body operator, see Eq.~(\ref{eq:one-body-eff}).
In section \ref{sec:ucom} and \ref{sec:srg}  approximations
in the calculations of $\Heff$ are introduced. 
The same approximations can also be
applied to calculate $\op{B}_\mathrm{eff}$.

%ssssssssssssssssssssssssssssssssssssssssssssssssssssssssssssssssssssssssss
\subsection{Jastrow ansatz}
%ssssssssssssssssssssssssssssssssssssssssssssssssssssssssssssssssssssssssss

An early attempt to incorporate short-range correlations in the many-body 
state was proposed by Jastrow \cite{Jast55,Irvi81}
%%%%%%%%%%%%%%%%%%%%%%%%%%%%%%%%%%%%%%%%%%%%%%%%%%%%%%%%%%%%%%%%%%%%%%%%%
\begin{equation}
\ket{\corr{\Phi},J}={\cal S}\;
\prod_{i<j}^A f(\vec{\op{r}}_{ij},\vec{\op{\sigma}}_i,\vec{\op{\sigma}}_j,
     \vec{\op{\tau}}_i,\vec{\op{\tau}}_j)\ \ket{\Phi}
\end{equation}
%%%%%%%%%%%%%%%%%%%%%%%%%%%%%%%%%%%%%%%%%%%%%%%%%%%%%%%%%%%%%%%%%%%%%%%%%
where in modern applications the Jastrow correlation functions $f$ contain besides
the relative distances $\vec{\op{r}}_{ij}=\vec{\op{x}}_i-\vec{\op{x}}_j$, also operators 
depending on spins and isospins to account for correlations other than the short-range
repulsion, like the tensorial ones.  For distances $|\vec{r}_{ij}|$
much larger than the range of the interaction $f(\vec{r}_{ij},\dots)$ approaches one. 
As the symmetrized product runs over all particle pairs,
the transformation from the uncorrelated state $\ket{\Phi}$ to the correlated 
$\ket{\corr{\Phi},J}$ is an $A$-body operator, like in all other approaches. 
Even three body products 
$f(\vec{\op{x}}_i,\vec{\op{x}}_j,\vec{\op{x}}_k,\dots)$ are being used \cite{PiWi01,PiWi04,Irvi81}.

Although very intuitive, the Jastrow ansatz implies a huge numerical effort in calculating
expectation values and matrix elements. Even the norm of $\ket{\corr{\Phi},J}$ cannot
be calculated analytically. As the transformation is not unitary, configuration mixing 
calculations encounter even greater numerical challenges.

%ssssssssssssssssssssssssssssssssssssssssssssssssssssssssssssssssssssssssssssssssssssssssss
\subsection{Projective approaches}
%ssssssssssssssssssssssssssssssssssssssssssssssssssssssssssssssssssssssssssssssssssssssssss

In the past and still nowadays many effective interactions are based on projection methods.
In this case one is looking for an effective Hamiltonian that 
should have the same eigenvalues $E_n$ as the full Hamiltonian $\op{H}$ and its eigenstates 
should be the projections of the exact eigenstates $\ket{\corr{\Psi}_n}$
onto the model space:
%%%%%%%%%%%%%%%%%%%%%%%%%%%%%%%%%%%%%%%%%%%%%%%%%%%%%%%%%%%%%%%%%%%%
\begin{equation}\label{eq:Heff-proj}
\op{P}\,\Heff \ket{{\Phi}_n}\stackrel{!}{=}E_n \ket{{\Phi}_n}\ \
\textnormal{and}\ \ \ket{{\Phi}_n}=\op{P}\,\ket{\corr{\Psi}_n}\ .
\end{equation}
%%%%%%%%%%%%%%%%%%%%%%%%%%%%%%%%%%%%%%%%%%%%%%%%%%%%%%%%%%%%%%%%%%%%%%%%%%
To simplify the equations one writes the Hamiltonian $\op{H}=\op{H}_0+\op{V}$ 
and the effective Hamiltonian $\Heff=\op{H}_0+\op{V}_\mathrm{eff}$
as a sum of $\op{H}_0$ and a residual interaction $\op{V}$ and $\op{V}_\mathrm{eff}$,
respectively. The formal solution for $\Heff$ that fulfills the requirement 
(\ref{eq:Heff-proj}) is actually energy-dependent and given by 
%-----------------------------------------------------------------------
\begin{equation}\label{eq:Veff-proj}
\op{V}_\mathrm{eff}(E)=\op{V}+\op{V}\frac{\op{Q}}{(E-\op{H}_0)}\,\op{V}_\mathrm{eff}(E)\ ,
\end{equation}
%-----------------------------------------------------------------------
where the energy $E$ has to be taken as the unknown eigenvalue $E_n$ of the full
problem. This means that each eigenstate in principle corresponds to a different effective
Hamiltonian. As the difference between $E$ and eigenvalues of $\op{H}_0$ 
that belong to the Q-space enters Eq.~(\ref{eq:Veff-proj}), the energy dependence
will be weak if the lowest Q-space eigenvalue is far away from the considered $E_n$.
Note that $\op{V}_\mathrm{eff}(E)$ in this equation is not
restricted to the P-space but connects P- and Q-space. However, the effective interaction
is actually $\op{P}\op{V}_\mathrm{eff}(E)\op{P}$. The full $\op{V}_\mathrm{eff}(E)$
is needed when one calculates other effective observables (see below).
Again Eq.~(\ref{eq:Veff-proj}) yields an effective Hamiltonian that contains
many-body operators. 

When solving this equation in two-body space one
obtains the Brueckner G-matrix
%-----------------------------------------------------------------------
\begin{equation}\label{eq:G-matrix}
\op{G}(\omega)=\op{V}+\,\op{V}\frac{\op{q}}{(\omega-\op{H}_0)}\,\op{G}(\omega) \ ,
\end{equation}
%-----------------------------------------------------------------------
where the choice of the starting energy $\omega$ and the Pauli projection
operator $\op{q}$ acting in two-body space is not uniquely defined.
After having solved the G-matrix equation (\ref{eq:G-matrix}) one 
uses $\op{G}(\omega)$ in the $A$-body space, which constitutes a low density 
approximation.

Although Eq.~(\ref{eq:Veff-proj}) appears to be suited for a perturbation 
expansion, this is not possible if the strong short-range correlations of the 
nuclear interaction have to be renormalized. At least infinite partial summations 
are needed. The Brueckner G-matrix is such a partial summation, namely over 
the so called ladder diagrams.

Another task (often not considered) is that not only the Hamiltonian should be
transformed to an effective one, but also all other observables $\op{B}$
such that 
%-----------------------------------------------------------------------
\begin{equation}\label{eq:B_eff}
\frac{\matrixe{{\Phi}_n}{\op{B}_\mathrm{eff}}{{\Phi}_m}}
         {\braket{\Phi_n}{\Phi_n}^{1/2}\braket{\Phi_m}{\Phi_m}^{1/2}}
=\frac{\matrixe{\corr{\Psi}_n}{\op{P}\op{B}_\mathrm{eff}\op{P}}{\corr{\Psi}_m}}
         {\braket{\Phi_n}{\Phi_n}^{1/2}\braket{\Phi_m}{\Phi_m}^{1/2}}
\stackrel{!}{=}
\matrixe{\corr{\Psi}_n}{\op{B}}{\corr{\Psi}_m}
\end{equation}
%----------------------------------------------------------------------- 
for the low lying many-body states of interest. This could in principle be done
by using the $A$-body Bethe-Goldstone equation
%-----------------------------------------------------------------------
\begin{equation}\label{eq:Bethe-Goldstone}
\ket{\corr{\Psi}_n}=\Big(\ \op{1}+\frac{\op{Q}}{(E_n-\op{H}_0)}\,
               \op{V}_\mathrm{eff}(E_n)\Big)\ket{{\Phi}_n}\ , 
\end{equation}
%----------------------------------------------------------------------- 
which also follows from Eq.~(\ref{eq:Heff-proj}). It reconstructs the
correlated many-body state $\ket{\corr{\Psi}_n}$ from its projected
part $\ket{\Phi_n}=\op{P}\,\ket{\corr{\Psi}_n}$. 
When inserting Eq.~(\ref{eq:Bethe-Goldstone}) into Eq.~(\ref{eq:B_eff})
to obtain the effective operator one does not need much imagination to see 
that this is a rather impracticable method and 
susceptible to errors in approximations
of the matrix elements of $\op{V}_\mathrm{eff}(E_n)$ that
connect P- and Q-space.
Within the G-matrix approximation
one uses the analoguous Bethe-Goldstone equation, where $\op{V}_\mathrm{eff}(E)$
is replaced by $\op{G}(\omega)$ and $Q$ by $q$. 
The G-matrix method is not discussed further in this contribution.
 
Unitary approaches are more transparent and not energy dependent.
A unitary approach is more robust as there is some freedom in the choice of $\op{U}$,
because only a few low lying eigenvalues and eigenstates of 
$\op{P}\Heff\op{P}=\op{P}\op{U}^{-1}\op{H}\,\op{U}\op{P}$ are needed.
A not so optimal choice of $\op{U}$ can be compensated by a larger 
P-space.
Unitary approaches can even generate effective interactions that are
by construction phase-shift equivalent with the original interaction.

%ssssssssssssssssssssssssssssssssssssssssssssssssssssssssssssssssssssssssssssssssssssssssss
\subsection{Plan of the review}
%ssssssssssssssssssssssssssssssssssssssssssssssssssssssssssssssssssssssssssssssssssssssssss

In Section \ref{sec:ucom}, the concept of the Unitary Correlation Operator Method (UCOM) is
laid out. Central and tensor correlators are introduced and their properties
are illustrated with descriptive examples. Section \ref{sec:srg} discusses the relation to
the Similarity Renormalization Group (SRG) approach, which is more general, but in its
application to nuclear effective interactions it is based on the same physical assumptions.    
From SRG we deduce a set of optimal correlation functions for UCOM and
compare the resulting effective low-momentum interactions. 
Section \ref{sec:ncsm} steps into the 
many-body Hilbert space using the No-Core Shell Model (NCSM) to probe the
effective interactions developed in the preceding sections. 
Properties, like masses, spectra, radii, magnetic dipole and quadrupole moments,  
are investigated for nuclei with mass numbers $A=3-7$.
In Section \ref{sec:hf} we move on to nuclei along the whole nuclear chart
by means of the Hartree-Fock approximation and low-order many-body perturbation
theory.
We show that the UCOM concept is very versatile not only for the understanding
of the nuclear many-body system but provides a general method to 
treat short-range correlations in a quantitative way. We also discuss evidence for
missing three-body forces or density dependences, which seem to be needed 
to obtain the correct nuclear saturation properties for large systems.
Finally we close with summarizing conclusions.

%%%%%%%%%%%%%%%%%%%%%%%%%%%%%%%%%%%%%%%%%%%%%%%%%%%%%%%%%%%%%%%%%%%%%%
%%%%%%%%%%%%%%%%%%%%%%%%%%%%%%%%%%%%%%%%%%%%%%%%%%%%%%%%%%%%%%%%%%%%%%
%%%%%%%%%%%%%%%%%%%%%%%%%%%%%%%%%%%%%%%%%%%%%%%%%%%%%%%%%%%%%%%%%%%%%%
\clearpage
\section{Unitary Correlation Operator Method (UCOM)}
\label{sec:ucom}

\begin{figure}
  \centering
  \includegraphics[width=0.36\textwidth]{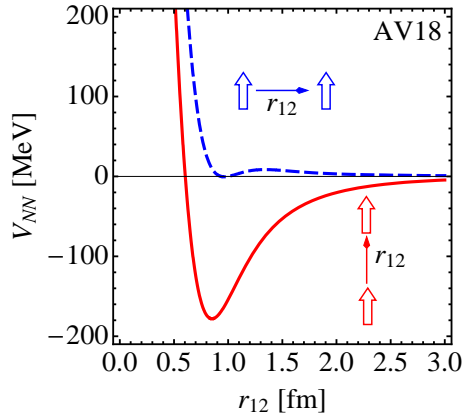}
  \caption{Argonne~V18 potential in the $S=1,T=0$ channel. The lower
    curve is obtained when the spins are aligned parallel to the
    distance vector of two nucleons whereas the upper curve is
    obtained when the spins are oriented perpendicular to the distance
    vector. This spin dependence of the potential is caused by the
    tensor force. The relative momentum $\qV$ is chosen to be zero,
    therefore, the spin-orbit force does not contribute.}
  \label{fig:AV18-projected}
\end{figure}

As discussed in the introduction, short-range central and tensor
correlations pose a major challenge for the solution of the nuclear
many-body problem. These correlations are induced by corresponding
features of realistic nucleon-nucleon forces that are illustrated in
Fig.~\ref{fig:AV18-projected}, where the Argonne~V18 potential is
plotted as a function of the distance of the nucleons and the
orientation of the spins relative to the distance vector. At short
distances, $r < 0.5\fm$, the potential is strongly repulsive, it has a
repulsive ``core'' which induces short-range central correlations. At
larger distances the potential shows a pronounced dependence on the
orientation of the spins. If the spins are aligned perpendicular to
the distance vector the potential is almost flat, whereas the
potential is attractive with a minimum at $r \approx 1.0\fm$ if the
spins are aligned parallel to the distance vector. This difference is
caused by the tensor force, which originates mainly from the one-pion
exchange part of the potential.

In the Unitary Correlation Operator Method
\cite{ucom05,ucom04,ucom03,ucom98} these correlations are imprinted
into a many-body state by means of a unitary correlation operator
$\CO$:
\eq{ \label{eq:corr_state}
  \ket{\corr{\Psi}} = \CO\; \ket{\Psi} \;.
}
Here the uncorrelated many-body state $\ket{\Psi}$ is a ``simple''
state that cannot represent the short-range correlations. This can be
a Hartree-Fock Slater determinant or a basis state of the No-Core
Shell Model.

Short-range correlations are important when the particles come
close. If the density is low enough, the probability for finding three
nucleons within the correlation volume defined by the correlation
range will be small. In that case the assumption that the short-range
correlations are of two-body nature will be a good approximation. At
nuclear saturation density $\rho_0 = 0.17\fm^{-3}$ the mean distance
between nucleons is $1.8\fm$ while the repulsive core sets in around
$0.5\fm$. Therefore, we assume that the correlations are essentially
independent of the environment and we use a state-independent ansatz
for the correlation operator $\CO$. This implies that the short-range
correlations for low-lying states are very similar in all nuclei.  The
unitary correlation operator $\CO$ describing this transformation is
given in an explicit operator form, independent of a particular
representation or model space. The correlation operator is optimized
for the lowest orbital angular momentum $L$ in each spin-isospin
channel but it does not explicitly depend on $L$. The UCOM approach
can, therefore, be also used in many-body approaches that do not use
basis states of good angular momentum. Alternatively, one could define
different correlation operators for each partial-wave channel. In the
$S$-wave UCOM approach proposed by Myo and Toki \cite{myo09} only
central correlations in $L=0$ channels are considered.

When calculating expectation values or matrix elements of some
operator $\AO$ we can either evaluate the bare operator $\AO$ in the
correlated states $\ket{\corr{\Psi}}$ or we can use a correlated
operator $\corr{\AO}$ defined through a similarity transformation
\eq{ \label{eq:corr_operator}
  \corr{\AO}
  = \CO^{-1}\AO\,\CO
  = \CCO\AO\,\CO
}
with the uncorrelated states.
\eq{ \label{eq:corr_matrixelem}
  \matrixe{\corr{\Psi}}{\AO}{\corr{\Psi}'}
  = \matrixe{\Psi}{\CCO\AO\CO}{\Psi'}
  = \matrixe{\Psi}{\corr{\AO}}{\Psi'} \;.
}
Due to the unitarity of $\CO$, the notions of correlated states and
correlated operators are equivalent and we may choose the form that is
technically more advantageous.

In the case of the nuclear many-body problem, the unitary correlation
operator $\CO$ has to account for short-range central and tensor
correlations as explained above. In the UCOM approach we explicitly
disentangle these different types of correlations and define the
correlation operator as a product of two unitary operators,
\eq{
  \CO 
  = \CO_{\Omega} \CO_{r} \;,
}
where $\CO_{\Omega}$ describes short-range tensor correlations and
$\CO_{r}$ central correlations. Each of these unitary operators is
expressed with hermitian two-body generators
\eq{ \label{eq:correlator}
  \CO_{\Omega} 
  =  \exp\!\Big[-\ii \sum_{i<j} \gO_{\Omega,ij} \Big] \;,\quad
  \CO_{r}
  = \exp\!\Big[-\ii \sum_{i<j} \gO_{r,ij} \Big]\;.
}
The details of the generators $\gO_r$ and $\gO_{\Omega}$ will depend
on the particular nucleon-nucleon interaction under consideration.

\subsection{Central correlations}

The central correlations are induced by the repulsive core of the
central part of the interaction which tries to keep the nucleons apart
from each other. The two-body density in the correlated many-body
state will be strongly suppressed at short interparticle distances,
i.e. in the range of the repulsive core, and it will be enhanced at
larger distances, where the potential is attractive. This can be
achieved by a distance-dependent shift in the relative wave function
for each pair of nucleons. The generator $\gO_r$ is constructed such
that it performs these shifts in a unitary way. The shifts are
generated by the projection of the relative momentum $\qOV =
\frac{1}{2} [\pOV_1 - \pOV_2]$ onto the distance vector $\rOV = \xOV_1
- \xOV_2$ of two nucleons:
\eq{
  \qO_r 
  = \frac{1}{2} \Big(\,\frac{\rOV}{\rO}\cdot\qOV 
    + \qOV\cdot\frac{\rOV}{\rO} \,\Big) 
}
with $\rO = |\rOV|$. The amplitude of the shift---large shifts at small distances within
the core, small or no shifts outside the core---is described by a
function $s_{ST}(r)$ for each spin-isospin channel. The detailed form
of the function $s_{ST}(r)$ will depend on the potential. Its
determination will be discussed in detail in
Sec.~\ref{sec:optcorr}. The full generator for the central
correlations is written in a hermitized form as
\eq{ \label{eq:central_generator}
  \gO_r 
  = \sum_{S,T} \frac{1}{2} 
    [ \qO_r\, s_{ST}(\rO) + s_{ST}(\rO)\, \qO_r]\, \PiO_{ST}\;,
}
where $\PiO_{ST}$ is the projection operator onto two-body spin $S$
and isospin $T$. Similar generators were already used by Ristig et al. \cite{Rist67,RiKi68,Rist70} for the description of the central correlations induced by hard-core potentials.

\subsection{Tensor correlations}

The correlations induced by the tensor force entangle the orientation
of the spins with the spatial distribution of the nucleons to
optimize the contribution of the tensor force. We construct the tensor
correlation operator in such a way that it will only act on the
orbital part of the relative wave function of two nucleons. This can
be achieved by using only the ``orbital part'' $\qOV_{\Omega}$ of the
relative momentum operator, which is obtained by subtracting the radial
part of the relative momentum operator (used in the generator
for the central correlations) from the full relative momentum operator
\eq{
  \qOV_{\Omega} 
  = \qOV - \frac{\rOV}{\rO}\cdot\qO_r
  = \frac{1}{2\rO^2}(\LOV\times\rOV - \rOV\times\LOV) 
}
with $\LOV = \rOV\times\qOV$. Like the tensor operator $\tensorRRO$ in the tensor force, the
generator for the tensor correlations is the scalar product of an
operator of rank 2 in coordinate space (constructed from the relative
distance vector $\rOV$ and the ``orbital momentum'' $\qOV_{\Omega}$)
with the operator of rank 2 in spin-space \cite{ucom03}. As for the
central correlator the amplitude of the shifts will depend on the
distance of the nucleons and the potential under consideration. The
amplitude is given by the tensor correlation function
$\vartheta_T(r)$.  In cartesian notation we can write the full
generator as
\eq{ \label{eq:tensor_generator}
  \gO_{\Omega} 
  = \sum_{T} \vartheta_T(r)\; \tensorRQO\; \PiO_{1T} 
}
using the general definition for a tensor operator of rank 2
\eq{
  \tensorO(\aOV,\bOV)
  =\tfrac{3}{2} \big[(\sigmaOV_1\!\cdot\aOV)(\sigmaOV_2\!\cdot\bOV)
     + (\sigmaOV_1\!\cdot\bOV)(\sigmaOV_2\!\cdot\aOV)\big] \\
  - \tfrac{1}{2}(\sigmaOV_1\!\cdot\!\sigmaOV_2) (\aOV\cdot\bOV +
    \bOV\cdot\aOV) \;.
}
%%

%%%%%%%%%%%%%%%%%%%%%%%%%%%%%%%%%%%%%%%%%%%%%%%%%%%%%%%%%%%%%%%%%%%%%%
%%%%%%%%%%%%%%%%%%%%%%%%%%%%%%%%%%%%%%%%%%%%%%%%%%%%%%%%%%%%%%%%%%%%%%
\bigskip
\subsection{Correlated wave functions}
\label{sec:corr_wavefunc}

As explained above, because the correlation operator acts only on the relative
motion of a nucleon pair, the center-of-mass motion of the nucleon
pair is not affected.  When discussing correlated two-body wave
functions we can, therefore, restrict ourselves to the relative wave
functions. For the uncorrelated relative wave function we assume
$LS$-coupled angular momentum eigenstates $\ket{\phi(LS)J M\; T M_T}$
with the radial wave function $\phi(r)$. The correlation operators do
not depend on $M$ and $M_T$ and we will omit these quantum numbers in
the following.

The central correlator $\CO_{r}=\exp(-\ii\, \gO_{r})$ %%
affects only the radial part of the wave function and leaves the
orbital part of the wave function and the spin and isospin unchanged. In
coordinate representation the correlated wave function can be
rewritten as a norm-conserving coordinate transformation \cite{ucom98}
\eqmulti{ \label{eq:corr_central_states} 
  \matrixe{r(L'S)JT}{\CO_r}{\phi(LS)JT}
  &= \frac{\Rm(r)}{r}\sqrt{\DRm(r)}\; \phi(\Rm(r)) \delta_{L'L} \\
  \matrixe{r(L'S)JT}{\CCO_r}{\phi(LS)JT}
  &= \frac{\Rp(r)}{r}\sqrt{\DRp(r)}\; \phi(\Rp(r)) \delta_{L'L} \;,
}
where $\Rp(r)$ and $\Rm(r)$ are mutually inverse,
$\Rpm(\Rmp(r))=r$. The correlation functions $\Rp(r)$ and $\Rm(r)$ are
related to the function $s(r)$ used in the generator
\eqref{eq:central_generator} through the integral equation
\eq{
  \int_{r}^{\Rpm(r)} \frac{\dd{\xi}}{s(\xi)} = \pm 1 \;.
}
For illustrative purposes the correlation functions can be
approximated as $\Rpm(r) \approx r \pm s(r)$.

In the $LS$-coupled basis the application of the tensor correlator
$\CO_{\Omega}$ can be expressed easily. The tensor operator
$\tensorRQO$ used in the generator has only
off-diagonal matrix elements in the $LS$-coupled basis
\eq{
  \matrixe{(J\pm1,1)JT}{\tensorRQO}{(J\mp1,1)JT} = \pm 3\ii \sqrt{J(J+1)} \;.
}
Within a subspace of fixed $J$ one can, therefore, calculate the matrix
exponential and thus the matrix elements of the full tensor correlator
$\CO_{\Omega}$.

The tensor correlation operator will have no effect for states with
$L=J$, whereas states with $L=J\pm1$ will be connected to states with $L=J\mp1$. The strength of the mixing is governed by $\vartheta(r)$
\eq{ \label{eq:corr_tensor_states3}
  \matrixe{r(L'S)JT}{\CO_{\Omega}}{\phi(LS)JT} =
  \begin{cases}
    \phi(r) & ; L'=L=J\\
    \cos \theta_J(r) \phi(r) & ; L'=L=J\pm1\\
    \pm \sin \theta_J(r) \phi(r) & ; L'=J\pm1,L=J\mp1
  \end{cases} \;,
}
where we use the abbreviation
\eq{
  \theta_J(r) 
  = 3 \sqrt{J(J+1)}\; \vartheta(r) \;.
}   

Combining central and tensor correlations we end up with the following
expression for the fully correlated wave function in coordinate space:
\eq{ \label{eq:corr_states} 
\matrixe{r(L'S)JT}{\CO_{\Omega}
    \CO_r}{\phi(LS)JT} =
  \begin{cases}
    \frac{\Rm(r)}{r}\sqrt{\DRm(r)} \; \phi(\Rm(r)) & ; L'=L=J\\
    \cos \theta_J(r) \frac{\Rm(r)}{r}\sqrt{\DRm(r)} \; \phi(\Rm(r)) & ; L'=L=J\pm1\\
    \pm \sin \theta_J(r) \frac{\Rm(r)}{r}\sqrt{\DRm(r)} \; \phi(\Rm(r)) & ; L'=J\pm1,L=J\mp1
  \end{cases} \;.
}

The wave function in momentum space can be obtained by Fourier
transformation of the coordinate space wave function
\eqref{eq:corr_states}
\eq{ \label{eq:corr_states_momentum} \matrixe{q(L'S)JT}{\CO_{\Omega}
    \CO_r}{\phi(LS)JT} = \sqrt{\frac{2}{\pi}} \ii^{L'} \int dr r^2
  j_{L'}(q r) \matrixe{r(L'S)JT}{\CO_{\Omega} \CO_r}{\phi(LS)JT} \: ,
} 
where we use momentum eigenstates normalized as
$\braket{q(LS)JT}{q'(L'S)JT}=\frac{1}{q^2}\delta(q-q')\delta_{L'L}$.
%%

%%%%%%%%%%%%%%%%%%%%%%%%%%%%%%%%%%%%%%%%%%%%%%%%%%%%%%%%%%%%%%%%%%%%%%
\begin{figure}[t]
\centering\includegraphics[width=0.85\textwidth]{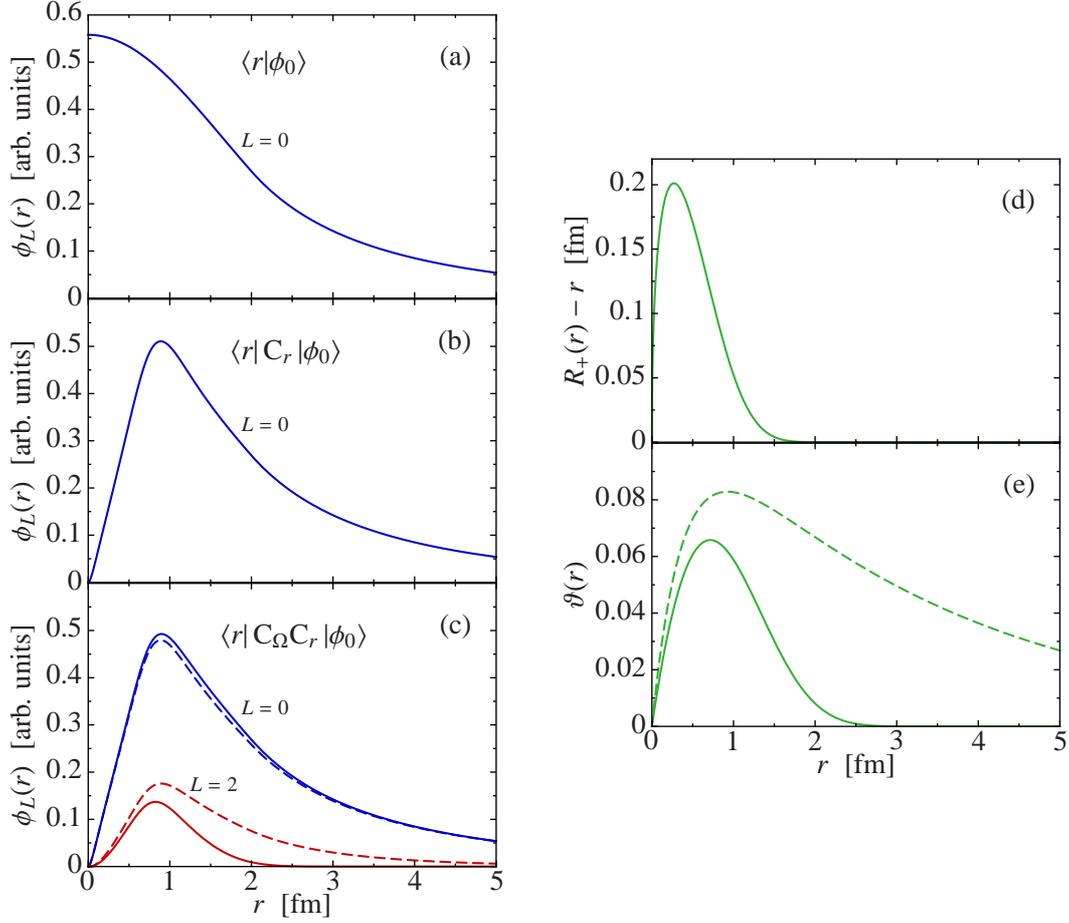}
\caption{Central and tensor correlations in the deuteron
    channel. Starting from an uncorrelated trial wave function (a), first the
    central (b) and then also the tensor correlated wave functions (c) are
    shown together with the corresponding central (d) and tensor
    correlation functions (e).}
\label{fig:deuteron_illu}
\end{figure}
%%%%%%%%%%%%%%%%%%%%%%%%%%%%%%%%%%%%%%%%%%%%%%%%%%%%%%%%%%%%%%%%%%%%%%

The deuteron wave function provides an illuminating example for the
role of central and tensor correlations. We will start from an
uncorrelated state $\ket{\phi_0(LS)JT} =\ket{\phi_0(01)10}$, which is a
pure $S$-wave state with the spin-isospin quantum numbers of the
deuteron. The radial wave function $\phi_0(r)$ shall not contain
short-range correlations induced by the repulsive
core. Figure \ref{fig:deuteron_illu} shows the uncorrelated $L=0$ radial
wave function. Applying the central correlator $\CO_r$ with the
correlation function $\Rp(r)$ leads to a wave function with a
correlation hole at short interparticle distances. The application of
the tensor correlation operator $\CO_{\Omega}$ in a second step
generates the $D$-wave component in the wave function, which depends
on the tensor correlation function $\vartheta(r)$. If we assume an
uncorrelated wave function that is purely $S$-wave, the entire
$D$-wave component has to be generated by the tensor correlator. This
can be achieved using a very long-ranged tensor correlation function
(as given by the dashed curves in Fig.~\ref{fig:deuteron_illu}). But
this is not the idea of the unitary correlation operator method, where
we only want to describe the state-independent short-range
correlations by means of the correlation operators. Long-range
correlations, like those governing the outer part of the deuteron wave
function, should be described explicitly by the many-body
approach. The solid curves show the $D$-wave component obtained with
such a short-range tensor correlator ($I_{\vartheta}=0.09 \fm^3$ as
discussed in Sec.~\ref{sec:optcorr}).

%%%%%%%%%%%%%%%%%%%%%%%%%%%%%%%%%%%%%%%%%%%%%%%%%%%%%%%%%%%%%%%%%%%%%%
%%%%%%%%%%%%%%%%%%%%%%%%%%%%%%%%%%%%%%%%%%%%%%%%%%%%%%%%%%%%%%%%%%%%%%
\bigskip
\subsection{Correlated operators}
\label{sec:corr_op}

\subsubsection*{Correlated Hamiltonian -- Central Correlations}
\label{sec:corr_op_central}

In the two-body system the unitary transformation with the central
correlator $\CO_r$ can be expressed analytically for any operator
that can be written as a function of relative distance $\rOV$ and
relative momentum $\qOV$.

The most important example is a Hamiltonian consisting of kinetic
energy and a realistic nucleon-nucleon interaction given in a generic
operator form
\eq{ \label{eq:corr_op_potgeneric}
  \VO
  = \sum_p \frac{1}{2} [V_p(\rO) \OO_p + \OO_p V_p(\rO)]
}
with 
\eqmulti{ \label{eq:corr_op_potgenericop}
  \OO_p 
  = \{&1,\;\sigmasigmaO,\;\qO_r^2,\;\qO_r^2 \sigmasigmaO,\;
    \LOV^2,\;\LOV^2 \sigmasigmaO,\\
    &\spinorbitO,\;\tensorRRO,\;\tensorLLO\} \otimes \{1,\; \tautauO\} \;.
}
Here we only consider non-local terms up to quadratic momentum
dependence. Such terms appear, e.g., in the Nijmegen \cite{StKl94} or
operator representations of the Bonn A/B potentials \cite{Mach89}. A
quadratic momentum dependence of the form $\qOV^2$ can be expressed by
the $\qO_r^2$ and $\orbitsqrO$ terms contained in
\eqref{eq:corr_op_potgeneric}. Charge dependent terms in the
interaction are not explicitly discussed, but are included in the
calculation of matrix elements and the many-body calculations based on
them.
 
For the formulation of the correlated Hamiltonian in two-body space,
we start with an initial Hamiltonian given by
\eq{
  \HO = \HO_\intr + \TO_\cm = \TO_\intr + \VO + \TO_\cm = 
  \TO_r + \TO_\Omega + \VO + \TO_\cm \;,
}  
where we have decomposed the kinetic energy operator $\TO$ into a
center of mass contribution $\TO_\cm$ and an intrinsic contribution
$\TO_\intr$ which in turn is written as a sum of a radial and an
angular part
\eq{
  \TO_{r}
  = \frac{1}{2\mu} \qO_r^2 \;,\quad
  \TO_{\Omega}
  = \frac{1}{2\mu} \frac{\orbitsqrO}{\rO^2} \;.
}
Applying the central correlator $\CO_r$ in two-body space leads to a
correlated Hamiltonian consisting of the one-body kinetic energy $\TO$
and two-body contributions for the correlated radial and angular
relative kinetic energy, $\corr{\TO}_r^{[2]}$ and
$\corr{\TO}_{\Omega}^{[2]}$, respectively, as well as the correlated
two-body interaction $\corr{\VO}^{[2]}$
\eq{
  \CCO_r\, \HO\, \CO_r
  = \TO +  \corr{\TO}_r^{[2]} + \corr{\TO}_{\Omega}^{[2]}
    + \corr{\VO}^{[2]} \;.
}
The explicit operator form of the correlated terms can be derived from
a few basic identities. The unitary transformation for the relative
distance operator $\rO$ results in the operator-valued function
$\Rp(\rO)$
\eq{
  \CCO_r\, \rO\,\CO_r
  = \Rp(\rO) \;.
}
Because of the unitarity of the correlation operator $\CO_r$ an
arbitrary function of $\rO$ transforms as
\eq{ \label{eq:corr_op_funcr}
  \CCO_r\, f(\rO)\,\CO_r
  = f(\CCO_r\, \rO\,\CO_r)
  = f(\Rp(\rO)) \;.
}
The interpretation of the unitary transformation in terms of a
norm-conserving coordinate transformation $r \mapsto \Rp(r)$ is
evident.  For the radial momentum operator $\qO_r$ one finds the
following correlated form \cite{ucom98}
\eq{ \label{eq:corr_op_reldist}
  \CCO_r\, \qO_r\,\CO_r
  = \frac{1}{\sqrt{\DRp(\rO)}}\;\qO_r\;\frac{1}{\sqrt{\DRp(\rO)}} \;.
}
With this we can express $\qO_r^2$ which enters the radial part of the
relative kinetic energy as
\eq{ \label{eq:corr_op_relmomsqr}
  \CCO_r\, \qO_r^2\,\CO_r
  = \frac{1}{2} \Big( \qO_r^2
    \frac{1}{\DRp(\rO)^2} + \frac{1}{\DRp(\rO)^2} \qO_r^2 \Big) + W(\rO) \; ,
}
which consists of a transformed momentum dependence plus an additional
local term depending only on the correlation function $\Rp(r)$
\eq{
  W(r)
  = \frac{7 \DDRp(r)^2}{4 \DRp(r)^4} - \frac{\DDDRp(r)}{2\DRp(r)^3} \;.
}   
All other basic operators, such as $\orbitsqrO$, $\spinorbitO$,
$\tensorRRO$ commute with the correlation operator $\CO_r$ and are,
therefore, unchanged by the central correlations.

Based on these elementary relations we can explicitly construct the
two-body contributions to the correlated kinetic energy. For the
radial part we obtain using \eqref{eq:corr_op_relmomsqr}
\eq{ \label{eq:corr_op_kinetic_rad}
  \corr{\TO}_r^{[2]} = \CCO_r \TO_r \CO_r - \TO_r
  = \frac{1}{2} \Big( \qO_r^2 \frac{1}{2\mu_r(\rO)} + \frac{1}{2\mu_r(\rO)} \qO_r^2 \Big)
    + \frac{1}{2\mu} W(\rO)
}
with a distance-dependent effective mass term
\eq{
  \frac{1}{2\mu_r(r)} 
  = \frac{1}{2\mu} \Big( \frac{1}{\DRp(r)^2} - 1 \Big) \;.
}
The two-body contribution to the correlated angular part of the
kinetic energy involves only the basic relation
\eqref{eq:corr_op_reldist} and gives
\eq{ \label{eq:corr_op_kinetic_ang}
  \corr{\TO}_{\Omega}^{[2]}
  = \CCO_r \TO_{\Omega} \CO_r - \TO_{\Omega}
  = \frac{1}{2\mu_{\Omega}(\rO)} \frac{\orbitsqrO}{\rO^2}
}
with a distance-dependent angular effective mass term
\eq{
  \frac{1}{2\mu_{\Omega}(r)} 
  = \frac{1}{2\mu} \Big( \frac{r^2}{\Rp(r)^2} - 1 \Big) \;.
}		   

The momentum dependent terms of the nucleon-nucleon interaction
\eqref{eq:corr_op_potgeneric} transform in a similar manner like the
kinetic energy. Using \eqref{eq:corr_op_reldist} and
\eqref{eq:corr_op_relmomsqr} we obtain
\eq{
  \CCO_r\; \frac{1}{2}\Big( \qO_r^2 V(\rO) + V(\rO) \qO_r^2 \Big) \; \CO_r
    = \frac{1}{2}\Big( \qO_r^2 \frac{V(\Rp(\rO))}{\DRp(r)^2} 
       + \frac{V(\Rp(\rO))}{\DRp(r)^2} \qO_r^2 \Big) +
  V(\Rp(\rO))\; W(\rO) - V'(\Rp(\rO)) \frac{\DDRp(r)}{\DRp(r)^2} \;.
}  
For all other terms of the NN-interaction
\eqref{eq:corr_op_potgeneric} the operators $\OO_p$ commute with the
generator $\qO_r$ and we only have to transform the radial
dependencies
\eq{  \label{eq:corr_op_localpot}
  \CCO_r\; V(\rO)\,\OO_p\; \CO_r
  = V(\Rp(\rO))\, \OO_p \;.
}
Many of the other relevant operators, e.g. the quadratic radius or
transition operators, can be transformed just as easily.

%%%%%%%%%%%%%%%%%%%%%%%%%%%%%%%%%%%%%%%%%%%%%%%%%%%%%%%%%%%%%%%%%%%%%%
\subsubsection*{Correlated Hamiltonian -- Tensor Correlations}
\label{sec:corr_op_tensor}

The transformation of the Hamiltonian with the tensor correlation
operator $\CO_{\Omega}$ is more involved. In general, it can be
evaluated via the Baker-Campbell-Hausdorff expansion
\eq{ \label{eq:corr_tensor_bch}
  \CCO_{\Omega}\, \AO\, \CO_{\Omega} =
    \AO + \ii [\gO_{\Omega}, \AO]
    + \frac{\ii^2}{2} [\gO_{\Omega},[\gO_{\Omega}, \AO]] 
    + \cdots \;.
}
In some cases the series expansion will terminate after a finite
number of terms.  A trivial case is the distance operator $\rO$, which
commutes with the tensor generator $\gO_{\Omega}$ and is thus
unchanged by the transformation
\eq{
  \CCO_{\Omega}\, \rO\, \CO_{\Omega} = \rO \;.
}
For the radial momentum operator $\qO_r$, the expansion
\eqref{eq:corr_tensor_bch} terminates after the first order
commutators and we obtain the simple expression
\eq{
  \CCO_{\Omega}\, \qO_r\, \CO_{\Omega} = \qO_r - \vartheta'(\rO)\,\tensorRQO \;.
}
For the tensor correlated quadratic radial momentum operator the
series terminates after the first two terms and we obtain
\eqmulti{ \label{eq:corrT_op_momentumsqr}
  \CCO_{\Omega}\, \qO_r^2\, \CO_{\Omega}
  = \qO_r^2 - [\vartheta'(\rO)\, \qO_r + \qO_r\, \vartheta'(\rO)]\, \tensorRQO 
  + [\vartheta'(\rO)\,\tensorRQO]^2 \;,
}
where $\tensorRQO^2 = 9 [\SOV^2 + 3\spinorbitO + \spinorbitO^2$]. By
applying the tensor correlator to the kinetic energy we have
generated momentum-dependent tensor operators as well as
``conventional'' spin-orbit and tensor terms (from the $\spinorbitO^2$
term). The correlated kinetic energy is no longer a central operator.

For all other operators of the interaction
\eqref{eq:corr_op_potgeneric} that depend on angular momentum, the
Baker-Campbell-Hausdorff series does not terminate. Through the
commutators additional tensor operators are generated. At first order
we obtain
\eqmulti{
  \comm{\gO_{\Omega}}{\tensorRRO}
  &= \ii \vartheta(\rO) [-24\, \PiO_1 - 18\, \spinorbitO + 3\, \tensorRRO] \\
  \comm{\gO_{\Omega}}{\spinorbitO}
  &= \ii \vartheta(\rO) [-\tensorbarQQO] \\
  \comm{\gO_{\Omega}}{\orbitsqrO}
  &= \ii \vartheta(\rO) [2\,\tensorbarQQO] \\
  \comm{\gO_{\Omega}}{\tensorLLO}
  &= \ii \vartheta(\rO) [7\,\tensorbarQQO] \;,
}
where we use the abbreviation
\eq{
  \tensorbarQQO
  = 2\rO^2 \tensorQQO + \tensorLLO - \tfrac{1}{2}\,\tensorRRO \;.
}
The next order generates terms of higher order in orbital angular
momentum, e.g., a $\orbitsqrO\spinorbitO$ term. In practice we have to
truncate the Baker-Campbell-Hausdorff expansion to some finite set of
operators \cite{ucom04}. In principle the contributions of the
higher-order operators will become more important with increasing
angular momenta. On the other hand the contributions of the correlated
Hamiltonian, which are of short range, will be overwhelmed by the
centrifugal barrier from the one-body kinetic energy for large angular
momenta.

Note that matrix elements of the correlated Hamiltonian using angular
momentum eigenstates are calculated by applying the tensor correlation
operator onto the basis states (see Sec.~\ref{sec:corr_wavefunc}), which
does not require approximations.

\subsection{Optimal correlation functions}
\label{sec:optcorr}

The central and tensor correlators depend on the correlation functions
$s(r)$ and $\vartheta(r)$ in the different spin- and
isospin-channels. We now have to determine these correlation functions
for a given nuleon-nucleon potential. One important question, that was
already raised in the discussion of the deuteron wave function, is the
separation between state-independent short-range correlations, which
we want to describe by the correlation operator, and long-range
correlations that should be described explicitly by the many-body
approach.

The most convenient procedure to determine the correlation functions
is based on an energy minimization in the two-body system
\cite{ucom05,ucom03}. For each combination of spin $S$ and isospin $T$
we compute the expectation value of the correlated energy using a
trial state with the lowest possible orbital angular momentum $L$. The
uncorrelated radial wave function should not contain any of the
short-range correlations, i.e., it should resemble the short-range
behavior of a non-interacting system. In the following we will use a
free zero-energy scattering solution $\phi_L(r)\propto r^L$. Other
uncorrelated trial wave functions, e.g. harmonic oscillator eigenfunctions, give very
similar results.

For practical reasons the correlation functions are represented by
parametrizations with typically three variational parameters. The
drop-off can be well-described by a double-exponential decay with
variable range.  For the short-range behavior, several different
parametrizations have been compared. For the Argonne~V18 potential,
the following two parametrizations for the central correlation
functions have proven appropriate:
\eqmulti{  \label{eq:ucom_central_param}
  \Rp^{\text{I}}(r)
  &= r + \alpha\, (r/\beta)^{\eta} \exp[-\exp(r/\beta)]  \;,\\
  \Rp^{\text{II}}(r)
  &= r + \alpha\, [1 - \exp(-r/\gamma)] \exp[-\exp(r/\beta)] \;.
}
Which of these parametrizations is best suited for a particular
channel will be decided on the basis of the minimal energy alone. All
parametrizations allow only outward shifts by construction. This is
different from the correlation functions that are obtained by the SRG
mapping procedure (see Sec.~\ref{sec:srg_ucomsrg_av18}). For the
tensor correlation functions the following parametrization is used
\eq{ \label{eq:ucom_tensor_param}
  \vartheta(r)
  = \alpha\, [1 - \exp(-r/\gamma)] \exp[-\exp(r/\beta)] \;.
}

The $S=0$ channels are only affected by the central correlators.
Their parameters are determined by minimizing the energy for the
lowest possible orbital angular momentum state, i.e., $L=1$ for $T=0$
and $L=0$ for $T=1$, respectively,
\eqmulti{
  E_{00} &= \matrixe{\phi_1 (10)1 0}{\CCO_r\, \HO_\intr\, \CO_r}{\phi_1 (10)1 0} \;,\\
  E_{01} &= \matrixe{\phi_0 (00)0 1}{\CCO_r\, \HO_\intr\, \CO_r}{\phi_0 (00)0 1} \;.\\
}
For $S=0, T=1$ the minimization of $E_{01}$ by variation of the
parameters of the central correlation function is straightforward. The
resulting parameters are summarized in
Table~\ref{tab:corr_centralpara}. For $S=0, T=0$ the potential is
purely repulsive and, therefore, the energy minimization leads to
central correlation functions of very long range. In order to avoid
this pathology we employ a constraint on the strength of the
correlation function given by
\eq{
  I_{\Rp}
  = \int\dd{r}\, r^2\, (\Rp(r)-r) \;.
}
The value of this constraint on the central correlation function for
the $S=0$, $T=0$ channel is fixed to $I_{\Rp}=0.1\text{fm}^{4}$ in
accordance with typical values in other channels.

%%%%%%%%%%%%%%%%%%%%%%%%%%%%%%%%%%%%%%%%%%%%%%%%%%%%%%%%%%%%%%%%%%%%%%
\begin{table}
  \centering\small
  \begin{tabular}{c c | c | c c c c}
    $S$ & $T$ & Param. &  $\alpha$ [fm] & $\beta$ [fm] & $\gamma$ [fm] & $\eta$ \\
    \hline
    0 & 0 & II &  0.7971 &  1.2638  &  0.4621  &  ---     \\
    0 & 1 & I  &  1.3793 &  0.8853  &  ---     &  0.3724 \\
    1 & 0 & I  &  1.3265 &  0.8342  &  ---     &  0.4471 \\
    1 & 1 & II &  0.5665 &  1.3888  &  0.1786  &  ---     \\
  \end{tabular}
  \caption{Parameters of the central correlation functions $\Rp(r)$ specified in \eqref{eq:ucom_central_param} for
    the Argonne~V18 potential obtained from two-body energy minimization.}
  \label{tab:corr_centralpara}
\end{table}
%%%%%%%%%%%%%%%%%%%%%%%%%%%%%%%%%%%%%%%%%%%%%%%%%%%%%%%%%%%%%%%%%%%%%%
%%%%%%%%%%%%%%%%%%%%%%%%%%%%%%%%%%%%%%%%%%%%%%%%%%%%%%%%%%%%%%%%%%%%%%
\begin{figure}
  \begin{center}
    \includegraphics[width=0.45\textwidth]{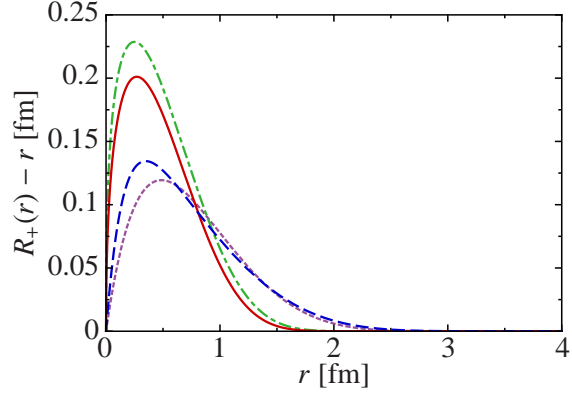}
  \end{center}
  \vskip-4ex
  \caption{Optimal central correlation functions
    $\Rp(r)-r$ for the Argonne~V18 potential according to the parameters
    given in Tab. \ref{tab:corr_centralpara}. The curves correspond
    to the different spin-isospin channels: $S=0,T=1$ 
    (\linemediumdashdot[FGGreen]), $S=1,T=0$ (\linemediumsolid[FGRed]),  $S=0,T=0$
    (\linemediumdotted[FGViolet]), and $S=1,T=1$ (\linemediumdashed[FGBlue]).}
  \label{fig:corr_central}
\end{figure}
%%%%%%%%%%%%%%%%%%%%%%%%%%%%%%%%%%%%%%%%%%%%%%%%%%%%%%%%%%%%%%%%%%%%%%

For $S=1$ we also have to consider tensor correlations and the
parameters of the central and the tensor correlation functions have to
determined simultaneously. For $T=0$ the energy is defined by the
matrix element with $L=0$ states
\eq{
  E_{10}
  = \matrixe{\phi_0 (01)1 0}{\CCO_r \CCO_{\Omega}\, \HO_\intr\, \CO_{\Omega}
      \CO_r}{\phi_0 (01)1 0} \;.
}
In the $T=1$ channel the lowest possible orbital angular momentum is
$L=1$. From angular momentum coupling we obtain $0,1,$ and $2$ as
possible values for $J$. Therefore, we define the energy functional
used for the minimization procedure by averaging over the angles,
which is the sum over all three possibilities with relative weights
given by $2J+1$
\eqmulti{
  E_{11}
  &= \tfrac{1}{9}\matrixe{\phi_1 (11)0 1}{\CCO_r\, \HO_\intr\, \CO_r}{\phi_1 (11)0 1} \\
    &+ \tfrac{3}{9}\matrixe{\phi_1 (11)1 1}{\CCO_r\, \HO_\intr\, \CO_r}{\phi_1 (11)1 1} \\
    &+ \tfrac{5}{9}\matrixe{\phi_1 (11)2 1}{\CCO_r\CCO_{\Omega}\, \HO_\intr\,
    \CO_{\Omega}\CO_r}{\phi_1 (11)2 1} \;.
}
%%

%%%%%%%%%%%%%%%%%%%%%%%%%%%%%%%%%%%%%%%%%%%%%%%%%%%%%%%%%%%%%%%%%%%%%%
\begin{table}
  \centering\small
  \begin{tabular}{c c c c | c c c c}
    \multicolumn{4}{c}{$T=0$} & \multicolumn{4}{c}{$T=1$} \\
    $I_{\vartheta}$ [fm${}^3$] &  $\alpha$ & $\beta$ [fm] & $\gamma$ [fm] &
    $I_{\vartheta}$ [fm${}^3$] &  $\alpha$ & $\beta$ [fm] & $\gamma$ [fm] \\
    \hline
    0.03 &  491.32  &  0.9793  &  1000.0  & -0.01 &  -0.1036  &  1.5869  &  3.4426  \\
    0.04 &  521.60  &  1.0367  &  1000.0  & -0.02 &  -0.0815  &  1.9057  &  2.4204  \\
    0.05 &  539.86  &  1.0868  &  1000.0  & -0.03 &  -0.0569  &  2.1874  &  1.4761  \\
    0.06 &  542.79  &  1.1360  &  1000.0  & -0.04 &  -0.0528  &  2.3876  &  1.2610  \\
    0.07 &  543.21  &  1.1804  &  1000.0  & -0.05 &  -0.0463  &  2.6004  &  0.9983  \\
    0.08 &  541.29  &  1.2215  &  1000.0  & -0.06 &  -0.0420  &  2.7984  &  0.8141  \\
    0.09 &  536.67  &  1.2608  &  1000.0  & -0.07 &  -0.0389  &  2.9840  &  0.6643  \\
    0.10 &  531.03  &  1.2978  &  1000.0  & -0.08 &  -0.0377  &  3.1414  &  0.6115  \\
    0.11 &  524.46  &  1.3333  &  1000.0  & -0.09 &  -0.0364  &  3.2925  &  0.5473  \\
    0.12 &  517.40  &  1.3672  &  1000.0  & -0.10 &  -0.0353  &  3.4349  &  0.4997  \\
    0.15 &  495.99  &  1.4610  &  1000.0 \\
    0.20 &  450.67  &  1.6081  &  1000.0 \\
    0.30 &  408.40  &  1.8240  &  1000.0 \\
  \end{tabular}
  \caption{Parameters of the tensor correlation functions $\vartheta(r)$ defined in \eqref{eq:ucom_tensor_param} for
    the Argonne~V18 potential with different values $I_{\vartheta}$ for the
    range constraint obtained from two-body energy minimization.}
  \label{tab:corr_tensorpara}
\end{table}

As mentioned earlier, the long-range character of the tensor force
leads to long-range tensor correlations. However, long-range tensor
correlation functions are not desirable for several reasons:
(\emph{i}) The optimal long-range behavior would be strongly
state-dependent. Therefore, our goal of extracting the
state-independent, universal correlations forbids long-range
correlation functions. (\emph{ii}) The two-body approximation would
not be applicable for long-range correlators. (\emph{iii})
Effectively, higher order contributions of the cluster expansion lead
to a screening of long-range tensor correlations between two nucleons
through the presence of other nucleons within the correlation range
\cite{ucom04}.  For these reasons, we constrain the range of the
tensor correlation functions in our variational procedure.  We use the
following integral constraint on the ``volume'' of the tensor
correlation functions
\eq{
  I_{\vartheta}
  = \int \dd{r}\, r^2\; \vartheta(r) \;.
}
The constrained energy minimization for the $S=1, T=0$ and the $S=1,
T=1$ channels with different values of the tensor correlation volume
$I_{\vartheta}$ leads to optimal parameters reported in Table
\ref{tab:corr_tensorpara}. The optimal parameters for the central
correlation functions change only marginally with the tensor
constraint.  Therefore, we adopt a fixed set of parameters for the
central correlators given in Table \ref{tab:corr_centralpara}.

%%%%%%%%%%%%%%%%%%%%%%%%%%%%%%%%%%%%%%%%%%%%%%%%%%%%%%%%%%%%%%%%%%%%%%
\begin{figure}
  \begin{center}
    \includegraphics[width=0.85\textwidth]{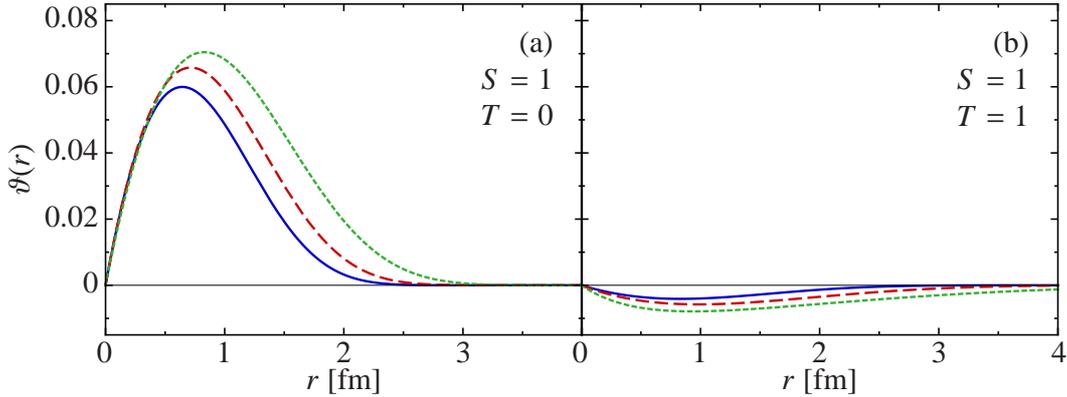}
  \end{center}
  \vskip-4ex
  \caption{Optimal tensor correlation functions
    $\vartheta(r)$ for different values of the range constraint
    $I_{\vartheta}$. (a) Correlation functions for $T=0$ with
    $I_{\vartheta}=0.06$ (\linemediumsolid[FGBlue]), $0.09$ (\linemediumdashed[FGRed]), and $0.15\fm^3$ (\linemediumdotted[FGGreen]). (b) Correlation
    functions for $T=1$ with $I_{\vartheta}=-0.01$ (\linemediumsolid[FGBlue]), $-0.03$ (\linemediumdashed[FGRed]), and
    $-0.09\fm^3$ (\linemediumdotted[FGGreen]).}
  \label{fig:corr_tensor}
\end{figure}
%%%%%%%%%%%%%%%%%%%%%%%%%%%%%%%%%%%%%%%%%%%%%%%%%%%%%%%%%%%%%%%%%%%%%%

The optimal central correlation functions for the Argonne~V18
potential are shown in Fig.~\ref{fig:corr_central}. In the even
channels, the correlation functions decrease rapidly and vanish beyond
$r\approx1.5 \fm$. The central correlators in the odd channels are
weaker and of slightly longer range due to the influence of the
centrifugal barrier. For the tensor correlation functions the
constraints on the range are important. Fig.~\ref{fig:corr_tensor}
shows the triplet-even (a) and triplet-odd (b) tensor correlation
functions $\vartheta(r)$ for different $I_{\vartheta}$. Because the
tensor interaction is significantly weaker for $T=1$ than for $T=0$,
the tensor correlator for this channel has a much smaller
amplitude. The relevant values for the constraint $I_{\vartheta}$ are
therefore smaller for the triplet-odd channel.

We stress that the range constraint for the tensor correlation
functions has an important physical and conceptual background. The
unitary correlation operator method is used to describe
state-independent short-range correlations only. Long-range
correlations of any kind have to be described by the model space
employed in the solution of the many-body problem. By constraining the
range of the tensor correlators we introduce a separation scale
between short-range and long-range correlations. The optimal value for
tensor constraints cannot be fixed in the two-body system alone, but
requires input from few-nucleon systems. We will come back to this
point in Sec.~\ref{sec:ncsm}.

%%%%%%%%%%%%%%%%%%%%%%%%%%%%%%%%%%%%%%%%%%%%%%%%%%%%%%%%%%%%%%%%%%%%%%
\subsection{Cluster expansion}
\label{sec:ucom_cluster}

The similarity transformation \eqref{eq:corr_operator} of an operator
$\AO$ gives a correlated operator that contains irreducible
contributions of higher orders in particle number as given by the
cluster expansion 
\eq{ \label{eq:clusterexp} \corr{\AO} = \CCO \AO \CO
  = \corr{\AO}^{[1]} + \corr{\AO}^{[2]} + \corr{\AO}^{[3]} + \cdots
  \;, }
where $\corr{\AO}^{[n]}$ denotes the irreducible $n$-body part
\cite{ucom98}. For a $k$-body operator $\AO$ there will only be
contributions $\corr{\AO}^{[n]}$ with $n \ge k$. For the correlated
Hamiltonian we will, therefore, have a one-body contribution (from the
kinetic energy), a two-body contribution (two-body part of the
correlated kinetic energy and correlated two-body potential),
three-body contributions and so on.

In practice it will not be possible to evaluate matrix elements of the
correlated operators to all orders. The importance of the higher-order
terms depends on the range of the central and tensor correlations
\cite{ucom04,ucom03,ucom98}. If the range of the correlation functions
is small compared to the mean interparticle distance, then three-body
and higher-order terms of the cluster expansion are expected to be
small. In the two-body approximation these higher-order contributions
are discarded
\eq{
  \corr{\AO}^{C2}
  = \corr{\AO}^{[1]} + \corr{\AO}^{[2]} \;.
}
In principle, the higher-order contributions to the cluster expansion
can be evaluated systematically \cite{roth:doktor}. However, for
many-body calculations the inclusion of those terms is an extreme
challenge and we restrict ourselves to the two-body approximation.

Within the two-body approximation the similarity transformation is
still unitary on the two-body level, e.g. the eigenvalues of the
Hamiltonian are conserved in two-body systems, but it is no longer
unitary on the many-body level.  The energy eigenvalues obtained in
exact many-body calculations using the correlated interaction in
two-body approximation will differ from the eigenvalues obtained in
exact calculations using the bare interaction. As will be discussed in
detail in Sec.~\ref{sec:ncsm} we can use exact solutions, e.g., in the
No-Core Shell Model framework, to estimate the size of the omitted
higher-order contributions.

Technically we can calculate matrix elements of correlated operators
in the two-body approximation of the cluster expansion in any
many-body approach using the density matrices $\rho^{(1)}_{m;k}$ and
$\rho^{(2)}_{mn;kl}$ of the uncorrelated states
\eq{
  \matrixe{\corr{\Phi}}{\AO}{\corr{\Phi}'}
  \stackrel{C2}{=}
  \sum_{km} \rho^{(1)}_{m;k} \; \matrixe{k}{\corr{\AO}^{[1]}}{m} +
  \sum_{k<l,m<n} \rho^{(2)}_{mn;kl} \; \matrixe{kl}{\corr{\AO}^{[2]}}{mn} \: .
}
Here the one- and two-body density matrices
\eq{
  \rho^{(1)}_{m;k} = \matrixe{\Phi}{\adj{\op{a}}_k \op{a}_m}{\Phi'} \:, \quad
  \rho^{(2)}_{mn;kl} =
  \matrixe{\Phi}{\adj{\op{a}}_k \adj{\op{a}}_l \op{a}_n \op{a}_m}{\Phi'}
}
are given in a generic single-particle basis ${\ket{k}}$. Typically we
will use harmonic oscillator basis states, as the harmonic oscillator
basis allows to expand the two-body states in products of relative and
center-of-mass harmonic oscillator states with the help of the
Talmi-Moshinsky transformation.

%%%%%%%%%%%%%%%%%%%%%%%%%%%%%%%%%%%%%%%%%%%%%%%%%%%%%%%%%%%%%%%%%%%%%%
\bigskip
\subsection{Correlated interaction $V_{\UCOM}$}

\begin{figure}
  \centering
  \includegraphics[width=0.6\textwidth]{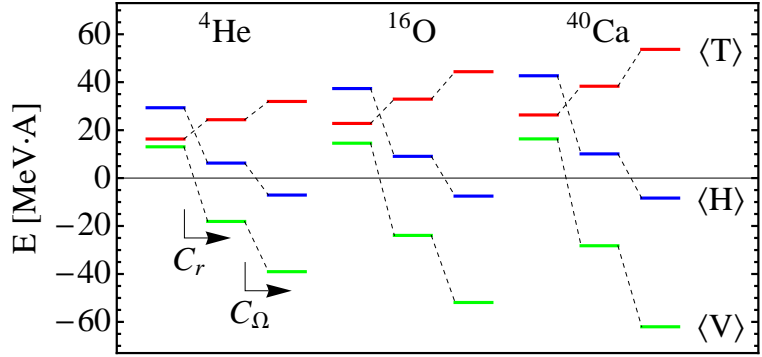}
  \caption{Kinetic $\expect{T}$, potential $\expect{V}$ and total
    energy $\expect{H}$ obtained with the bare Argonne~V18 interaction
    (left), including central correlations (middle) and with central
    and tensor correlations (right) for the doubly-magic nuclei
    \elem{He}{4}, \elem{O}{16}, and \elem{Ca}{40} using $0\hbar\Omega$
    shell-model wave functions.}
  \label{fig:correlations-energy}
\end{figure}

We define the UCOM interaction $\VO_{\UCOM}$ as the two-body part of
the correlated Hamiltonian
\eq{
  \corr{\HO} = \TO + \VO_{\UCOM} + \VO_{\UCOM}^{[3]} + \ldots \;.
}
It contains the contributions of the
correlated kinetic energy and of the correlated potential. The
three-body contribution of the correlated Hamiltonian
$\VO_{\UCOM}^{[3]}$ has not been evaluated explicitly yet. 

If we start from a realistic interaction that is given in an operator
representation, e.g. the Argonne~V18 potential, then the UCOM
correlated interaction can also be given in operator representation
\eq{ \label{eq:VUCOM}
  \VO_{\UCOM}
  = \sum_{p} \frac{1}{2} 
    \big[ \corr{V}_{p}(\rO)  \corr{\OO}_p + \corr{\OO}_p \corr{V}_{p}(\rO)
    \big] \;,
} 
where
\eqmulti{ \label{eq:VUCOM_ops}
  \corr{\OO}_p =
  \{ &1,\;\sigmasigmaO,\;\qO_r^2,\;\qO_r^2 \sigmasigmaO,\;
    \LOV^2,\;\LOV^2 \sigmasigmaO, \spinorbitO,\;\tensorRRO,\;\tensorLLO,\\
    &\tensorbarQQO,\;\qO_r\, \tensorRQO,\;\orbitsqrO\spinorbitO,
    \orbitsqrO \tensorbarQQO,\dots \}  \otimes \{1,\; \tautauO\} \;.
}
The dots indicate higher-order contributions of the
Baker-Campbell-Hausdorff expansion for the tensor transformation that have
been omitted. The terms shown above result from a truncation to
operators of up to fourth order in momentum. For most applications the
inclusion of these terms is sufficient \cite{ucom04,neff:diss}.

In Fig.~\ref{fig:correlations-energy} we show the effect of the
correlations on the kinetic and potential energy contributions for the
Argonne~V18 interaction. The uncorrelated many-body state is given by
the $0\hbar\Omega$ harmonic oscillator configuration (the oscillator
parameter is chosen to reproduce the experimental radius). Without any
short-range correlations the considered nuclei (\elem{He}{4},
\elem{O}{16}, and \elem{Ca}{40}) are not bound at all. Even the
potential contributions are repulsive. By including the central
correlations the potential contributions become attractive but the
nuclei are still unbound. Part of the gain by the potential
contributions has to be paid in form of a larger kinetic
energy. Altogether the central correlations increase the binding
energies by about $20-30\MeV$ per nucleon. With the closed-shell trial
wave functions used here, contributions from the tensor force are only
obtained when including tensor correlations. As can be seen this is
again a huge effect. The binding energies per nucleon increase by
about $15-20\MeV$ and a total binding of about $4\MeV$ per nucleon is
obtained on the $0\hbar\Omega$ level using the correlated interaction.

The existence of an operator representation of $\VO_{\UCOM}$ is
essential for many-body models that are not based on a simple
oscillator or plane-wave basis. One example is the Fermionic Molecular
Dynamics model \cite{fmd00,fmd08,bacca08} which uses a non-orthogonal
Gaussian basis and does not easily allow for a partial wave
decomposition of the relative two-body states. Nevertheless, it is
possible to evaluate the two-body matrix elements of $\VO_{\UCOM}$
analytically (radial dependencies are fitted by sums of Gaussians),
which facilitates efficient computations with this extremely versatile
basis \cite{ucom04}.

As we have emphasized already, the operators of all observables have
to be transformed consistently. The unitary transformation of
observables like quadratic radii, densities, momentum distributions,
or transition matrix elements is straightforward given the toolbox
acquired for the transformation of the Hamiltonian. The Unitary
Correlation Operator Method owes this simplicity to the explicit state
and representation-independent form of the correlation operators.  In
other approaches like the Lee-Suzuki transformation
\cite{SuLe80,NaVa00b,CaNa02} or the $V_{\mathrm{low}k}$
renormalization group method \cite{BoKu03}, it is not possible to
provide a closed form for the effective operators. A discussion of
effective operators in the Lee-Suzuki approach can be found in
\cite{StBa05,stetcu06}.

An important feature of $\VO_{\UCOM}$ results from the finite range of
the correlation functions $s_{ST}(r)$ and $\vartheta_T(r)$ entering
into the generators. Since the correlation functions are of finite
range, i.e., the correlation operator acts as a unit operator at large
$r$, asymptotic properties of a two-body wave function are
preserved. This implies that $\VO_{\UCOM}$ is by construction
phase-shift equivalent to the original NN-interaction. The unitary
transformation can, therefore, be viewed as a way to construct an
infinite manifold of realistic potentials, that all give identical
phase-shifts.

It is interesting to observe in which way the unitary transformation
changes the operator form of the interaction while preserving the
phase-shifts.  The central correlator reduces the short-range
repulsion in the local part of the interaction and, at the same time,
creates a non-local repulsion through the momentum-dependent
terms. The tensor correlator removes some strength from the local
tensor interaction and creates additional central contributions as
well as new momentum-dependent tensor terms.  Hence, the unitary
transformation exploits the freedom to redistribute strength between
local and non-local parts of the potential without changing the
phase-shifts. The non-local tensor terms establish an interesting
connection to the CD Bonn potential, which among the realistic
potentials is the only one including non-local tensor contributions
\cite{MaSl01}.

\subsection{Correlated densities}

As we have seen, the inclusion of short-range correlations is
essential for obtaining bound nuclei when using realistic
interactions. However binding energies and spectra provide no direct
information about these short-range correlations as they are hidden in
the correlated or effective interactions.

Other quantities are much better suited to provide insight into
correlations. The two-body density in coordinate space visualizes the
effect of the correlations directly, whereas the one-body density in
momentum space, the nucleon momentum distribution, allows a comparison
with experimental data, providing a direct proof for the existence of
short-range correlations in nuclei.

\subsubsection*{Two-body coordinate space density}

\begin{figure}
  \centering
  \includegraphics[width=0.7\textwidth]{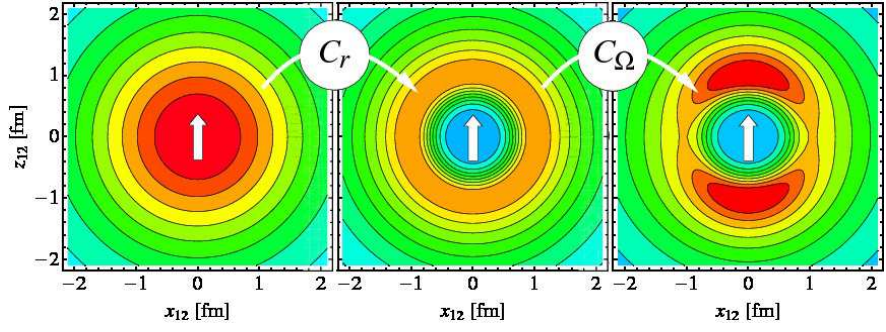}
  \caption{Two-body density $\rho^{(2)}_{S M_S, T M_T}(\vec{r}_{12})$
    in the $S=1,M_S=1$, $T=0,M_T=0$ channel of \elem{He}{4}. Left: for
    the uncorrelated trial state \ket{\Phi}, middle: including central
    correlations, right: including central and tensor
    correlations. The arrow indicates the orientation of the spin in
    the two-body channel.}
  \label{fig:correlationplot}
\end{figure}

In Fig.~\ref{fig:correlationplot} we show the two-body density
$\rho^{(2)}_{S M_S, T M_T}(\vec{r}_{12})$ in the deuteron channel. It
has been calculated with a $0\hbar\Omega$ wave function for
\elem{He}{4}. For the uncorrelated wave function the two-body density
has a maximum at $\vec{r}_{12} = 0$, where the potential is strongly
repulsive. This defect in the wave function is cured by the central
correlator which shifts the nucleons apart. For the centrally
correlated wave function we find now the largest density at
$|\vec{r}_{12}| \approx 1\fm$. At this distance the potential is most
attractive. The tensor force provides attraction if the spins are
aligned parallel with the distance vector. This is reflected in the
two-body density. After applying the tensor correlations the density
is enhanced at the ``poles'' and reduced at the ``equator''.

The correlated two-body density $\corr{\rho}^{(2)}_{S M_S, T
  M_T}(\vec{r}_{12})$ is calculated from the diagonal part of the
two-body density in two-body approximation by introducing relative and
center-of-mass variables $\vec{r}_{12} = \vec{x}_1 - \vec{x}_2$ and
$\vec{X}_{12} = \half (\vec{x}_1 + \vec{x}_2)$ and coupling the
single-particle spins and isospins to total spin $S$ and isospin $T$
\eq{
  \begin{split}
  \corr{\rho}^{(2)}_{S M_S, T M_T}(\vec{r}_{12},\vec{X}_{12}) = &
  \sum_{m_{s_1},m_{t_1},m_{s_2},m_{t_2}}
  \clebschgordan{\half}{\half}{S}{m_{s_1}}{m_{s_2}}{M_S}
  \clebschgordan{\half}{\half}{T}{m_{t_1}}{m_{t_2}}{M_T} \times \\
  & \corr{\rho}^{(2)}_{m_{s_1},m_{t_1},m_{s_2},m_{t_2};m_{s_1},m_{t_1},m_{s_2},m_{t_2}}(\vec{X}_{12} + \half \vec{r}_{12}, \vec{X}_{12} - \half \vec{r}_{12};
   \vec{X}_{12} + \half \vec{r}_{12}, \vec{X}_{12} - \half \vec{r}_{12})
  \end{split} \;.
} 
The dependence on the center-of-mass coordinate is removed by
integrating over $\vec{X}_{12}$
\eq{ \corr{\rho}^{(2)}_{S M_S, T M_T}(\vec{r}_{12}) = \int \dd^3X_{12}\;
  \; \corr{\rho}^{(2)}_{S M_S, T M_T}(\vec{r}_{12},\vec{X}_{12}) \; .}

The correlated two-body density is very similar to the results from
microscopic calculations using the bare Argonne~V8 interactions by
Suzuki and Horiuchi \cite{suzuki09}.

\subsubsection*{Momentum distributions}

Momentum distributions also directly reflect the existence of
short-range correlations. Without short-range correlations there would
be no high-momentum components in the nuclear many-body state. To
illustrate this, we calculate the correlated one-body momentum
distributions $\corr{n}(k)$ for the doubly-magic nuclei \elem{He}{4}
and \elem{O}{16} in two-body approximation. The uncorrelated wave
functions $\ket{\Phi}$ are again $0\hbar\Omega$ harmonic oscillator
configurations, where the oscillator parameters have been adjusted to
reproduce the experimental radii. The momentum distribution is given
by
\eq{   \label{eq:momentumdistr}
  \corr{n}(\vec{k})= \sum_{m_s,m_t}
  \matrixe{\corr{\Phi}}{\adj{\op{a}}_{m_s,m_t} (\vec{k}) \op{a}_{m_s,m_t}(\vec{k})}{\corr{\Phi}}
  \stackrel{C2}{=} \sum_{m_s,m_t}
  \matrixe{\Phi}{[\adj{\CO} \adj{\op{a}}_{m_s,m_t}(\vec{k})
    \op{a}_{m_s,m_t}(\vec{k}) \CO]^{C2}}{\Phi} \;.
}
The evaluation of this expression is straightforward but lengthy as it
requires an integration over single-particle coordinates for
correlated wave functions that are expressed in relative and
center-of-mass coordinates.

\begin{figure}
  \centering
  \includegraphics[width=0.35\textwidth]{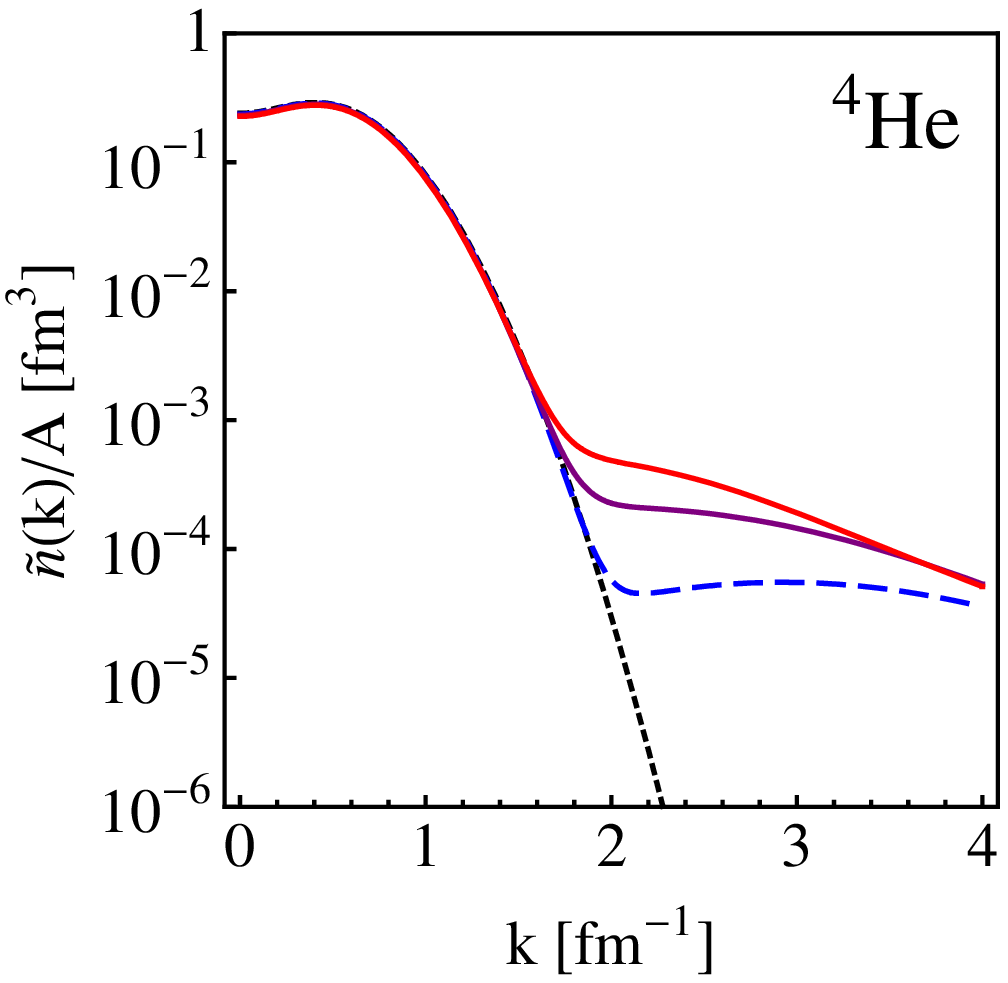}\hfil
  \includegraphics[width=0.35\textwidth]{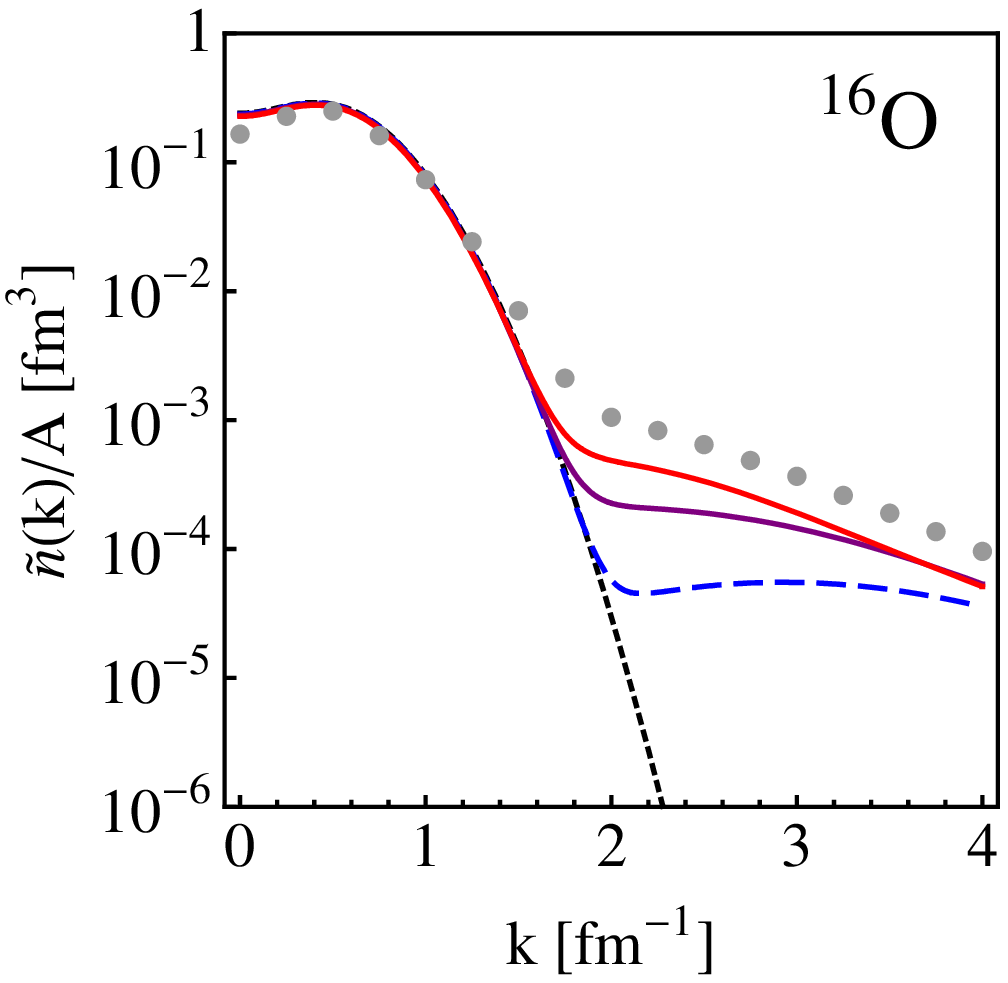}
  \caption{Nucleon momentum distribution $\corr{n}(k)$ calculated with closed-shell 
    wave functions for \elem{He}{4} and \elem{O}{16}. Results
    are shown for the uncorrelated wave functions (\linemediumdotted[FGBlack]), for wave functions
    including central correlations (\linemediumdashed[FGBlue]), and including central and tensor
    correlations with $I_{\vartheta}=0.10\,\text{fm}^3$ (\linemediumsolid[FGViolet]) and $I_{\vartheta}=0.20\,\text{fm}^3$ (\linemediumsolid[FGRed]). The gray dots indicate the Variational Monte Carlo results from Ref. \cite{pieper92}.}
  \label{fig:momentum-distributions}
\end{figure}

The results for the momentum distributions are shown in
Fig.~\ref{fig:momentum-distributions}. Without short-range
correlations the momentum distributions have no high-momentum
components. With only the central correlations included, we observe a
high-momentum tail, which is almost constant as a function of
momentum. The contributions from the tensor correlations to the
high-momentum tails are remarkable. The momentum-distribution from the
Fermi surface at about $2 \fm^{-1}$ up to about $4 \fm^{-1}$ is
dominated by tensor correlations. Only for very high momenta the
central correlations become more important. We also observe a strong
dependence of the momentum distributions on the range of the tensor
correlator, especially for smaller momenta close to the Fermi
surface. This strong dependence on the range of the tensor correlator
is caused mainly by the simplified uncorrelated wave function used in
the present calculation. The effect of long-range correlations induced
by the long-range part of the tensor force is here only included when
using a long-range tensor correlator. In more realistic calculations
these long-range correlations can also be expressed within the
many-body model space. Other long-range correlations will lead to a
further softening of the Fermi surface.

The dominating role of the tensor correlations was also found in other
microscopic calculations \cite{pieper92,schiavilla07,alvioli08} and
has been confirmed experimentally by comparing $(e,e'pp)$ with
$(e,e'pn)$ cross sections at high momentum transfer
\cite{egiyan06,subedi08} and $(p,pp)$ with $(p,ppn)$ cross sections
\cite{piasetzky06}.
  
\subsection{Correlated transition operators}

Short-range correlations have to be considered also for the evaluation
of transition matrix elements. The transition operators are correlated
using the same techniques that have been explained for the Hamiltonian
in Sec.~\ref{sec:corr_op}. In case of the radius operator
\eq{ \RO_\mathrm{ms} = \frac{1}{A} \sum_i (\xOV_i - \XOV_\cm)^2 =
  \frac{1}{A^2} \sum_{i<j} |\rOV_{ij}|^2 }
the correlated radius operator in two-body approximation is simply
given by
\eq{
  \corr{\RO}_\mathrm{ms} \stackrel{C2}{=} \frac{1}{A^2} \sum_{i<j} \left[ \Rp(|\rOV_{ij}|) \right]^2 \; .
}
The radius like the quadrupole operator and other long-range operators
are only weakly affected by the short-range correlations
\cite{PaPa06}.

But there are other cases, where short-range correlations are very
important. One prominent example is the neutrinoless double
$\beta$-decay. The effective neutrino potentials appearing here
connect to high momenta and are therefore sensitive to short-range
correlations in the many-body state. UCOM is nowadays one of the
standard approaches for including the short-range correlations in the
transition matrix elements \cite{kortelainen07,simkovic08,menendez09}.

%%%%%%%%%%%%%%%%%%%%%%%%%%%%%%%%%%%%%%%%%%%%%%%%%%%%%%%%%%%%%%%%%%%%%%
\bigskip
\subsection{Correlated matrix elements}

For many-body calculations the matrix elements of the correlated
interaction $\VO_{\UCOM}$ and possibly of other operators are
needed. For the many-body calculations presented in this review we
employ the harmonic oscillator basis. Here the two-body matrix
elements can be decomposed into a relative and a center-of-mass matrix
element by means of the Talmi-Moshinsky transformation. In the end the
relative matrix elements of the correlated interaction
\eqmultline{ \label{eq:corr_me_VUCOM}
  \matrixe{n(LS)J M T M_T}{\VO_{\UCOM}}{n'(L'S)J M T M_T} = \\
  \matrixe{n(LS)J M T M_T}{\CCO_{r}\CCO_{\Omega}\,\HO_\intr\,\CO_{\Omega}\CO_{r}
    - \TO_\intr}{n'(L'S)J M T M_T}
}
have to be calculated.  We use here relative $LS$-coupled basis states
$\ket{n (LS) J M \; T M_T}$, where $n$ is the radial quantum
number. The corresponding wave function will be denoted as
$\phi_{n,L}(r)$, the radial wave function as $u_{n,L}(r)$,
\eq{
  \braket{r(LS)J M T M_T}{n(LS)J M T M_T} = \phi_{n,L}(r) = \frac{u_{n,L}(r)}{r} \;,
}
where $L$ is the relative orbital angular momentum, $S$ spin, $J$ total
angular momentum, and $T$ isospin. The interaction also contains
Coulomb and other isospin- and charge-symmetry breaking terms so that
the matrix elements also depend explicitly on $M_T$. In the following
we will omit the $M$ and $M_T$ quantum numbers to simplify the
notation.

In the UCOM approach matrix elements can be calculated in different
ways. It is possible to expand the correlation operators in the basis
states $\ket{n (LS) J T}$ and use the matrix elements of the
uncorrelated interaction.

An alternative approach is to use the operator representation of
$\VO_{\UCOM}$ and evaluate the matrix elements directly. If one
expands the radial dependencies of the individual operator channels in
a sum of Gaussians, all radial integrals can be calculated
analytically.  The matrix elements of the additional tensor operators
contained in $\VO_{\UCOM}$ can be given in closed form as
well. However, this direct approach relies on the truncation of the
Baker-Campbell-Hausdorff expansion \eqref{eq:corr_tensor_bch}.

This can be avoided by applying the tensor correlations to the basis
states which can be done exactly. For interactions in operator
representation the central correlations will still be applied to the
Hamiltonian as we have a simple and exact expression for the centrally
correlated Hamiltonian (cf. Sec.~\ref{sec:corr_op_central}). This approach
requires a rewriting of the correlated matrix elements by
interchanging the order of central and tensor correlation operators
using the identity
\eqmulti{
  \CCO_{r} \CCO_{\Omega}\, \HO_\intr\, \CO_{\Omega} \CO_{r}
  &= (\CCO_{r} \CCO_{\Omega} \CO_{r})\, \CCO_{r}\, \HO_\intr\, \CO_{r}\,
     (\CCO_{r} \CO_{\Omega} \CO_{r}) \\
  &=  \corr{\CO}^{\dag}_{\Omega} \CCO_{r} \, \HO_\intr\, \CO_{r} \corr{\CO}_{\Omega}
}
with the ``centrally correlated'' tensor correlation operator
\eq{
  \corr{\CO}_{\Omega}
  = \CCO_{r} \CO_{\Omega} \CO_{r}
  = \exp[-\ii\, \vartheta(\Rp(\rO))\, \tensorRQO] \;.
}
The central correlator commutes with $\tensorRQO$ and transforms
therefore only $\vartheta(\rO)$, see Eq.~\eqref{eq:corr_op_funcr}. The
tensor correlator $\corr{\CO}_{\Omega}$ acts on $LS$-coupled two-body
states with $L=J$ like the identity operator and couples states with
$L=J\pm1$ with states $L=J\mp1$ (cf. Sec. \ref{sec:corr_wavefunc})
\eq{ \label{eq:tensor_corr_me}
  \matrixe{r(L'S)JT}{\corr{\CO}_{\Omega}}{n(LS)JT} =
  \begin{cases}
    \phi_{n,L}(r) & ; L'=L=J\\
    \cos \corr{\theta}_J(r) \phi_{n,L}(r) & ; L'=L=J\pm1\\
    \pm \sin \corr{\theta}_J(r) \phi_{n,L}(r) & ; L'=J\pm1,L=J\mp1
  \end{cases} \;,
}
where
\eq{
  \corr{\theta}_J(\rO) 
  = 3 \sqrt{J(J+1)}\, \vartheta(\Rp(\rO)) \;.
}   
Using these relations we can calculate the correlated two-body matrix
elements exactly.

\subsubsection*{Matrix elements using correlator expansion}
\label{sec:corr_me_expansion}

This approach is very general as it works for any unitary
transformation of the basis states and for any two-body operator. To
calculate the matrix element of the correlated interaction
\eqmultline{ \label{eq:corr_me_VUCOM2}
  \matrixe{n(LS)J M T M_T}{\VO_{\UCOM}}{n'(L'S)J M T M_T} = \\
  \matrixe{n(LS)J M T M_T}{\CCO_{r}\CCO_{\Omega}\,\HO_\intr\,\CO_{\Omega}\CO_{r}
    - \TO_\intr}{n'(L'S)J M T M_T}
}
we evaluate the correlator in the basis states 
\eq{
  C^{(J\bar{L}LST)}_{\bar{n},n} =
  \matrixe{\bar{n}(\bar{L}S)JT}{\CO_{\Omega}\CO_{r}}{n(LS)JT} \: .  }
The tensor correlation operator acts as the identity operator in $L=J$
channels and couples $L=J\mp1$ with $L'=J\pm1$ states:
\eq{
  \label{eq:correlator_me}
  C^{(J\bar{L}LST)}_{\bar{n},n} =
  \begin{cases}
    \int\!\dd{r}\, \conj{u}_{\bar{n},\bar{L}}(r)  \frac{\Rm(r)}{r} \sqrt{\DRm(r)} u_{n,L}(\Rm(r))
  & ; \bar{L}=L=J\\
    \int\!\dd{r}\, \conj{u}_{\bar{n},\bar{L}}(r)  \cos \theta_J(r) \frac{\Rm(r)}{r} \sqrt{\DRm(r)} u_{n,L}(\Rm(r))
  & ; \bar{L}=L=J\pm1\\
    \pm \int\!\dd{r}\, \conj{u}_{\bar{n},\bar{L}}(r)  \sin \theta_J(r) \frac{\Rm(r)}{r} \sqrt{\DRm(r)} u_{n,L}(\Rm(r))
  & ; \bar{L}=J\pm1,L=J\mp1
  \end{cases}
}
The correlated matrix elements of the interaction are given with these
correlator matrix elements as
\eqmultline{
  \matrixe{n(LS)J T}{\VO_{\UCOM}}{n'(L'S)J T} \\
  \begin{split}
  &= \matrixe{n(LS)J T}{\CCO_{r}\CCO_{\Omega}\,\HO_\intr\,\CO_{\Omega}\CO_{r}
    - \TO_\intr}{n'(L'S)J T }\\
  &= \sum_{\bar{n}\bar{L},\bar{n}'\bar{L}'}^{n_\mathrm{max}}
  \conj{C^{(JL\bar{L}ST)}_{n,\bar{n}}}
  \matrixe{\bar{n}(\bar{L}S)JT}{\HO_\intr}{\bar{n}'(\bar{L}'S)J T}
  C^{(J\bar{L'}L'ST)}_{\bar{n}',n'} -
  \matrixe{n(LS)J T}{\TO_\intr}{n'(L'S)J T}
  \end{split}
}
Besides the correlator matrix elements \eqref{eq:correlator_me} the
matrix elements of the bare interaction are needed
\eq{ \matrixe{\bar{n}(\bar{L}S)JT}{\HO_\intr}{\bar{n}'(\bar{L}'S)J T} \; . }
Here $\bar{n}$ and $\bar{n}'$ will now run up to some cut-off
$n_\mathrm{max}$. Convergence is reached only if $n_\mathrm{max}$ is
chosen large enough. A hard interaction like Argonne~V18 will connect
low momentum states with high momentum states up to about $15
\fm^{-1}$. For typical oscillator constants of $10 \MeV$ the summation
therefore has to extend to $n_\mathrm{max} \approx 300$.

%%%%%%%%%%%%%%%%%%%%%%%%%%%%%%%%%%%%%%%%%%%%%%%%%%%%%%%%%%%%%%%%%%%%%%
\subsubsection*{Matrix elements for interactions in operator representation}
\label{sec:corr_me_operator}

For a bare potential given in the generic operator representation
\eqref{eq:corr_op_potgeneric} the matrix elements can be evaluated
using the closed form for the centrally correlated interaction. For
the tensor correlations only the correlated kinetic energy is given in
a closed form without approximations. For the potential contributions
the tensor correlations will be applied to the wave functions.

We start with the matrix elements for the local contributions of the
form $V(\rO) \OO$ with $[\rO,\OO] = [\qO_r,\OO] = 0$, which includes
all operators of the set \eqref{eq:corr_op_potgenericop} except for
the $\qO_r^2$ terms. The matrix elements for $L=L'=J$ are not affected by the tensor
correlations, only the central correlators act according to
\eqref{eq:corr_op_localpot}.  In coordinate representation we obtain
\eq{ \label{eq:corr_me_local1}
  \matrixe{n(JS)JT}{\CCO_{r} \CCO_{\Omega}\, V(\rO) \OO\,
    \CO_{\Omega}\CO_{r}}{n'(JS)JT} =
  \int\!\dd{r}\,u_{n,J}^{\star}(r)\, u_{n',J}(r)\;
     \corr{V}(r)\; \matrixe{(JS)JT}{\OO}{(JS)JT} \;,
}
where $\corr{V}(r) = V(\Rp(r))$ is the transformed radial dependence
of the potential. For the diagonal matrix elements with $L=L'=J\mp 1$
we get
\eqmultline{ \label{eq:corr_me_local2}
  \matrixe{n(J\mpS 1,1) J T}{\CCO_{r} \CCO_{\Omega}\, V(\rO) \OO\,
    \CO_{\Omega}\CO_{r}}{n'(J\mpS 1,1) J T} = \\
  \begin{split}
  \int\!\dd{r}\,u_{n,J\mpS 1}^{\star}(r)\, u_{n',J\mpS 1}(r)\;
     \corr{V}(r)
  \big[ \: & \matrixe{(J\mpS 1,1) J T}{\OO}{(J\mpS 1,1) J T}\,
    \cos^2 \corr{\theta}_J(r) \\[-1ex]
    + & \matrixe{(J\pmS 1,1) J T}{\OO}{(J\pmS 1,1) J T}\,
     \sin^2 \corr{\theta}_J(r) \\
  \pm & \matrixe{(J\mpS 1,1) J T}{\OO}{(J\pmS 1,1) J T}\,
    2 \cos\corr{\theta}_J(r) \sin\corr{\theta}_J(r) \big]
  \end{split}
}
with $\corr{\theta}_J(r) = \theta_J(\Rp(r))$. Finally, the
off-diagonal matrix elements for $L=J\mp 1$ and $L'=J\pm 1$ are
obtained as
\eqmultline{ \label{eq:corr_me_local3}
  \matrixe{n(J\mpS 1,1) J T}{\CCO_{r} \CCO_{\Omega}\, V(\rO) \OO\,
    \CO_{\Omega}\CO_{r}}{n'(J\pmS 1,1) J T} = \\
  \begin{split}
   \int\!\dd{r}\, u_{n,J\mpS 1}^{\star}(r)\, u_{n',J\pmS 1}(r)\;
     \corr{V}(r)
   \big[ \: & \matrixe{(J\mpS 1,1) JT}{\OO}{(J\pmS 1,1) JT}\,
    \cos^2 \corr{\theta}_J(\rO) \\[-1ex]
   - & \matrixe{(J\pmS 1,1) JT}{\OO}{(J\mpS 1,1) JT}\,
    \sin^2 \corr{\theta}_J(\rO)\\
  \mp & \matrixe{(J\mpS 1,1) JT}{\OO}{(J\mpS 1,1) JT}\,
    \cos\corr{\theta}_J(\rO) \sin\corr{\theta}_J(\rO)\\
  \pm & \matrixe{(J\pmS 1,1) JT}{\OO}{(J\pmS 1,1) JT}\,
    \sin\corr{\theta}_J(\rO) \cos\corr{\theta}_J(\rO) \big] \;.
  \end{split}
}
Apart from the integration involving the radial wave functions, the
matrix elements of the operators $\OO$ in $LS$-coupled angular
momentum states are required. Only for the standard tensor operator
$\OO=\tensorRRO$ the off-diagonal terms on the right hand side of Eqs.
\eqref{eq:corr_me_local2} and \eqref{eq:corr_me_local3}
contribute. For all other operators in \eqref{eq:corr_op_potgenericop}
the off-diagonal matrix elements vanish, and the above equations
simplify significantly.

The effect of the tensor correlator is reflected in the structure of
the correlated matrix elements \eqref{eq:corr_me_local2} and
\eqref{eq:corr_me_local3}. It admixes components with $\Delta L =
\pm2$ to the states. Therefore, the correlated matrix element consists
of a linear combination of diagonal and off-diagonal matrix elements
$\matrixe{(LS)JT}{\OO}{(L'S)JT}$. In this way even simple operators,
like $\orbitsqrO$ or $\spinorbitO$ acquire non-vanishing off-diagonal
\emph{correlated} matrix elements \eqref{eq:corr_me_local3}.

A closed form is available for the momentum dependent terms of the
potential \eqref{eq:corr_op_potgeneric}. For the tensor correlated
form of the operator
\eq{
  \VO_{qr}
  = \frac{1}{2}\big[\qO_r^2 V(\rO) + V(\rO) \qO_r^2 \big]
}
we obtain 
\eqmulti{
  \CCO_{\Omega} \VO_{qr} \CO_{\Omega}
  &= \frac{1}{2}\big[ \qO_r^2 V(\rO) + V(\rO) \qO_r^2 \big]
    + V(\rO) [\vartheta'(\rO) \tensorRQO]^2 \\
  &- \big[\qO_r V(\rO)\vartheta'(\rO) +  \vartheta'(\rO)V(\rO) \qO_r \big]
    \tensorRQO 
}
by using Eq. \eqref{eq:corrT_op_momentumsqr}. Subsequent inclusion of
the central correlations leads to the following expression for the
diagonal matrix elements with $L=L'=J$ in coordinate representation:
\eqmulti{ \label{eq:corr_me_momentum1}
  \matrixe{n(JS) JT}{\CCO_{r} \CCO_{\Omega} \VO_{qr} \CO_{\Omega}\CO_{r}}
    {n'(JS) JT} = 
  \int\!\dd{r}\,\bigg\{ & u_{n,J}^{\star}(r)\, u_{n',J}(r)\;
    \bigg[\corr{V}(\rO)\; W(\rO) - \corr{V}'(\rO)
    \frac{\DDRp(r)}{\DRp(r)^2}\bigg] \\
    & -\frac{1}{2}\big[u_{n,J}^{\star}(r)\, u''_{n',J}(r)
    +u_{n,J}^{\prime\prime\star}(r)\, u_{n',J}(r) \big] 
    \frac{\corr{V}(r)}{\DRp(r)^2} \bigg\} \;,
}
where $\corr{V}'(r) = V'(\Rp(r))$. As before, the tensor correlator
does not affect these matrix elements and only the central
correlations have to be considered. For the diagonal matrix elements
with $L=L'=J\mp1$ the tensor terms contribute and we obtain
\eqmultline{ \label{eq:corr_me_momentum2}
  \matrixe{n(J\mpS 1,1) JT}{\CCO_{r} \CCO_{\Omega} \VO_{qr}
     \CO_{\Omega}\CO_{r}}{n'(J\mpS 1,1) JT} = \\
  \begin{split}
    \int\!\dd{r}\,\bigg\{ & u_{n,J\mpS 1}^{\star}(r)\, u_{n',J\mpS 1}(r)
    \bigg[\corr{V}(r)\; W(r) + \corr{V}(r)\;\corr{\theta'}_J(r)^2
    - \corr{V}'(r) \frac{\DDRp(r)}{\DRp(r)^2}\bigg]\\
    & -\frac{1}{2} \big[ u_{n,J\mpS 1}^{\star}(r)\, u''_{n',J\mpS 1}(r) + u_{n,J\mpS 1}^{\prime\prime\star}(r)\, u_{n',J\mpS 1}(r) \big]
    \frac{\corr{V}(r)}{\DRp(r)^2} \bigg\}
  \end{split}
}
with $\corr{\theta'}_J(r) = \theta'_J(\Rp(r))$. Likewise, we find 
\eqmultline{ \label{eq:corr_me_momentum3}
  \matrixe{n(J\mpS 1,1) JT}{\CCO_{r} \CCO_{\Omega} \VO_{qr}
    \CO_{\Omega}\CO_{r}}{n'(J\pmS 1,1) JT} = \\
  \pm \int\!\dd{r}\,
    \big[u_{n,J\mpS 1}^{\star}(r)\, u'_{n',J\pmS 1}(r)
     - u_{n,J\mpS 1}^{\prime\star}(r)\, u_{n',J\pmS 1}(r) \big]
  \frac{\corr{V}(r)\, \corr{\theta'}_J(r)}{\DRp(r)}
}
for the off-diagonal matrix elements with $L=J\mp1$ and $L'=J\pm1$.

The matrix elements for the correlated radial and angular kinetic
energy can be constructed as special cases of the interaction matrix
elements discussed above. By setting $V(r) = 1/(2\mu_r(r))$ in Eqs.
\eqref{eq:corr_me_momentum1} to \eqref{eq:corr_me_momentum3} we obtain
the matrix elements for the effective mass part of the correlated
radial kinetic energy \eqref{eq:corr_op_kinetic_rad}. The matrix
elements of the additional local potential in
\eqref{eq:corr_op_kinetic_rad} and the angular kinetic energy
\eqref{eq:corr_op_kinetic_ang} follow directly from Eqs.
\eqref{eq:corr_me_local1} to \eqref{eq:corr_me_local3}.
%%%%%%%%%%%%%%%%%%%%%%%%%%%%%%%%%%%%%%%%%%%%%%%%%%%%%%%%%%%%%%%%%%%%%%
%%%%%%%%%%%%%%%%%%%%%%%%%%%%%%%%%%%%%%%%%%%%%%%%%%%%%%%%%%%%%%%%%%%%%%
%%%%%%%%%%%%%%%%%%%%%%%%%%%%%%%%%%%%%%%%%%%%%%%%%%%%%%%%%%%%%%%%%%%%%%
\section{Similarity Renormalization Group (SRG)}
\label{sec:srg}

Unlike the UCOM framework, the concept of the Similarity Renormalization Group (SRG) is universal and not tied to the specific correlations relevant in the nuclear many-body problem. The renormalization of the Hamiltonian through flow-equations was proposed by G\l{}azek and Wilson \cite{GlWi93} in the context of light-front field theory and was further developed by Perry \emph{et al.} \cite{SzPe00,GlPe08}. Independently, Wegner \cite{Wegn94,Wegn01} proposed flow equations for the renormalization of Hamiltonians in the context of condensed matter physics. A summary of these developments is given in Refs. \cite{Kehr06,Wegn06} and in Ref.\cite{ScFu10} in this volume. 

Already these initial publications on the SRG contain all formal elements relevant for the application in the nuclear physics context, even the specific choice of the generator for the SRG flow evolution that will be used in the following was discussed by Szpigel and Perry \cite{SzPe00}. The first application of the SRG for the transformation of a nuclear Hamiltonian was presented by Bogner et al. \cite{BoFu07,BoFu07b} and the SRG in connection to the UCOM approach was first discussed in Refs. \cite{HeRo07,RoRe08}.    

The general concept of all implementations of the SRG is the transformation of the Hamiltonian to a band- or block-diagonal structure with respect to a specific basis by a continuous unitary evolution determined via renormalization-group flow equations. The particular physical system and application under consideration determines which basis and generator is used in the flow evolution. In this respect the SRG approach is very flexible and can be adapted to all kinds of band- or block-diagonalizations in any basis of choice \cite{Wegn06,AnBo08}. This flexibility is an advantage of the SRG scheme as compared to the UCOM transformation, which is tailored for a very specific type of correlations. Moreover, the computational simplicity of the SRG-evolution on the level of matrix elements opens a clear path towards a consistent evolution of many-body forces beyond the level of the two-body cluster approximation (cf. Sec. \ref{sec:ucom_cluster}). Recently, the SRG evolution of a nuclear Hamiltonian has been performed on the level of three-body matrix elements \cite{JuNa09}, thus demonstrating the feasibility and power of this scheme.

%%%%%%%%%%%%%%%%%%%%%%%%%%%%%%%%%%%%%%%%%%%%%%%%%%%%%%%%%%%%%%%%%%%%%%
%%%%%%%%%%%%%%%%%%%%%%%%%%%%%%%%%%%%%%%%%%%%%%%%%%%%%%%%%%%%%%%%%%%%%%
\subsection{SRG flow equations}
\label{sec:srg_floweq}

The basic idea of the SRG in the formulation of Wegner \cite{Wegn94,Wegn01,Wegn06,Kehr06} is to transform the initial Hamiltonian $\op{H}$ of a many-body system into a diagonal form with respect to a given basis. The renormalization group flow equation governing the evolution of the Hamiltonian is given by 
\eq{ \label{eq:srg_flow}
  \frac{d\op{H}_\alpha}{d\alpha}
  = \comm{\op{\eta}_\alpha}{\op{H}_\alpha} \;,
}
where $\alpha$ is a formal flow parameter and $\op{H}_{\alpha}$ the evolved Hamiltonian. Here we use a general operator form of the flow equation in many-body space. The central quantity is the anti-hermitian generator $\op{\eta}_\alpha$ which determines the physics of the flow evolution. Formally, this is an initial value problem with the original Hamiltonian $\op{H}$ as initial condition $\op{H}_{\alpha=0}=\op{H}$. Analogous equations can be formulated for the evolution of operators $\op{B}_{\alpha}$ of all observables one is interested in, 
\eq{ \label{eq:srg_flow_op}
  \frac{d\op{B}_\alpha}{d\alpha}
  = \comm{\op{\eta}_\alpha}{\op{B}_\alpha} \;.
}
Apart from trivial cases, the generator $\op{\eta}_\alpha$ will depend on the evolved Hamiltonian $\op{H}_{\alpha}$ itself. Therefore, the flow equation for an observable $\op{B}_{\alpha}$ cannot be solved independently from the flow equation of the Hamiltonian, they have to be solved simultaneously. Formally, we can integrate these flow equations defining a unitary operator $\op{U}_\alpha$ of the explicit transformations   
\eqmulti{ \label{eq:srg_unitrafo}
  \op{H}_\alpha
  &= \adj{\op{U}}_\alpha\op{H}\op{U}_\alpha \;, \\
  \op{B}_\alpha
  &= \adj{\op{U}}_\alpha\op{B}\op{U}_\alpha \;. 
}
Note that we chose to define the unitary operator $\op{U}_\alpha$ such that the adjoint operator appears on the left in the similarity transformation---this is consistent with convention introduced for the UCOM transformation (cf. Eq. \eqref{eq:corr_operator}), but different from many other discussions of the SRG. From Eqs. \eqref{eq:srg_unitrafo} and \eqref{eq:srg_flow} we obtain a differential equation for the unitary operator $\op{U}_\alpha$, 
\eq{ \label{eq:srg_unidgl}
  \frac{d\op{U}_\alpha}{d\alpha}
  = - \op{U}_\alpha \op{\eta}_\alpha \;,
}
which describes an initial value problem with the trivial initial condition $\op{U}_{\alpha=0} = \op{1}$ for the unitary operator. If the generator would be independent of the flow parameter, e.g. $\op{\eta}_{\alpha} \equiv \ii \op{g}$ with a hermitian generator $\op{g}$, then this differential equation could be readily integrated, yielding the standard exponential form of the unitary transformation operator $\op{U}_{\alpha} = \exp(-\ii \alpha \op{g})$. However, typical generators used in the SRG have a non-trivial $\alpha$-dependence, such that the formal solution for the unitary operator does not yield a simple exponential but rather a Dyson series. In practical applications it is, therefore, much easier to solve the flow equations  \eqref{eq:srg_flow} and \eqref{eq:srg_flow_op} directly, without reference to the explicit unitary operator.

The ansatz for the non-hermitian generator $\op{\eta}_\alpha$ originally used by Wegner \cite{Wegn94,Wegn01} has a quite intuitive structure. It is written as a commutator of the diagonal part of the evolved Hamiltonian, $\mathrm{diag}(\op{H}_\alpha)$, with the full Hamiltonian $\op{H}_\alpha$,   
\eq{ \label{eq:srg_generator0}
  \op{\eta}_\alpha 
  = \comm{\text{diag}(\op{H}_\alpha)}{\op{H}_\alpha} \;.
}
Obviously, the definition of $\text{diag}(\op{H}_\alpha)$ presumes the choice of a basis---this is the basis with respect to which the Hamiltonian shall be diagonalized. By transferring the generator \eqref{eq:srg_generator0} and the flow equation \eqref{eq:srg_flow} into a matrix representation for this basis, two properties of the flow become evident: First, the diagonal form of the Hamiltonian provides a fix point of the flow evolution, since the generator $\op{\eta}_{\alpha}$ and thus the right-hand side of the flow equation vanish in this case. Second, the off-diagonal matrix elements of the Hamiltonian are continuously suppressed throughout the flow evolution, the sum of their squares decreases monotonically \cite{Wegn94,Wegn01}. Hence the diagonal form is a trivial attractive fixpoint of the SRG flow equation.

So far this approach is generic and independent of the properties of the particular physical system, the Hamiltonian, or the basis under consideration. If considering an $A$-body system, then all the aforementioned relations refer to the operators in $A$-body space. One of the consequences is that even a simple initial Hamiltonian, containing two-body operators at most, acquires up to $A$-body terms in the course of the evolution. For practical applications of the SRG approach in the nuclear structure context one, therefore, has to simplify the scheme by confining the evolution to two or three-body space, thus discarding higher-order contributions in the evolved interaction. Furthermore, instead of using the diagonal part of the Hamiltonian in the definition of the generator, one can use the operator that defines the eigenbasis with respect to which the Hamiltonian shall be diagonalized. In this way, we depart from the original goal of Wegner's ansatz to diagonalize a Hamiltonian via a flow evolution with $\alpha\to\infty$ and rather aim at the derivation of tamed few-body interactions for intermediate values of the flow parameter $\alpha$ that are pre-diagonalized with respect to a certain basis.

A simplified scheme along these lines was suggested by Szpigel and Perry \cite{SzPe00} and applied by Bogner and others \cite{BoFu07,HeRo07}. It confines the evolution to two-body space and uses the generator   
\eq{ \label{eq:srg_generator}
  \op{\eta}_\alpha
  = (2\mu)^2\; \comm{\op{T}_{\text{int}}}{\op{H}_\alpha}
  = 2\mu\; \comm{\vec{\op{q}}^2}{\op{H}_\alpha}\,,
}
containing the intrinsic kinetic energy $\op{T}_{\text{int}} = \frac{1}{2\mu} \vec{\op{q}}^2$ in the two-body system. The prefactor of the commutator is chosen such that the dimension of the flow parameter $\alpha$ is $[\text{momentum}]^{-4}$ or $[\text{length}]^4$. It is also common to specify the parameter $\lambda=\alpha^{-4}$, which has the dimension of momentum, instead of the flow parameter $\alpha$. The square of the two-body relative momentum operator can be decomposed into a radial and an angular part, 
\eq{ \label{eq:srg_momentumop}
  \vec{\op{q}}^2 
  = \op{q}_r^2 + \frac{\vec{\op{L}}^2}{\op{r}^2}
  \;,\qquad
  \op{q}_r 
  = \frac{1}{2}\Big( \vec{\op{q}} \cdot \frac{\vec{\op{r}}}{\op{r}} 
                   + \frac{\vec{\op{r}}}{\op{r}} \cdot \vec{\op{q}} \Big) \;.
}
Thus an obvious fix point of the evolution with the generator \eqref{eq:srg_generator} is a two-body Hamiltonian $\op{H}_\alpha$ that commutes with $\op{q}_r^2$ and $\vec{\op{L}}^2/\op{r}^2$. Hence, this generator drives the matrix elements of the Hamiltonian towards a band-diagonal structure with respect to relative momentum $(q,q')$ and orbital angular momentum $(L,L')$, i.e., with respect to a partial-wave momentum space representation.

Though we will only use the generator \eqref{eq:srg_generator} in the following, one should note that there are many other possible choices for $\op{\eta}_\alpha$. An evident alternative is to use the single-particle Hamiltonian of the harmonic oscillator instead of the kinetic energy in the generator. In this way, the Hamiltonian is driven towards a diagonal form in the harmonic oscillator basis. Using various projection operators one can design generators that drive the Hamiltonian towards a block-diagonal structure in a given basis \cite{AnBo08}. This flexibility of the SRG technique holds great potential for further refinements and applications of the approach.

%%%%%%%%%%%%%%%%%%%%%%%%%%%%%%%%%%%%%%%%%%%%%%%%%%%%%%%%%%%%%%%%%%%%%%
%%%%%%%%%%%%%%%%%%%%%%%%%%%%%%%%%%%%%%%%%%%%%%%%%%%%%%%%%%%%%%%%%%%%%%
\subsection{Evolution of two-body matrix elements}
\label{sec:srg_evolint}

Starting from an initial two-body Hamiltonian $\op{H}$ composed of relative kinetic energy $\op{T}_{\text{rel}}$ and two-body interaction $\op{V}$ it is convenient to decompose the SRG-evolved Hamiltonian $\op{H}_\alpha$ in a similar way
\eq{ \label{eq:srg_definitiov}
  \op{H}_{\alpha} = \op{T}_{\text{rel}} + \op{V}_{\alpha} \;.
}
All flow-dependence is absorbed in the SRG-evolved two-body interaction $\op{V}_\alpha$ defined by this relation. Rewriting of the flow equation \eqref{eq:srg_flow} using the generator \eqref{eq:srg_generator} explicitly for the evolved interaction $\op{V}_\alpha$ leads to 
\eq{ \label{eq:srg_flowv}
  \frac{d\op{V}_\alpha}{d\alpha}
  = \comm{\op{\eta}_\alpha}{\op{T}_{\text{rel}} + \op{V}_\alpha} 
  = (2\mu)^2\; \comm{ \comm{\op{T}_{\text{rel}}}{\op{V}_\alpha}}{\op{T}_{\text{rel}} + \op{V}_\alpha}  \;.
}

Even in this simplified form a direct solution of the operator equation is far from trivial. For practical applications we, therefore, work on the level of matrix elements. Though any basis in two-body space can be used to define this matrix representation, it is convenient to use the eigenbasis of the operator entering into the ansatz for the generator \eqref{eq:srg_generator}. In our case, this is the $\vec{\op{q}}^2$ operator and it is most convenient to adopt the partial-wave momentum eigenbasis $\ket{q(LS)JT}$, where the projection quantum numbers $M$ and $M_T$ have been omitted for brevity.

In this basis the flow equation \eqref{eq:srg_flowv} translates into a set of coupled integro-differential equations for the matrix elements 
\eq{
  V^{(JLL'ST)}_{\alpha}(q,q')
  = \matrixe{q(LS)JT}{\op{V}_{\alpha}}{q'(L'S)JT} \;.
}
In a generic form, the resulting evolution equation reads: 
\eqmulti{ \label{eq:srg_flowvme}
  \frac{d}{d\alpha} V_{\alpha}(q,q') 
  &= - (q^2-q'^2)^2\; V_{\alpha}(q,q') \\
  &+ 2\mu \int dQ\, Q^2\; (q^2 + q'^2 -2 Q^2)\; 
    V_{\alpha}(q,Q) V_{\alpha}(Q,q') \;.
}
For non-coupled partial waves with $L=L'=J$, the matrix elements entering this equation are simply 
\eq{ \label{eq:srg_vme_noncoupled}
  V_{\alpha}(q,q') 
  = V^{(JJJST)}_{\alpha}(q,q') \;.
}
For coupled partial waves with $L,L'=J\pm1$, the $V_{\alpha}(q,q')$ are understood as $2\times2$ matrices of the matrix elements for the different combinations of the orbital angular momenta $L = J-1$ and $L'=J+1$
\eq{ \label{eq:srg_vme_coupled}
  V_{\alpha}(q,q') 
  = \begin{pmatrix}
    V^{(JLLST)}_{\alpha}(q,q')  & V^{(JLL'ST)}_{\alpha}(q,q') \\
    V^{(JL'LST)}_{\alpha}(q,q')  & V^{(JL'L'ST)}_{\alpha}(q,q') \\
  \end{pmatrix} \;.
}
Each non-coupled partial wave and each set of coupled partial waves evolves independently of the other channels of the interaction. This is a direct consequence of the choice of the generator --- the evolution towards a diagonal in momentum space is done in an optimal way for each individual partial wave.

As mentioned earlier, analogous evolution equations have to be solved for all observables in order to arrive at a consistent set of effective operators. The evolution of these operators, e.g. the multipole operators necessary for the evaluation of transition strengths or the one-body density operators employed for the computation of the momentum distribution, is coupled to the evolution of the Hamiltonian via the generator $\eta_\alpha$. Hence we have to solve these evolution equations simultaneously. 

An alternative approach is to determine the matrix elements of the unitary operator $\op{U}_\alpha$ explicitly by solving \eqref{eq:srg_unidgl}. The evolved matrix elements of all observables can then be obtained by a simple matrix transformation using the same unitary transformation matrix. In the case of the momentum-space partial-wave matrix elements of the unitary transformation operator,
\eq{
  U^{(JLL'ST)}_{\alpha}(q,q')
  = \matrixe{q(LS)JT}{\op{U}_{\alpha}}{q'(L'S)JT} \;,
}
the operator equation \eqref{eq:srg_unidgl} leads to a coupled set of integro-differential equations 
\eq{ \label{eq:srg_flowume}
  \frac{d}{d\alpha} U_{\alpha}(q,q') 
  = 2\mu \int dQ\, Q^2\; (q'^2 - Q^2)\; 
    U_{\alpha}(q,Q) V_{\alpha}(Q,q')\;,
}
where we assume that the evolution equation \eqref{eq:srg_flowvme} is solved simultaneously providing the $V_{\alpha}(q,q')$. The generic notation defined in \eqref{eq:srg_vme_noncoupled} and \eqref{eq:srg_vme_coupled} for non-coupled and coupled partial waves, respectively, applies here as well. This differential equation provides direct access to the matrix elements of the unitary operator, which maps the initial operators onto any particular point of the flow trajectory.

%%%%%%%%%%%%%%%%%%%%%%%%%%%%%%%%%%%%%%%%%%%%%%%%%%%%%%%%%%%%%%%%%%%%%%
%%%%%%%%%%%%%%%%%%%%%%%%%%%%%%%%%%%%%%%%%%%%%%%%%%%%%%%%%%%%%%%%%%%%%%
\subsection{Evolved interactions and wave functions}
\label{sec:srg_mewavefunc}

%%%%%%%%%%%%%%%%%%%%%%%%%%%%%%%%%%%%%%%%%%%%%%%%%%%%%%%%%%%%%%%%%%%%%%
\begin{figure}[p]
\centering\includegraphics[width=0.95\textwidth]{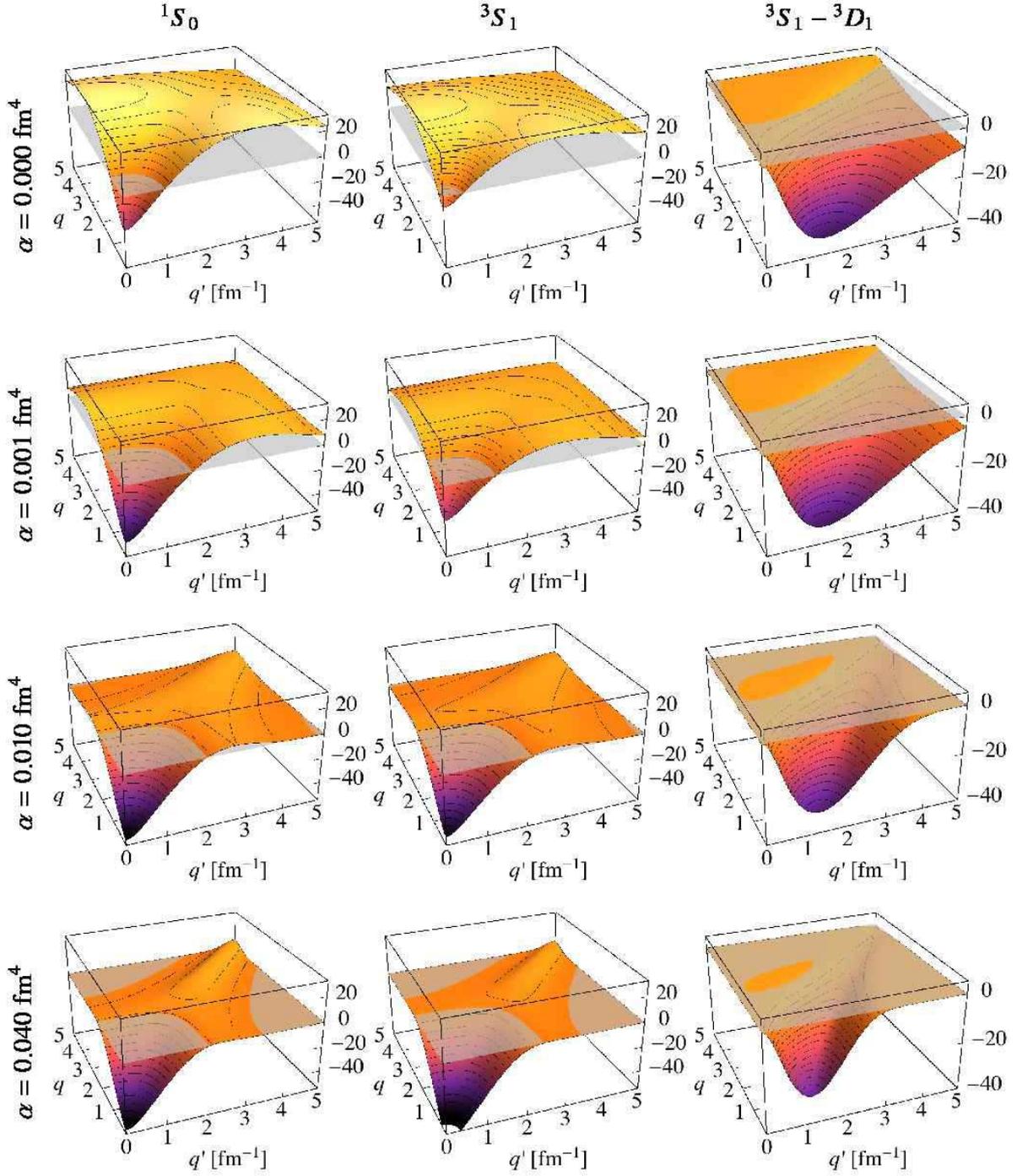}
\caption{Snapshots of the momentum-space matrix elements (in units of MeV fm$^3$) of the SRG-evolved Argonne V18 potential (charge independent parts only) in the $^{1}{S}_0$, $^{3}{S}_1$, and $^{3}{S}_1$-$^{3}{D}_1$ partial waves (from left to right) for the flow parameters $\alpha=0.0\,\text{fm}^4$, $0.001\,\text{fm}^4$, $0.01\,\text{fm}^4$, and $0.04\,\text{fm}^4$ (from top to bottom).}
\label{fig:srg_matrixelem_evolution}
\end{figure}
%%%%%%%%%%%%%%%%%%%%%%%%%%%%%%%%%%%%%%%%%%%%%%%%%%%%%%%%%%%%%%%%%%%%%%
\afterpage{\clearpage}

The concept of the SRG transformation becomes very transparent when looking at the flow evolution of the momentum-space matrix elements of the SRG-transformed interaction $\op{V}_{\alpha}$. In Fig. \ref{fig:srg_matrixelem_evolution} we show the matrix elements $V^{(JLL'ST)}_{\alpha}(q,q')$ obtained for the Argonne V18 potential in the three most important partial waves: the ${}^{1}{S}_0$ partial wave, i.e. matrix elements $V_{\alpha}^{(00001)}(q,q')$, the ${}^{3}{S}_1$ partial wave, i.e. $V_{\alpha}^{(10010)}(q,q')$, and the ${}^{3}{S}_1-{}^{3}{D}_1$ partial wave, i.e. $V_{\alpha}^{(10210)}(q,q')$. We start with the matrix elements of the initial Argonne V18 potential at $\alpha=0\,\text{fm}^4$ and display snapshots of the SRG evolution at $\alpha=0.001\;\text{fm}^4$, $\alpha=0.01\;\text{fm}^4$, and $\alpha=0.04\;\text{fm}^4$.

The initial matrix elements show the characteristic features that are responsible for the emergence of strong correlations in the many-body system: The strong off-diagonal matrix elements that couple low-momentum components with high-momentum components of the wave function. In the $^1S_0$ and $^3S_1$ partial waves these off-diagonal high-momentum matrix elements are generated by the short-range repulsion of the Argonne V18 potential, while the off-diagonal matrix elements in the ${}^3S_1-{}^3D_1$ partial wave are solely due to the tensor interaction. 

Already in the early phase of the flow evolution, i.e. for $\alpha\lesssim0.01\,\text{fm}^4$, the matrix elements far off the diagonal in the $S$-wave channels are suppressed quickly. The plateau of positive high-momentum matrix elements is pushed towards the zero-plane and the negative low-momemtum matrix elements are enhanced. Later in the flow evolution, the residual high-momentum matrix elements are pushed towards the diagonal and the matrix elements in the low-momentum are further enhanced. For the tensor-dominated ${}^3S_1-{}^3D_1$ partial wave the matrix elements far-off the diagonal are depleted successively. Over all, the SRG-evolution with the generator \eqref{eq:srg_generator} leads to a transformed interaction with band-diagonal matrix elements in momentum space. The pre-diagonalization of the interaction and thus the Hamiltonian matrix elements through the SRG transformation is evident.

%%%%%%%%%%%%%%%%%%%%%%%%%%%%%%%%%%%%%%%%%%%%%%%%%%%%%%%%%%%%%%%%%%%%%%
\begin{figure}[t]
\centering\includegraphics[width=0.85\textwidth]{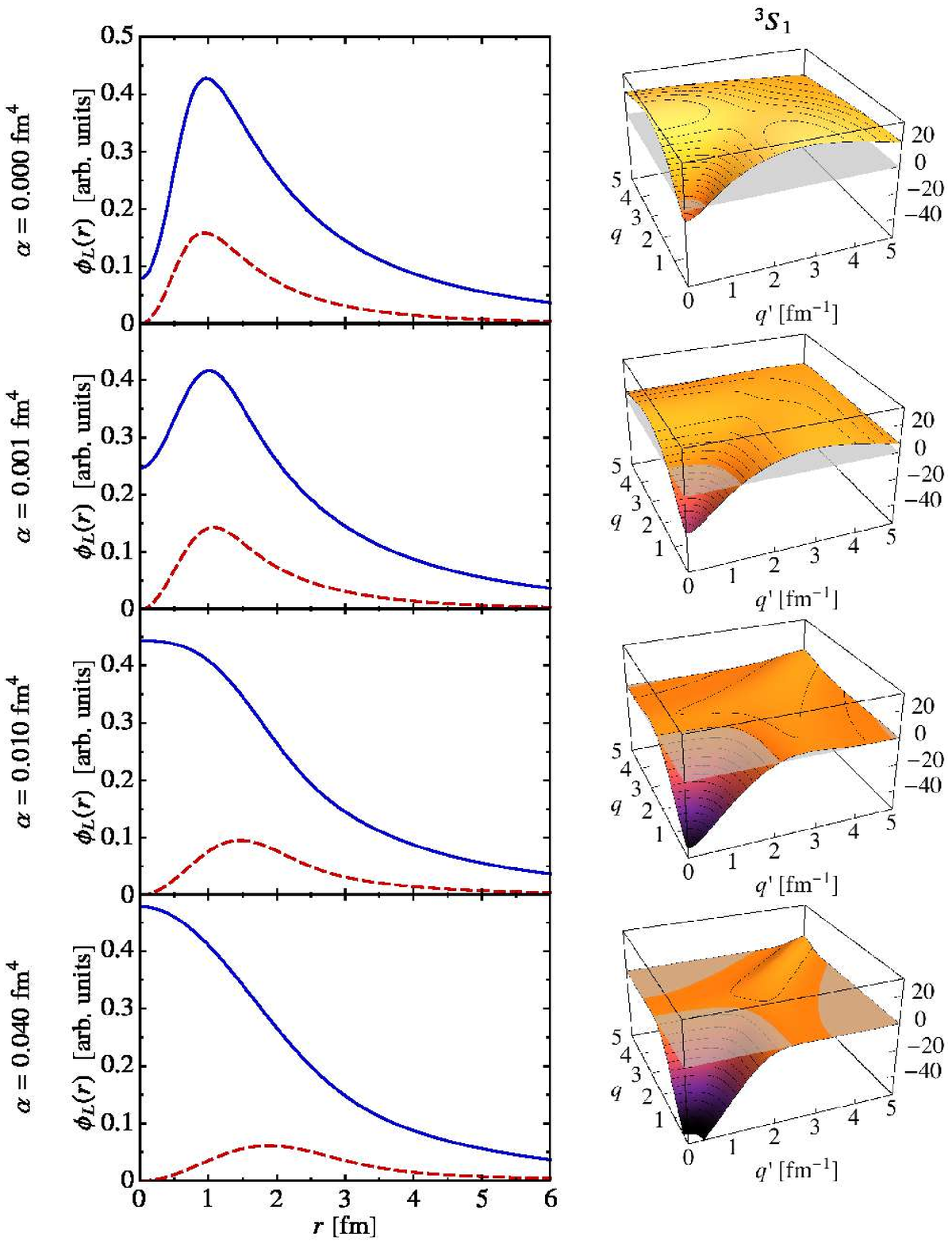}
\caption{Snapshots of the deuteron wave function obtained from the SRG-evolved Argonne V18 potential for the flow parameters $\alpha=0.0\,\text{fm}^4$, $0.001\,\text{fm}^4$, $0.01\,\text{fm}^4$, and $0.04\,\text{fm}^4$ (from top to bottom). The main panels show the radial wave functions $\phi_L(r)$ for $L=0$ (\linemediumsolid[FGBlue]) and $L=2$ (\linemediumdotted[FGRed]). The 3D plots show the corresponding momentum-space matrix elements in the $^3S_1$ partial wave for orientation (cf. Fig. \ref{fig:srg_matrixelem_evolution}).}
\label{fig:srg_wavefunction_evolution}
\end{figure}
%%%%%%%%%%%%%%%%%%%%%%%%%%%%%%%%%%%%%%%%%%%%%%%%%%%%%%%%%%%%%%%%%%%%%%

In order to assess the effect of the flow-evolution on the structure of wave functions, we again consider the deuteron ground state as an example. Using the SRG-evolved momentum-space matrix elements we solve the two-body problem in the deuteron channel numerically. The dependence of the resulting coordinate-space wave functions on the flow parameter $\alpha$ provides an intuitive picture of the effect of the SRG evolution in coordinate space. This is illustrated in Fig.~\ref{fig:srg_wavefunction_evolution}. 

The deuteron wave function for the initial interaction shows the signatures of strong short-range central and tensor correlations, i.e., the suppression of the wave function at small inter-particle distances and the presence of the $D$-wave admixture, respectively. Already early in the flow evolution, i.e. for $\alpha\lesssim0.01\,\text{fm}^4$, the short-range dip in the $S$-wave component is removed---this effect is connected to the suppression of the plateau of high-momentum matrix elements in the $^3S_1$ partial wave. As a result the deuteron wave function obtained with the SRG-evolved Argonne V18 potential for $\alpha=0.01\,\text{fm}^4$ has lost any signature of a strong short-range repulsion in the interaction. The total strength of the $D$-wave admixture is reduced and pushed towards larger inter-particle distances. Thus, the short-range wave function evolves from being dominated by short-range central and tensor correlations to an almost uncorrelated pure $S$-wave. The long-range behavior, all asymptotic properties, and the deuteron binding energy are not affected by the SRG-evolution.

The behavior of the matrix elements and of the two-body wave functions highlights the relation between the UCOM transformation and the SRG evolution regarding the pre-diagonalization of the Hamiltonian and the description of short-range correlations. Both approaches describe the same physics.

%%%%%%%%%%%%%%%%%%%%%%%%%%%%%%%%%%%%%%%%%%%%%%%%%%%%%%%%%%%%%%%%%%%%%%
%%%%%%%%%%%%%%%%%%%%%%%%%%%%%%%%%%%%%%%%%%%%%%%%%%%%%%%%%%%%%%%%%%%%%%
\subsection{UCOM from an SRG perspective}
\label{sec:srg_ucomgen}

Because the UCOM and the SRG transformations have the same effect on the matrix elements and wave functions,
one might ask for the connection of both approaches on the underlying formal level. The properties of both unitary transformations are governed by their generators: the dynamical generator $-\ii\op{\eta}_{\alpha}$ in the case of the SRG and the static generators $\op{g}_{r}$ and $\op{g}_{\Omega}$ in the case of the UCOM transformation. 

There is a non-trivial relation between these generators that provides an insight into the formal relation and the differences of the  two approaches \cite{HeRo07}. This becomes evident by evaluating the SRG generator \eqref{eq:srg_generator} for a typical nuclear Hamiltonian in two-body space. We assume a simplified local two-nucleon interaction composed of a central, a spin-orbit and a tensor part. The operator for this interaction is given by
\eq{
  \op{V}=\sum_p V_p(\op{r})\,\op{O}_p 
}
with $\op{O}_p \in \{\op{1}, \sigmasigmaO, \spinorbitO, \tensorRRO \} \otimes \{\op{1}, \tautauO \}$. By evaluating the commutator defining the SRG generator \eqref{eq:srg_generator} explicitly for $\alpha=0$ using this interaction operator we obtain  
\eq{ \label{eq:srg_ucom_gen}
  -\ii \op{\eta}_0
  =  \bigg[ \frac{1}{2} \big(\op{q}_r\,\op{S}(\op{r}) + \op{S}(\op{r})\,\op{q}_r \big) + \Theta(\op{r})\, \op{S}_{12}(\vec{\op{r}},\vec{\op{q}}_{\Omega}) \bigg] \;.
}
The operator-valued functions $\op{S}(\op{r})$ and $\op{\Theta}(\op{r})$ contain the radial dependencies of the different terms of the interaction
\eq{ \label{eq:srg_ucom_corrfunc}
  \op{S}(\op{r}) = -\frac{1}{\mu} \bigg(\sum_p V'_p(\op{r})\, \op{O}_p \bigg)  \;,\qquad 
  \op{\Theta}(\op{r}) = -\frac{2}{\mu}\frac{V_t(\op{r})}{\op{r}^2}  \,. 
}
If the functions $\op{S}(\op{r})$ and $\op{\Theta}(\op{r})$ were functions of the relative distance $\op{r}$ alone, then the structure of the initial SRG generator $\ii\op{\eta}_0$ would be identical to the UCOM generators $\op{g}_r$ and $\op{g}_\Omega$ that were constructed based on the physical picture of central and tensor correlations \cite{HeRo07}. The explicit dependence of $\op{S}(\op{r})$ on the operator set $\op{O}_p$ indicates that the SRG transformation acts differently in different partial waves. In the UCOM terminology, this dependence encodes a partial-wave dependence of the central correlation functions.

This formal connection shows that both approaches address the same physics of short-range correlations, although starting from quite different backgrounds. Moreover, it proves that the set of UCOM generators covers the most relevant terms. Although there are other operators appearing in the initial interaction, e.g. the spin-orbit operator, they do not require separate generators---their effect on the correlations is absorbed in the operator-valued function $\op{S}(\op{r})$.   

Regarding the partial-wave dependence of the correlations, the standard formulation of the UCOM approach uses a simplified picture. The correlation functions $s(r)$ and $\vartheta(r)$ are chosen to depend on spin $S$ and isospin $T$ only, they do not depend on orbital and total angular momentum. Formally, one could drop this restriction and work with separate correlation functions for each partial wave and thus mimic the flexibility of the SRG generator. 

A more fundamental difference between the UCOM and the SRG transformations results from the fact that SRG uses a dynamical generator, whose operator structure changes throughout the flow evolution, whereas UCOM is based on a static generator. Though the UCOM generator and the SRG generator share the same basic operator structure at $\alpha=0$, the SRG generator acquires a more complicated form involving higher-order momentum and momentum-dependent tensor operators at later stages of the evolution. Therefore, the SRG generator is more flexible and adapts to the behavior of the matrix elements during the flow evolution, leading to a non-trivial flow trajectory in an operator-space representing the generator. The UCOM transformation, in contrast, consists of a one-step transformation along a linear trajectory confined to a subspace of the operator-space spanned by the SRG generator. It is, therefore, not surprising that the matrix elements of the UCOM-transformed interaction do not exhibit the same perfect band-diagonal structure as the SRG-evolved interaction. However, the leading operator contributions to the generator are also present in the UCOM approach and allow for an efficient pre-diagonalization.

The dynamic nature of the SRG generator is also the reason, why the optimal UCOM correlation function cannot be determined directly from Eqs. \eqref{eq:srg_ucom_corrfunc}. We have to consider the whole SRG flow trajectory up to a certain value of $\alpha$ to extract meaningful UCOM correlation functions. One option to do so is discussed in the following section.

%%%%%%%%%%%%%%%%%%%%%%%%%%%%%%%%%%%%%%%%%%%%%%%%%%%%%%%%%%%%%%%%%%%%%%
%%%%%%%%%%%%%%%%%%%%%%%%%%%%%%%%%%%%%%%%%%%%%%%%%%%%%%%%%%%%%%%%%%%%%%
\subsection{UCOM correlation functions extracted from SRG}
\label{sec:srg_ucomcorr}

As an alternative to the variational determination of the UCOM correlation functions $R_+(r)$ and $\vartheta(r)$ discussed in Sec. \ref{sec:optcorr} we can use the SRG approach to generate optimized UCOM correlation functions. Our aim is to construct a UCOM transformation that uses the result of the SRG-evolution of a Hamiltonian in two-body space from $\alpha=0$ to a fixed finite value of the flow parameter to determine $R_+(r)$ and $\vartheta(r)$. In contrast to the dynamical SRG evolution, the UCOM correlations have to map the initial Hamiltonian onto the evolved Hamiltonian for a specific $\alpha$ using a single explicit unitary transformation. Obviously, the correlation operator $\op{C}$ is not flexible enough to allow for an exact mapping of all matrix elements in all partial waves---even if we would allow for different correlation operators $\op{C}$ for each partial wave. 

One could consider an approximate mapping of the matrix elements as one possible scheme to determine the UCOM correlation functions. Here, we use a different strategy, which is rooted in the interpretation of the UCOM transformation as a tool to imprint short-range correlations into the many-body state. Instead of considering the initial and evolved two-body matrix elements, we consider two-body eigenstates of the initial and evolved Hamiltonian for different partial waves. The optimal UCOM correlation functions are then required to map a selected two-body eigenstate of the SRG-evolved Hamiltonian onto the corresponding eigenstate of the initial Hamiltonian. This wave-function mapping defines the so-called SRG-generated UCOM correlation functions.

The procedure for the construction of SRG-generated UCOM correlation functions consists of three steps: (\emph{i}) We solve the SRG evolution equations for a given initial interaction up to a flow parameter $\alpha$, obtaining the momentum space matrix elements $V_\alpha(q,q')$ for a certain partial wave. (\emph{ii}) Using the evolved matrix elements the two-body problem is solved, leading to a set of coordinate-space wave functions. (\emph{iii}) The UCOM correlation functions $R_+(r)$ and $\vartheta(r)$ are determined such that they map a selected two-body eigenstate of the SRG evolved interaction onto the corresponding two-body state of the initial interaction in the respective partial wave.

The steps (\emph{i}) and (\emph{ii}), i.e., the evolved momentum-space matrix elements and the wave functions of the corresponding two-body eigenstates, respectively, have already been illustrated for the deuteron channel in Sec. \ref{sec:srg_mewavefunc}. Step (\emph{iii}) is discussed in the following.
  
Consider two eigenstates $\ket{\varphi^{(0)}}$ and $\ket{\varphi^{(\alpha)}}$ with the same energy eigenvalue resulting from the solution of the two-body problem for the initial and the SRG-evolved potential, respectively, in a given coupled or non-coupled partial wave. We can define a UCOM correlation operator $\op{C}$ that maps the two states onto each other 
\eq{ \label{eq:srg_srgcorr_definition}
  \ket{\varphi^{(0)}} 
  = \op{C} \ket{\varphi^{(\alpha)}}
  = \op{C}_\Omega \op{C}_r \ket{\varphi^{(\alpha)}} \;.
}
Based on this formal definition we can derive equations that determine the correlation functions $R_-(r)$ and $\vartheta(r)$ that characterize the correlation operator.

For non-coupled partial waves with $L=J$ only the central correlator is relevant. With the two-body solutions 
\eqmulti{
  \ket{\varphi^{(0)}} 
  &= \ket{\phi^{(0)} (LS)JT} \\
  \ket{\varphi^{(\alpha)}} 
  &= \ket{\phi^{(\alpha)} (LS)JT}  
}
for the initial and the SRG-evolved interaction, respectively, we obtain from \eqref{eq:srg_srgcorr_definition} and \eqref{eq:corr_central_states} a relation connecting the known radial wave functions $\phi^{(0)}(r)$ and $\phi^{(\alpha)}(r)$ via a yet unknown correlation function $R_-(r)$:
\eq{ \label{eq:srgcorr_noncoupledpw_pre}
  \phi^{(0)}(r) 
  = \frac{R_-(r)}{r} \sqrt{R'_-(r)}\; \phi^{(\alpha)}(R_-(r)) \;.
}
Here and in the following we assume real-valued wave functions. The relation \eqref{eq:srgcorr_noncoupledpw_pre} can be viewed as a differential equation for the correlation function $R_-(r)$. After formal integration we arrive at an implicit integral equation for $R_-(r)$
\eq{  \label{eq:srgcorr_noncoupledpw}
  [R_-(r)]^3 
  = 3 \int_0^{\,r} d\xi\; \xi^2 \frac{[\phi^{(0)}(\xi)]^2}{[\phi^{(\alpha)}(R_-(\xi))]^2} \;,
}
which can be solved easily in an iterative fashion. We end up with a discretized representation of the correlation function $R_-(r)$ for the partial wave under consideration. By construction it maps a selected SRG-evolved two-body state onto the corresponding initial state. In general, $R_-(r)$ will depend on the pair of states, e.g. the ground states or a pair of excited states, we have selected. We will show later on that this dependence is very weak.

For coupled partial waves with $L=J-1$ and $L'=J+1$ central and tensor correlators act simultaneously. Using the two-body eigenstates 
\eqmulti{
  \ket{\varphi^{(0)}} 
  &= \ket{\phi^{(0)}_{L} (LS)JT} + \ket{\phi^{(0)}_{L'} (L'S)JT} \\
  \ket{\varphi^{(\alpha)}} 
  &= \ket{\phi^{(\alpha)}_{L} (LS)JT} + \ket{\phi^{(\alpha)}_{L'} (L'S)JT} 
}
of the initial interaction and the evolved interaction, respectively, we can extract a unique set of central and tensor correlation functions. After multiplying the mapping equation \eqref{eq:srg_srgcorr_definition} with $\bra{r(LS)JT}$ and $\bra{r(L'S)JT}$, respectively, and using Eq. \eqref{eq:corr_states}, we obtain a system of coupled equations
\eq{ \label{eq:srgcorr_coupledpw}
  \begin{pmatrix}
  \phi^{(0)}_{L}(r) \\
  \phi^{(0)}_{L'}(r)
  \end{pmatrix}
  =
  \frac{R_-(r)}{r} \sqrt{R'_-(r)} \;
  \begin{pmatrix}
  \cos\theta_J(r) && \sin\theta_J(r) \vphantom{\phi^{(\alpha)}_{L}} \\
  -\sin\theta_J(r) && \cos\theta_J(r) \vphantom{\phi^{(\alpha)}_{L'}}
  \end{pmatrix}\;
  \begin{pmatrix}
  \phi^{(\alpha)}_{L}(R_-(r)) \\
  \phi^{(\alpha)}_{L'}(R_-(r))
  \end{pmatrix} \;,
}
from which the correlation functions  $R_-(r)$ and $\vartheta(r)$ can be determined.

Because the central correlation function acts on both orbital components in the same way and because the transformation matrix in \eqref{eq:srgcorr_coupledpw} has to be unitary, we can determine the central correlation function $R_-(r)$ independently of the tensor correlations function $\vartheta(r)$. By considering the sum of the squares of the two orbital components we obtain from \eqref{eq:srgcorr_coupledpw} the identity
\eq{
  [\phi^{(0)}_{L}(r)]^2 + [\phi^{(0)}_{L'}(r)]^2 
  = \frac{[R_-(r)]^2}{r^2} R'_-(r)\; 
  \big( [\phi^{(\alpha)}_{L}(R_-(r))]^2 + [\phi^{(\alpha)}_{L'}(R_-(r))]^2 \big) \;.
}
which corresponds to \eqref{eq:srgcorr_noncoupledpw_pre} for the non-coupled case. The correlation function $R_-(r)$ can then be determined iteratively from the integral equation
\eq{
  [R_-(r)]^3 
  = 3 \int_0^{\,r} d\xi\; \xi^2 \frac{[\phi^{(0)}_{L}(\xi)]^2 + [\phi^{(0)}_{L'}(\xi)]^2} 
    {[\phi^{(\alpha)}_{L}(R_-(\xi))]^2 + [\phi^{(\alpha)}_{L'}(R_-(\xi))]^2} \;.
}
Once $R_-(r)$ is known, the system \eqref{eq:srgcorr_coupledpw} reduces to a set of two nonlinear equations for $\theta_J(r) = 3 \sqrt{J(J+1)}\; \vartheta(r)$, which can be solved numerically for each $r$. 

In practice this mapping scheme can be easily implemented using discretized wave functions. Typically, the SRG evolution of the Hamiltonian in two-body space for a given partial wave is performed on a sufficiently large grid in momentum space. Using the discretized momentum-space matrix elements of the evolved Hamiltonian we solve the two-body problem for the respective partial wave on the same momentum-space grid. The ground state wave-functions are then transformed to coordinate space, where the mapping equation \eqref{eq:srgcorr_coupledpw} is solved. In this way, we obtain discretized correlation functions $R_-(r)$ and, by numerical inversion, $R_+(r)$ as well as $\vartheta(r)$ for each partial wave. In contrast to the UCOM correlation functions determined variationally, there are no parameterizations of the correlation functions necessary, which might induce artifacts due to their limited flexibility.

%%%%%%%%%%%%%%%%%%%%%%%%%%%%%%%%%%%%%%%%%%%%%%%%%%%%%%%%%%%%%%%%%%%%%%
%%%%%%%%%%%%%%%%%%%%%%%%%%%%%%%%%%%%%%%%%%%%%%%%%%%%%%%%%%%%%%%%%%%%%%
\subsection{SRG-generated UCOM correlation functions}
\label{sec:srg_ucomsrg_av18}

As an example for the determination of UCOM correlations functions through a mapping of SRG-evolved wave functions, we again consider the Argonne V18 potential. Note, however, that the mapping procedure is completely generic and can be used with any other interaction, be it local or non-local.

In order to stay within the framework set by the variational UCOM correlators, we do not consider separate correlation functions for each partial wave, which could be done easily, but only distinguish different channels of two-body spin $S$ and isospin $T$. As in the variational scheme, we optimize the correlators for a given $(S,T)$-channel using the lowest $J$ partial wave, since the low-$J$ and thus low-$L$ partial waves are affected most by short-range correlations. For partial waves with higher $J$ and $L$ the impact of short-range correlations and of the UCOM transformation is reduced due to the angular momentum barrier, thus the non-optimal correlators for these partial waves do not have a big impact.  

For determining the central correlation functions $R_+(r)$ in the spin-singlet channels, we use the ${}^1S_0$ partial wave for the singlet-even ($S=0$, $T=1$) and the $^1P_1$ partial wave for the single-odd ($S=0$, $T=0$) channel. The central and tensor correlation functions in the triplet-even ($S=1$, $T=0$) channel are extracted from the deuteron solution in the coupled ${}^3S_1-{}^3D_1$ partial wave. For the triplet-odd channel ($S=1$, $T=1$) we encounter the same ambiguity as in the variational treatment: the lowest possible orbital angular momentum allowed by antisymmetry is $L=1$ for which $J$ can be $0$, $1$, or $2$ and only for $J=2$ the tensor correlator does contribute. One possible recipe for handling this channel is to use only the coupled ${}^3P_2-{}^3F_2$ partial wave to fix the triplet-odd central and tensor correlation functions, as done in Ref.~\cite{RoRe08}. The central correlation functions obtained in this way are not well adapted for the lower-$J$ partial waves. Therefore, a scheme that includes all possible $J$ for the determination of the central correlation function seems more appropriate. Thus, we determine the central correlation function through a mapping of wave functions for a pseudo interaction obtained by averaging the ${}^3P_0$, ${}^3P_1$, and ${}^3P_2$ partial waves with a relative weight $2J+1$. This recipe comes closest to the energy-average used for the variational determination in this channel (cf. Sec. \ref{sec:optcorr}). 

In each of the partial waves we use the energetically lowest pair of states, which is bound in the case of the triplet-even channel and unbound otherwise, to determine the correlation functions via the mapping \eqref{eq:srg_srgcorr_definition}. In principle one could use any other pair of two-body states obtained for the initial and the SRG-evolved Hamiltonian with the same energy. As was shown in Ref.~\cite{RoRe08} the correlation functions do not change significantly when using one of the low-lying excited two-body states instead of the ground state.     

%%%%%%%%%%%%%%%%%%%%%%%%%%%%%%%%%%%%%%%%%%%%%%%%%%%%%%%%%%%%%%%%%%%%%%
\begin{figure}[p]
\centering\includegraphics[width=0.85\textwidth]{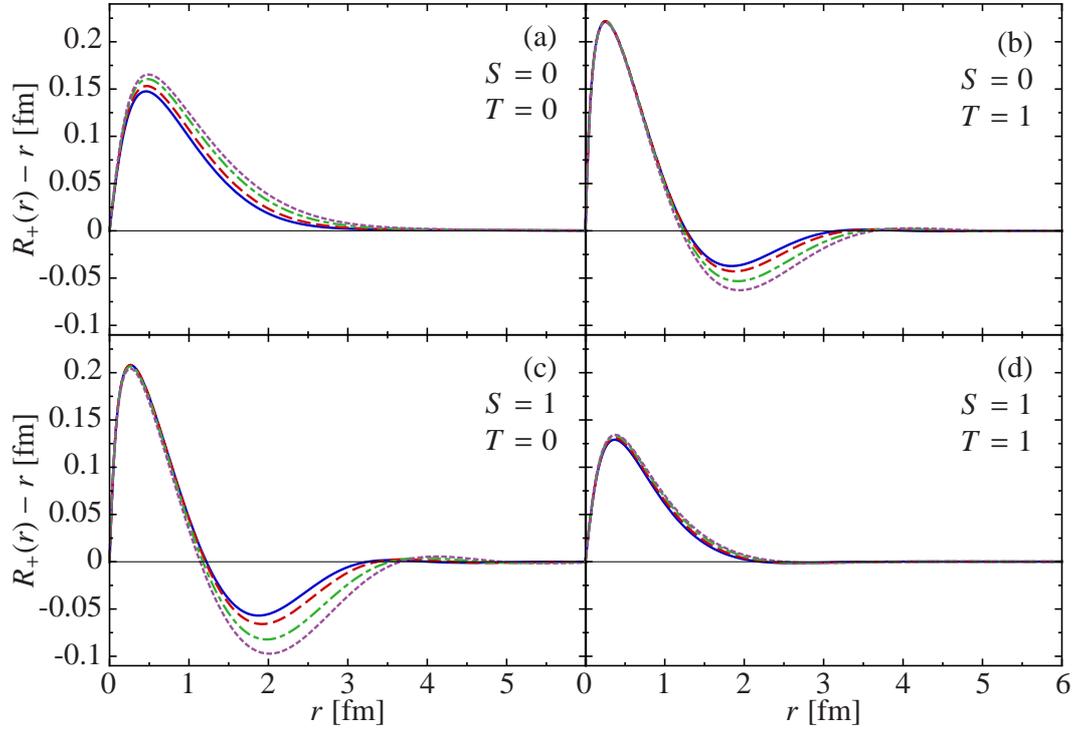}
\caption{Illustration of the dependence of the SRG-generated central correlation functions $R_+(r)-r$ for the Argonne V18 potential on the flow parameter $\alpha$ for the different spin $S$ and isospin $T$ channels. The curves correspond to $\alpha=0.03\text{fm}^4$ (\linemediumsolid[FGBlue]), $\alpha=0.04\text{fm}^4$ (\linemediumdashed[FGRed]), $\alpha=0.06\text{fm}^4$ (\linemediumdashdot[FGGreen]), and $\alpha=0.08\text{fm}^4$ (\linemediumdotted[FGViolet]).}
\label{fig:srg_correlator_Rp_alpha}
\end{figure}
%%%%%%%%%%%%%%%%%%%%%%%%%%%%%%%%%%%%%%%%%%%%%%%%%%%%%%%%%%%%%%%%%%%%%%
%%%%%%%%%%%%%%%%%%%%%%%%%%%%%%%%%%%%%%%%%%%%%%%%%%%%%%%%%%%%%%%%%%%%%%
\begin{figure}[p]
\centering\includegraphics[width=0.85\textwidth]{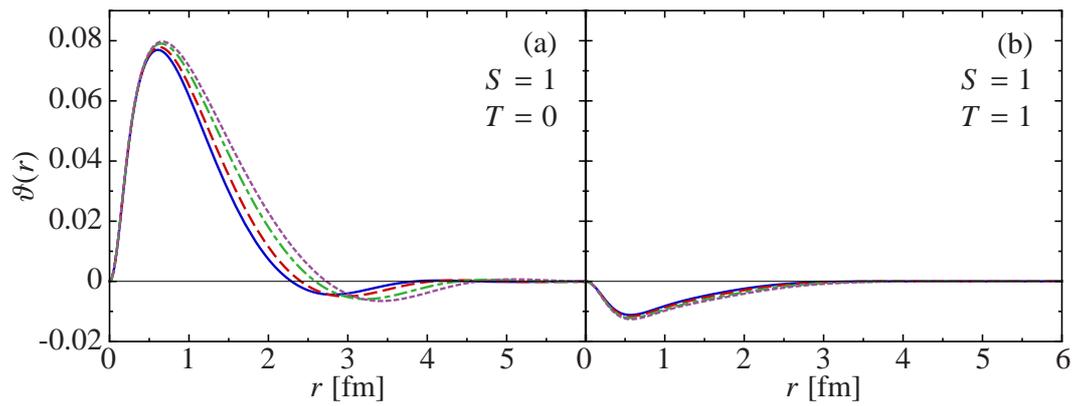}
\caption{Illustration of the dependence of the SRG-generated tensor correlation functions $\vartheta(r)$ for the Argonne V18 potential on the flow parameter $\alpha$ for the different spin $S$ and isospin $T$ channels. The curves correspond to $\alpha=0.03\text{fm}^4$ (\linemediumsolid[FGBlue]), $\alpha=0.04\text{fm}^4$ (\linemediumdashed[FGRed]), $\alpha=0.06\text{fm}^4$ (\linemediumdashdot[FGGreen]), and $\alpha=0.08\text{fm}^4$ (\linemediumdotted[FGViolet]).}
\label{fig:srg_correlator_Th_alpha}
\end{figure}
%%%%%%%%%%%%%%%%%%%%%%%%%%%%%%%%%%%%%%%%%%%%%%%%%%%%%%%%%%%%%%%%%%%%%%

A crucial advantage of the SRG-generated correlation functions is that there is no need for artificial constraints to control the range of the correlations functions---the SRG flow-parameter $\alpha$, which enters through the evolved two-body eigenstates, is the only control parameter. In contrast to the ad-hoc integral constraints formulated for central and tensor correlations functions in Sec.~\ref{sec:optcorr} the flow parameter $\alpha$ is a physically motivated control parameter that enters the central and tensor correlation function in a consistent way.

The dependence of the correlation functions on the flow-parameter $\alpha$ is illustrated in Fig. \ref{fig:srg_correlator_Rp_alpha} for the central correlation functions and in Fig. \ref{fig:srg_correlator_Th_alpha} for the tensor correlation functions obtained for the Argonne V18 potential through the SRG mapping. Evidently, the over-all range of the correlation functions is directly controlled by the flow-parameter $\alpha$: Larger $\alpha$ result in correlation functions with longer range. This is in-line with our observations on the evolution of the momentum-space matrix elements and the two-body wave functions. Initially, the SRG flow-evolution affects only the high-momentum matrix elements and thus the short-distance behavior of the wave functions. Throughout the flow evolution, i.e., with increasing $\alpha$, the wave functions are modified at increasingly larger distances, which results in an increasing range of the associated correlation functions. This localized action of the SRG transformation on coordiante-space wave functions is also responsible for the fact that the SRG-generated correlation functions automatically have finite range. This property is not imposed by the mapping scheme, but results from the structure of the two-body wave functions alone\footnote{We note that the short range of the correlation functions depends on the initial interaction. If the interaction is such that the SRG evolution affects also long-range components of the wave function, then the UCOM correlators obtained by the mapping will be long-ranged as well. This is the case for the chiral N3LO interaction \cite{EnMa03}, for example.}.
 
Closer inspection of the structure of the correlation functions in Figs. \ref{fig:srg_correlator_Rp_alpha} and \ref{fig:srg_correlator_Th_alpha} reveals an interesting new aspect as compared to the correlation functions discussed in Sec. \ref{sec:ucom}. In the even channels the central correlation functions $R_{+}(r)-r$ exhibit a sign-change at $r\approx1.1\,\text{fm}$. At shorter ranges $R_{+}(r)-r$ is positive, indicating an outward shift in a transformed two-body wave function, and turns negative at larger distances, inducing an inward shift in a transformed wave function. Pictorially speaking, the UCOM transformation attempts to exploit the attractive parts of the central potential in the even channels by moving probability amplitudes from small and large inter-particle distances into the attractive region. The triplet-even tensor correlation function $\vartheta(r)$ exhibits a similar structure, though the negative contribution is much weaker than the positive part. In the odd channels the correlation functions do not show this sign change, which can be explained by the lack of a sufficiently strong attraction in the central interaction. 

The details of the $\alpha$-dependence are different for the different types of correlation functions. For the dominant central correlators in the even channels, the short-range positive component is practically independent of $\alpha$ in the range covered in Fig. \ref{fig:srg_correlator_Rp_alpha}---these are the generic short-range correlations induced by the strong short-range repulsion of the Argonne V18 potential. Only the negative long-range part shows a sizable $\alpha$-dependence affecting its range and strength. All other correlation functions show a smooth increase of the over-all range with increasing $\alpha$, only a very short distances the curves are independent of $\alpha$.

%%%%%%%%%%%%%%%%%%%%%%%%%%%%%%%%%%%%%%%%%%%%%%%%%%%%%%%%%%%%%%%%%%%%%%
\begin{figure}[p]
\centering\includegraphics[width=0.85\textwidth]{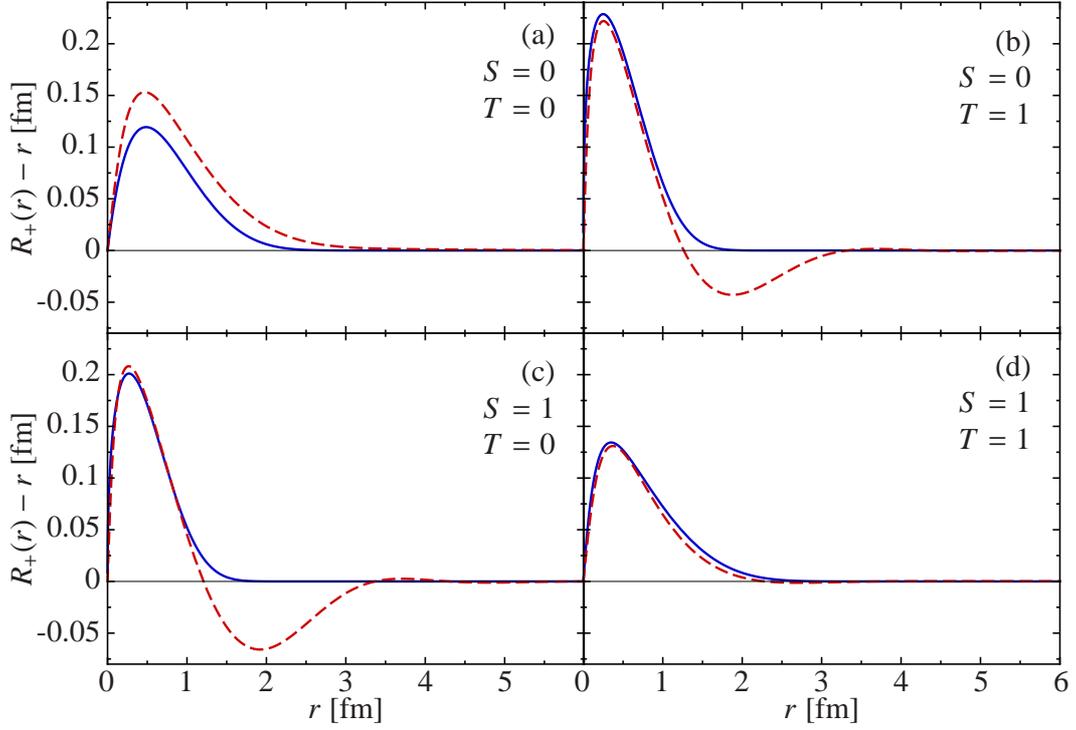}
\caption{Comparison of the central correlation functions $R_+(r)-r$ for the Argonne V18 potential obtained via variational optimization with constraints $I_{\vartheta}^{(10)}=0.09\,\text{fm}^3$, and $I_{\vartheta}^{(11)}=-0.03\,\text{fm}^3$ (\linemediumsolid[FGBlue]) and via SRG-mapping with $\alpha=0.04\text{fm}^{4}$ (\linemediumdashed[FGRed]).}
\label{fig:srg_correlator_Rp_comp}
\end{figure}
%%%%%%%%%%%%%%%%%%%%%%%%%%%%%%%%%%%%%%%%%%%%%%%%%%%%%%%%%%%%%%%%%%%%%%
%%%%%%%%%%%%%%%%%%%%%%%%%%%%%%%%%%%%%%%%%%%%%%%%%%%%%%%%%%%%%%%%%%%%%%
\begin{figure}[p]
\centering\includegraphics[width=0.85\textwidth]{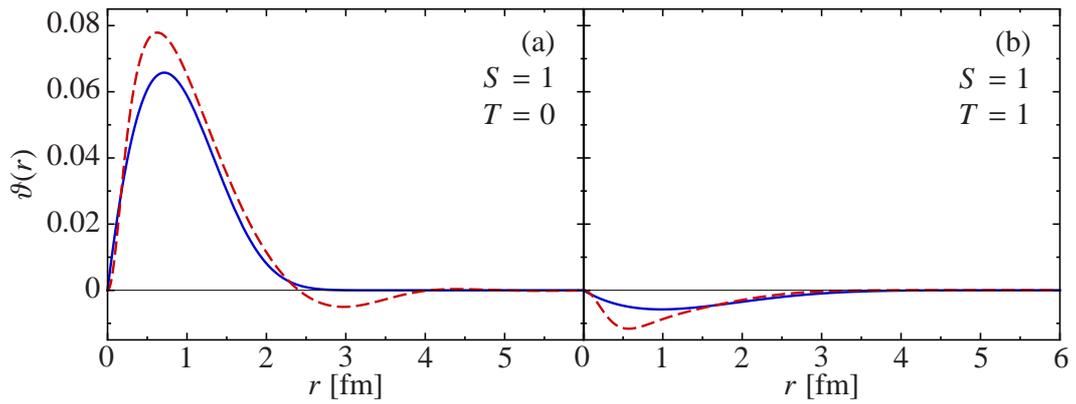}
\caption{Comparison of the tensor correlation functions $\vartheta(r)$ for the Argonne V18 potential obtained via variational optimization with constraints $I_{\vartheta}^{(10)}=0.09\,\text{fm}^3$, and $I_{\vartheta}^{(11)}=-0.03\,\text{fm}^3$ (\linemediumsolid[FGBlue]) and via SRG-mapping with $\alpha=0.04\,\text{fm}^{4}$ (\linemediumdashed[FGRed]).}
\label{fig:srg_correlator_Th_comp}
\end{figure}
%%%%%%%%%%%%%%%%%%%%%%%%%%%%%%%%%%%%%%%%%%%%%%%%%%%%%%%%%%%%%%%%%%%%%%

The direct comparison of the SRG-generated correlation functions with the correlators determined variationally in Sec.~\ref{sec:optcorr} is also quite instructive. In Figs. \ref{fig:srg_correlator_Rp_comp} and \ref{fig:srg_correlator_Th_comp} we compare the central and tensor correlations functions, respectively, of the SRG-generated correlator for $\alpha=0.04\,\text{fm}^4$ and the standard variational correlator with constraints $I_{\vartheta}=0.09\,\text{fm}^3$ for the tensor correlator in $T=0$, and $I_{\vartheta}=-0.03\,\text{fm}^3$ for the tensor correlator in $T=1$. Both sets of correlators yield approximately the same ground-state energy of ${}^4$He in a No-Core Shell Model calculation, as discussed in Sec.~\ref{sec:ncsm}. 

In the even channels, the short-range parts of the central correlation functions of both sets agree very well---another indication that these dominant short-range correlations are truly generic and independent of the methodology used to determine the correlation functions. The negative contributions in $R_{+}(r)-r$ appearing in the SRG-generated correlators is absent in the variational correlators, simply because the parameterizations used for the latter did not allow for such a structure. The triplet-odd central correlators also agree very well, thus providing additional justification for the treatment of this channel in the mapping procedure. 

In the singlet-odd channel the central correlation functions exhibit the same shape, but the variational correlator is suppressed due to the explicit range constraint. However, the variational correlation function agrees very well with the SRG-generated central correlators for smaller values of $\alpha$. The tensor correlations functions determined in the variational scheme are also subject to explicit range constraints that affect their shape. Over-all the correlation functions $\vartheta(r)$ of both sets are similar, but the agreement is not as good as for the channels without ad hoc constraints.

For the following discussion, we will identify the SRG-generated UCOM correlation functions and the resulting UCOM-transformed potential with the abbreviation ``UCOM(SRG)''. The UCOM correlation functions determined from a variational calculation (cf. Sec. \ref{sec:optcorr}) are termed ``UCOM(var.)''. Finally, the purely SRG-transformed interactions are labelled ``SRG''.

%%%%%%%%%%%%%%%%%%%%%%%%%%%%%%%%%%%%%%%%%%%%%%%%%%%%%%%%%%%%%%%%%%%%%%
%%%%%%%%%%%%%%%%%%%%%%%%%%%%%%%%%%%%%%%%%%%%%%%%%%%%%%%%%%%%%%%%%%%%%%
\subsection{Comparison of matrix elements}
\label{sec:srg_matrixelem}

%%%%%%%%%%%%%%%%%%%%%%%%%%%%%%%%%%%%%%%%%%%%%%%%%%%%%%%%%%%%%%%%%%%%%%
\begin{figure}[t]
\centering\includegraphics[width=0.95\textwidth]{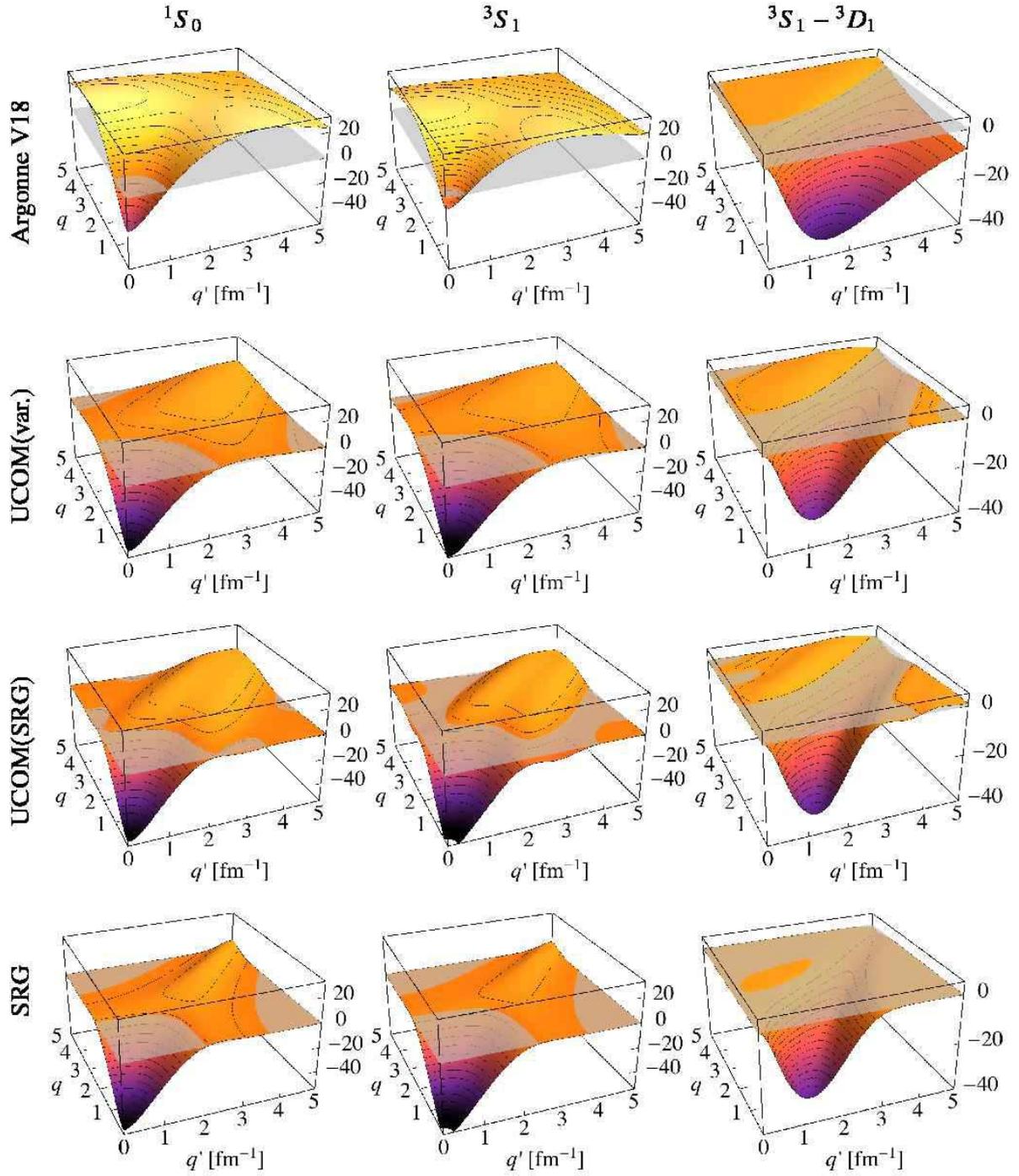}
\caption{Momentum-space matrix elements (in units of MeV fm$^3$) of the UCOM and SRG-transformed Argonne V18 potential in the $^{1}{S}_0$, $^{3}{S}_1$, and $^{3}{S}_1$-$^{3}{D}_1$ partial waves. First row: initial Argonne V18 matrix elements (charge independent terms only). Second row: UCOM transformed matrix elements using UCOM(var.) correlation functions for $I_{\vartheta}^{(10)}=0.09\,\text{fm}^3$. Third row: UCOM transformed matrix elements using UCOM(SRG) correlation functions for $\alpha=0.04\,\text{fm}^4$. Fourth row: SRG-transformed matrix elements for $\alpha=0.03\,\text{fm}^4$.}
\label{fig:srg_matrixelem_comp}
\end{figure}
%%%%%%%%%%%%%%%%%%%%%%%%%%%%%%%%%%%%%%%%%%%%%%%%%%%%%%%%%%%%%%%%%%%%%%

To conclude the discussion of the transformed interactions resulting in the UCOM and in the SRG scheme, we consider the momentum-space matrix elements once more. In Fig. \ref{fig:srg_matrixelem_comp} we compare the matrix elements in the dominant $S$-wave channels obtained for the bare Argonne V18 potential, the UCOM-transformed Argonne V18 potential using correlation functions obtained variationally as well as via the SRG mapping, and the pure SRG-evolved interaction.

All similarity transformed interactions show a strong suppression of the off-diagonal matrix elements, i.e., a decoupling of low-momentum and high-momentum states, and an enhancement of the low-momentum matrix elements. This leads to a significant improvement of the convergence in No-Core Shell Model calculations for light nuclei, as will be discussed in detail in Sec. \ref{sec:ncsm}. There are, however, distinctive differences in the behavior of the matrix elements for the $^1S_0$ and the $^3S_1$ partial waves in the high-momentum sector. 

The SRG evolution leads to a transformed interaction with a perfect band-diagonal structure in momentum space, i.e., in the high-momentum regime the matrix elements drop to zero rapidly with increasing distance from the diagonal. For the UCOM-transformed interaction, the domain of non-vanishing high-momentum matrix elements extends further out. In the case of the UCOM(var.) correlation functions there is a plateau of non-vanishing matrix elements in the high-momentum sector, which falls off slowly when leaving the diagonal. As a result, the band-diagonal structure is far less pronounced than in the case of the SRG-evolved interaction. For the UCOM(SRG) interaction there appears a broad band of non-vanishing high-momentum matrix elements. The far-off diagonal matrix elements outside of this band are more suppressed than for the UCOM with variationally determined correlators, but compared to the SRG-transformed interactions the band is significantly broader. 

The apparent differences between the SRG- and the UCOM(SRG)-transformed matrix elements show that, despite the construction of the UCOM correlation functions using input from the SRG evolution, the transformed interactions are very similar for $\sqrt{q^2+q'^2}\lesssim2\,\text{fm}^{-1}$, but are quite different above. The origin of this differences is the limited flexibility of the UCOM generator, which is determined from the eigenstates with the lowest energy and does not allow for a perfect band-diagonalization of the high-momentum matrix elements. However, it does allow for a decoupling of low- and high-momentum modes, which will be important for the convergence of shell-model-like many-body calculation as discussed in the following section.

%%%%%%%%%%%%%%%%%%%%%%%%%%%%%%%%%%%%%%%%%%%%%%%%%%%%%%%%%%%%%%%%%%%%%%
%%%%%%%%%%%%%%%%%%%%%%%%%%%%%%%%%%%%%%%%%%%%%%%%%%%%%%%%%%%%%%%%%%%%%%
%%%%%%%%%%%%%%%%%%%%%%%%%%%%%%%%%%%%%%%%%%%%%%%%%%%%%%%%%%%%%%%%%%%%%%
\clearpage
\section{No-Core Shell Model}
\label{sec:ncsm}

The No-Core Shell Model (NCSM) is a powerful and well established many-body technique that has been used successfully in a wide range of nuclear structure calculations in light nuclei \cite{NaQu09,MaVa09,CaMa05}. It provides a perfect framework for assessing the properties of nuclear interactions, both from the technical perspective, e.g., regarding their convergence properties, and from the experimental view, e.g., regarding the agreement of predicted observables with experiment. 

In this section we discuss NCSM calculations using the
UCOM interaction with correlation functions determined variationally
(cf. Sec.~\ref{sec:ucom}) and the UCOM interaction using SRG-generated
correlation functions (cf. Sec.~\ref{sec:srg}). For comparison we also
show results with the SRG interaction. All UCOM and SRG interactions
are derived from the Argonne~V18 interaction. The matrix elements are
calculated without approximation as explained in
Sec.~\ref{sec:corr_me_operator} and Sec.~\ref{sec:srg_evolint} including
all electromagnetic and charge dependent terms. 

Using exact calculations in the three- and four-body system, we
investigate the convergence properties and the role of induced three-body
interactions of the UCOM and SRG interactions as a function
of tensor correlation range or flow parameter. The binding energies in
the three- and four-body system map out the so-called Tjon-line.  By
choosing particular values for the tensor correlation range and the
flow parameters, respectively, we obtain UCOM and SRG two-body
interactions that provide binding energies very close to the
experimental results in the three- and four-body system.

The UCOM and SRG interactions selected by this choice are then used in
NCSM calculations of \elem{He}{6}, \elem{Li}{6} and
\elem{Li}{7}. Although we are not able to reach full convergence,
binding energies can be obtained by extrapolation. A better
convergence behavior is found for the excitation energies. The spectra
provide some insight about the spin dependence of the UCOM and SRG
interactions. Additional hints are provided by electromagnetic
properties like radii, magnetic dipole moments and electric
quadrupole moments.

%%%%%%%%%%%%%%%%%%%%%%%%%%%%%%%%%%%%%%%%%%%%%%%%%%%%%%%%%%%%%%%%%%%%%%
%%%%%%%%%%%%%%%%%%%%%%%%%%%%%%%%%%%%%%%%%%%%%%%%%%%%%%%%%%%%%%%%%%%%%%
\subsection{Benchmarking $V_{\UCOM}$ in ab initio few-body calculations} 

\newcommand{\Nmax}{\ensuremath{N_\mathrm{max}}}
\newcommand{\hbo}{\ensuremath{\hbar\Omega}}

For the NCSM calculations in the three- and four-body system we use
the \textsc{ManyEff} code by Petr Navr{\'a}til \cite{NaKa00}. It
employs a translationally invariant oscillator basis in Jacobi
coordinates. The model space is defined by the oscillator frequency
$\hbo$ and the total number $\Nmax$ of excitations with respect to the
$0\hbo$ configuration. The interaction is provided in form of relative
harmonic-oscillator matrix elements. In the standard NCSM approach an
effective interaction adapted to the model space ($\Nmax$, $\hbo$) is
derived using the Lee-Suzuki transformation. For the calculations
presented here we directly use the ``bare'' UCOM and SRG matrix elements, which
do not depend on the model space size $\Nmax$. Within this procedure,
the NCSM provides a variational approach for the energy. The energy eigenvalues will converge from above to the exact solution for sufficiently large model spaces.
The converged results should also be
independent of the oscillator frequency $\hbo$, although the rate of
convergence will be different for different oscillator frequencies.

\begin{figure}
  \centering
  \includegraphics[width=0.31\textwidth]{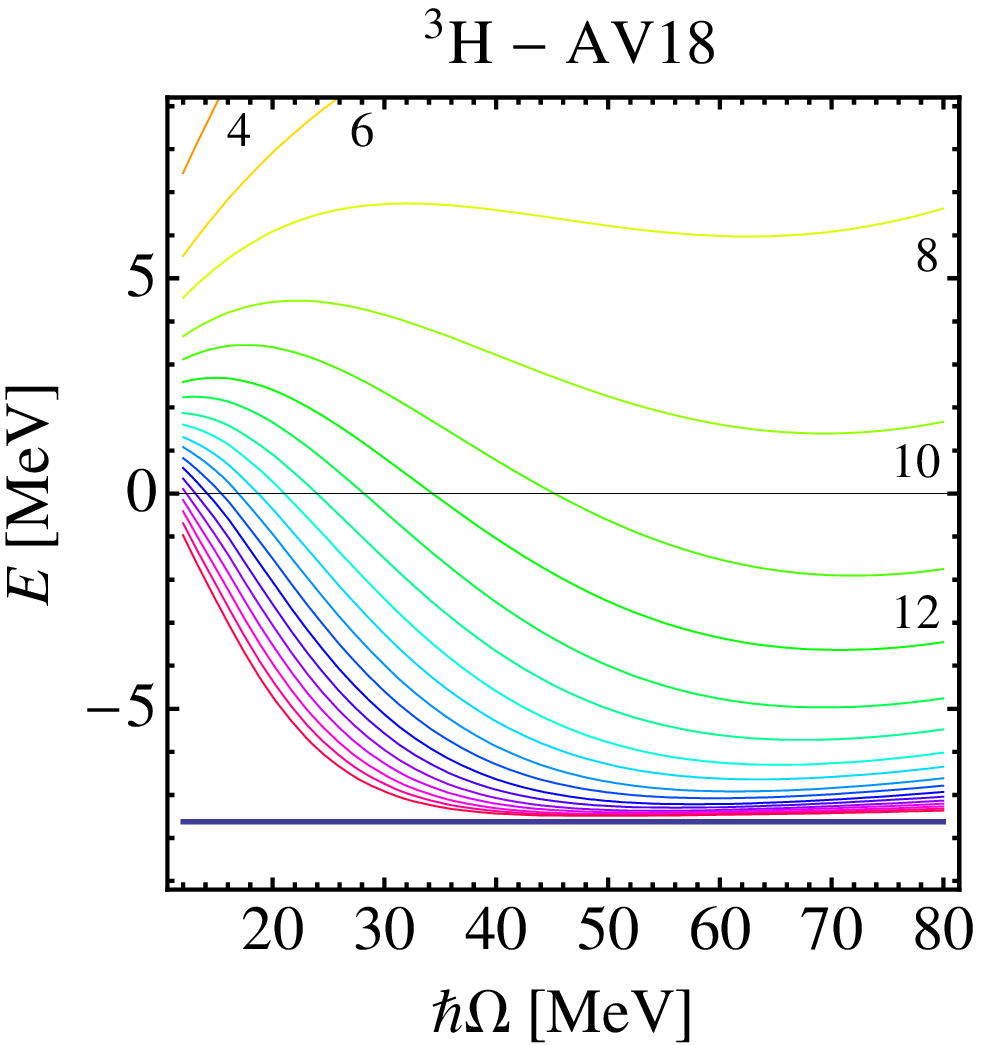}\hspace{0.1\textwidth}
  \includegraphics[width=0.31\textwidth]{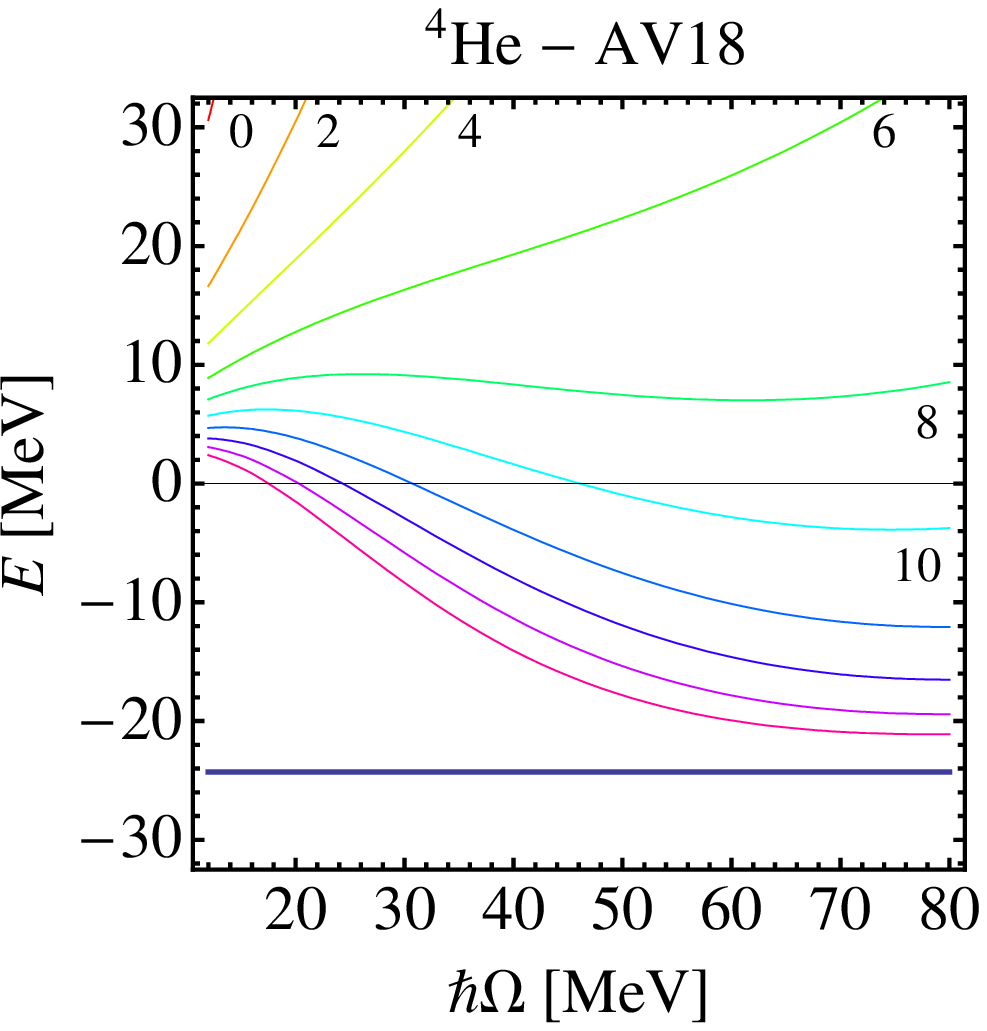}
  \caption{NCSM calculations with the bare Argonne~V18 interaction for
    \elem{H}{3} and \elem{He}{4}. The ground state energy is
    calculated in model space with $\Nmax = 0, 2, \ldots, 40$ for \elem{H}{3} and with
    $\Nmax = 0, 2, \ldots, 18$ for \elem{He}{4}. The exact binding energies
    \cite{nogga00} are indicated by horizontal lines.}
  \label{fig:ncsm-av18}
\end{figure}

Before presenting results with UCOM and SRG interactions we illustrate
that it is not possible to reach full convergence with the bare
Argonne~V18 interaction, even within the huge model spaces possible for
\elem{H}{3} and \elem{He}{4}. As can be seen in
Fig.~\ref{fig:ncsm-av18} it is already very hard to obtain a bound
nucleus. One can also observe that the lowest energies are obtained
for very large oscillator frequencies. With these narrow oscillator wave
functions it becomes eventually possible to explicitly describe the
short-range correlations for very large $\Nmax$.

\begin{figure}
  \includegraphics[width=0.31\textwidth]{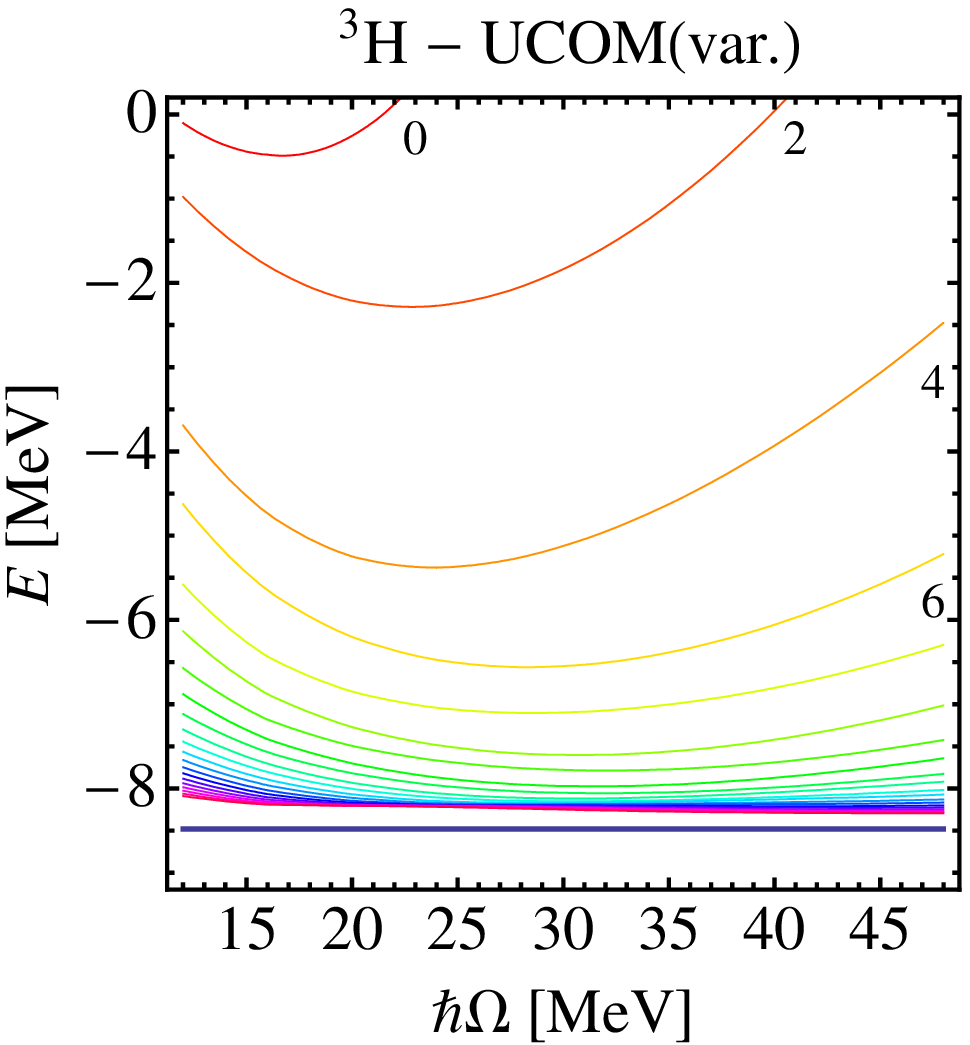}\hfil
  \includegraphics[width=0.31\textwidth]{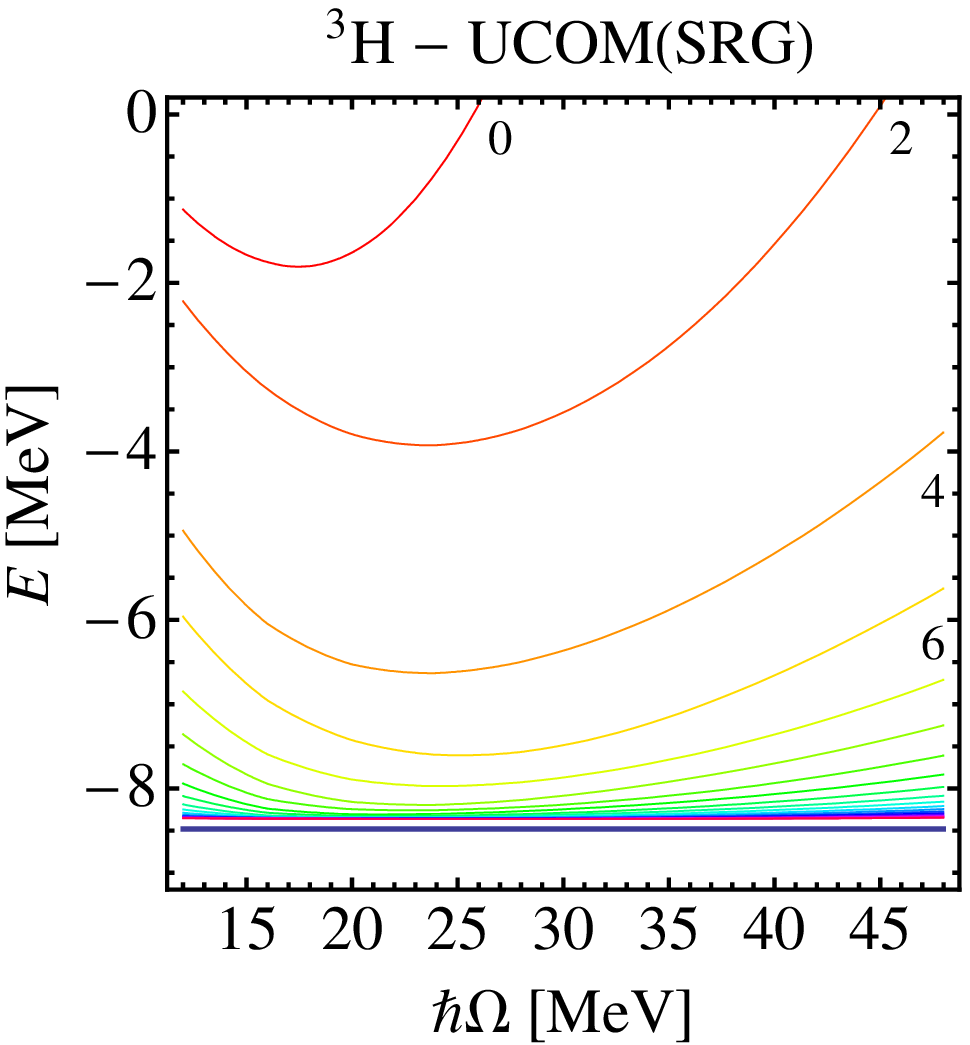}\hfil
  \includegraphics[width=0.31\textwidth]{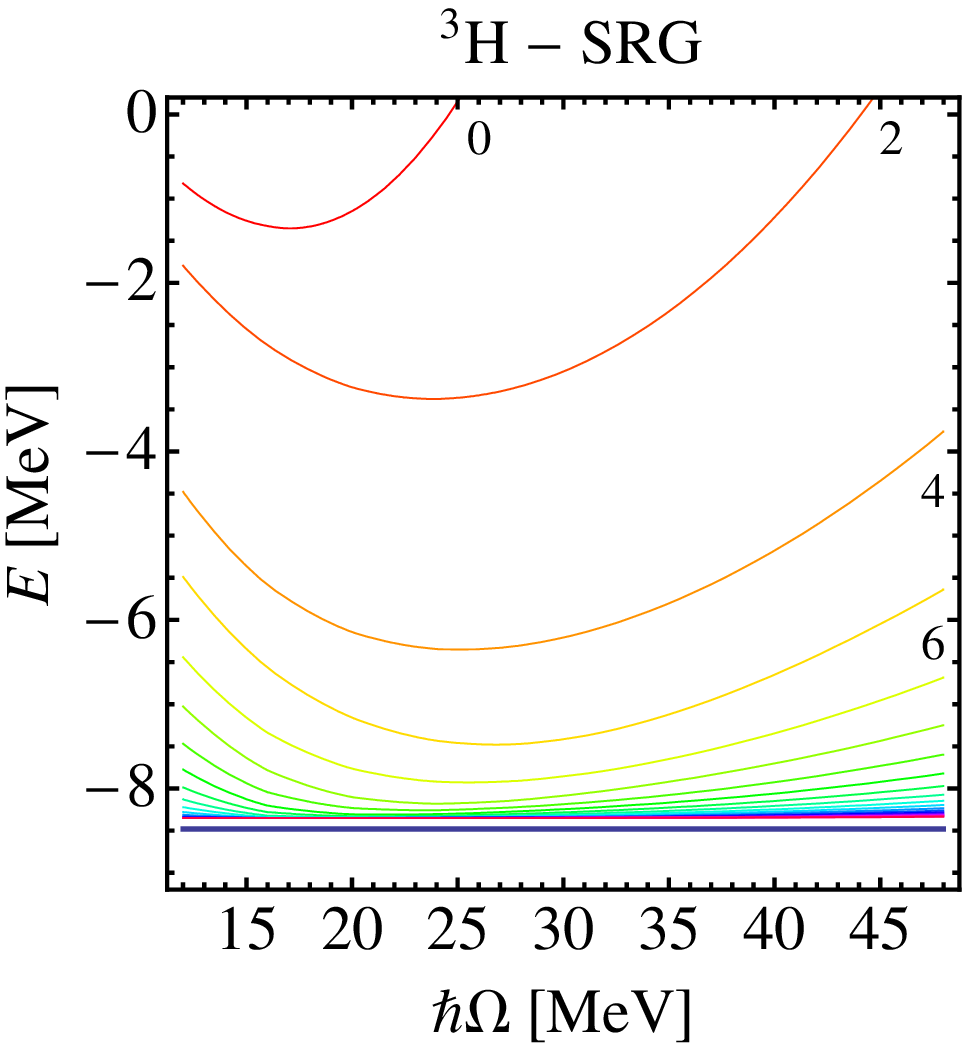}
  \caption{NCSM calculations for the ground-state energy of \elem{H}{3} in model spaces with $\Nmax = 0, 2, \ldots, 40$. The tensor correlation range for the UCOM(var.) interaction is $I_{\vartheta} = 0.09\fm^3$. The flow parameters are $\alpha=0.04 \fm^4$ for the UCOM(SRG) and $\alpha=0.03 \fm^4$ for the SRG interaction.
  The experimental binding energy is indicated by a horizontal line.}
  \label{fig:ncsm-h3}
\end{figure}

\begin{figure}
  \includegraphics[width=0.31\textwidth]{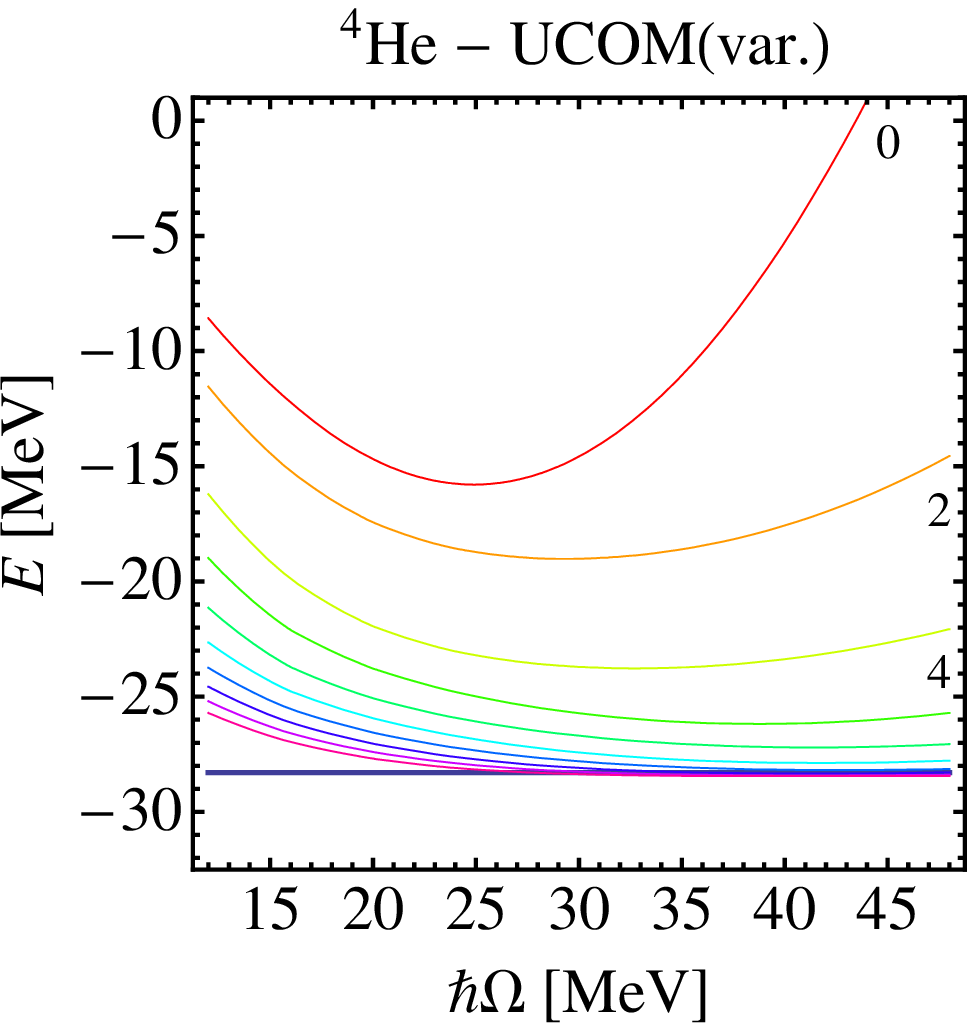}\hfil
  \includegraphics[width=0.31\textwidth]{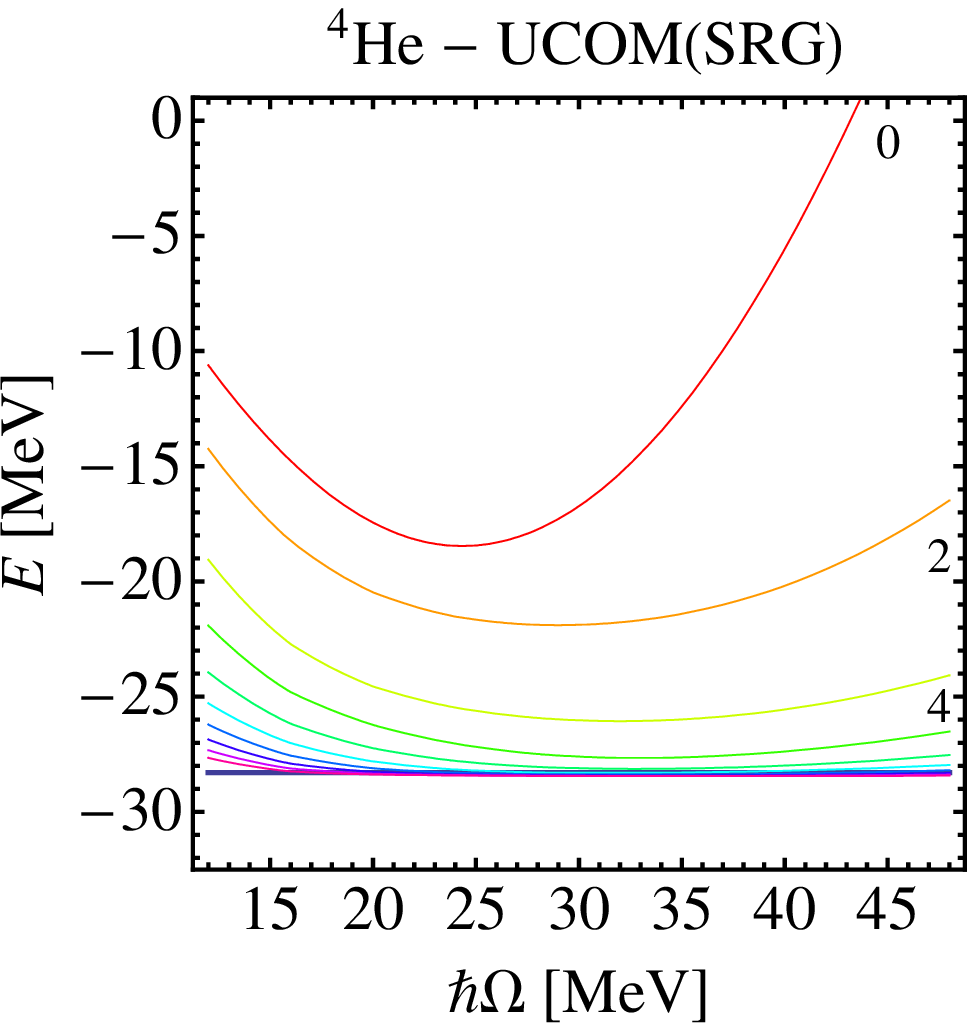}\hfil
  \includegraphics[width=0.31\textwidth]{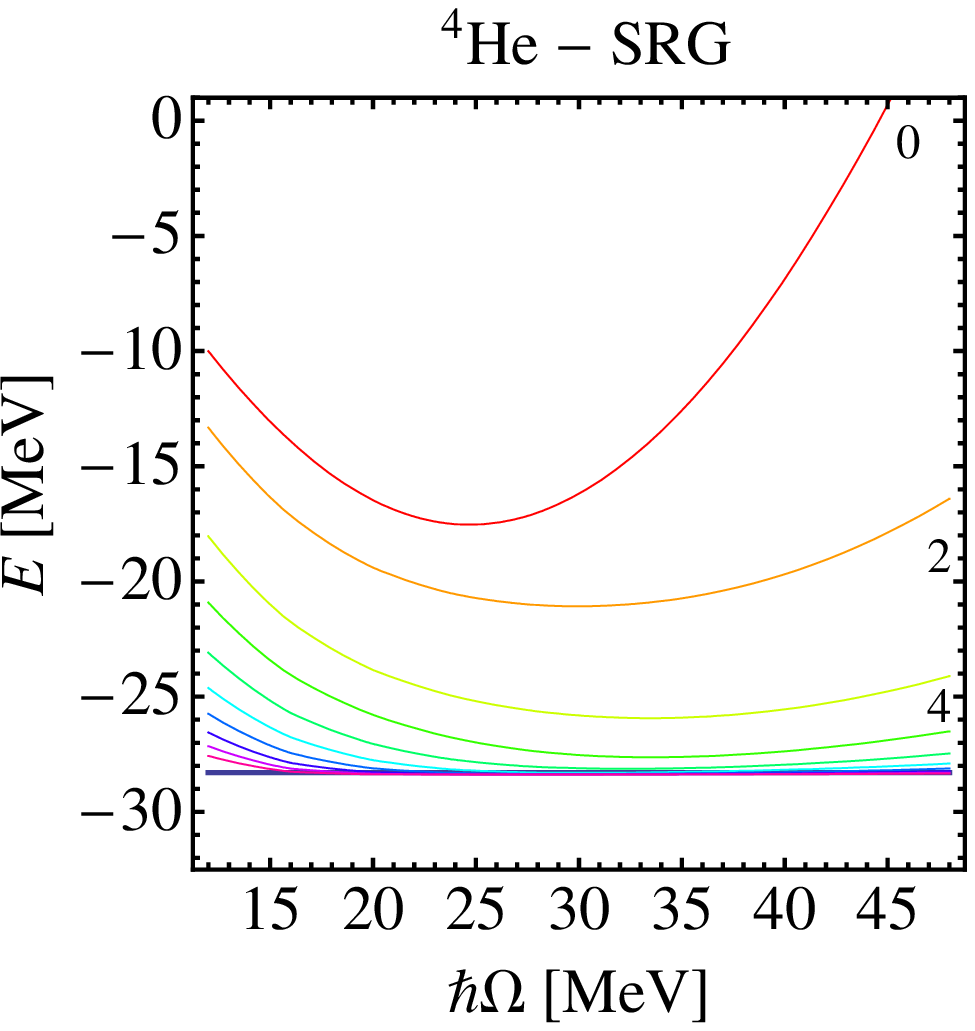}
  \caption{NCSM calculations for the ground-state energy of \elem{He}{4} in model spaces with $\Nmax = 0, 2, \ldots, 18$. The tensor correlation range for the UCOM(var.) interaction is $I_{\vartheta} = 0.09\fm^3$. The flow parameters are $\alpha=0.04 \fm^4$ for the UCOM(SRG) and $\alpha=0.03 \fm^4$ for the SRG interaction.
  The experimental binding energy is indicated by a horizontal line.}
  \label{fig:ncsm-he4}
\end{figure}

We can compare these results with calculations with the UCOM and SRG
interactions as shown in Fig.~\ref{fig:ncsm-h3} and
Fig.~\ref{fig:ncsm-he4}. We use our ``standard'' choices given in \eqref{eq:optpar} for the
tensor correlation range $I_\vartheta$ and flow parameters $\alpha$ as
explained later in this section. For all interactions we find
a bound minimum already in the $0\hbo$ space. The minima are at
oscillator frequencies that correspond roughly to the experimental
sizes of the nuclei. With increasing model-space size $\Nmax$ we
observe a fast convergence for all interactions. In direct comparison,
the UCOM(SRG) and SRG interactions converge faster than the UCOM(var.)
interaction with respect to both, model space size $\Nmax$ and
oscillator frequency $\hbo$.  It is important to note that the
converged energy, while being close to the experimental binding
energy, is lower than the exact result for the bare Argonne~V18
interaction by about $0.7\MeV$ in case of \elem{H}{3} and by about
$4.1\MeV$ in case of \elem{He}{4}. This overbinding of UCOM and SRG
interactions with respect to the bare interaction is caused by the
missing three- and four-body contributions -- UCOM and SRG
interactions are only calculated in two-body approximation. This point
will be discussed in detail in Sec.~\ref{sec:ncsm-tjon-line}.

\begin{figure}
  \includegraphics[width=0.31\textwidth]{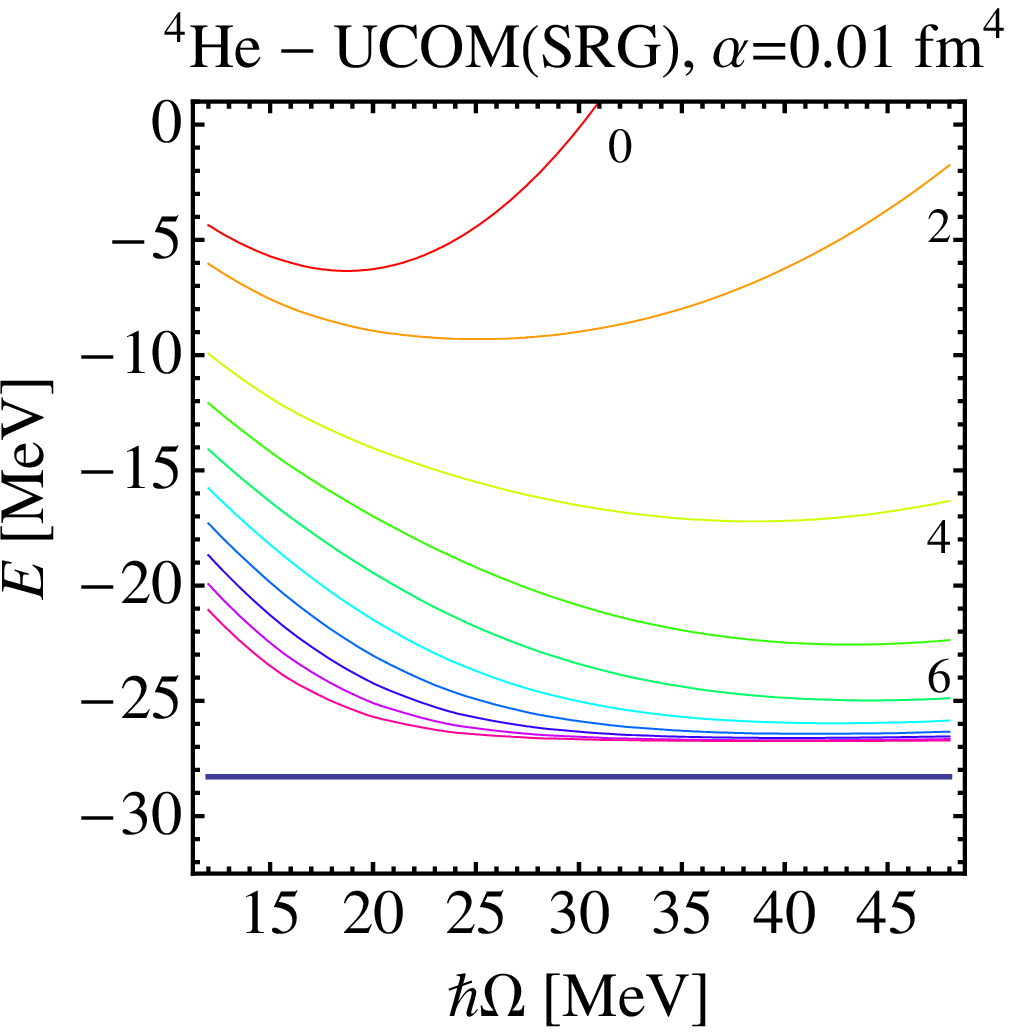}\hfil
  \includegraphics[width=0.31\textwidth]{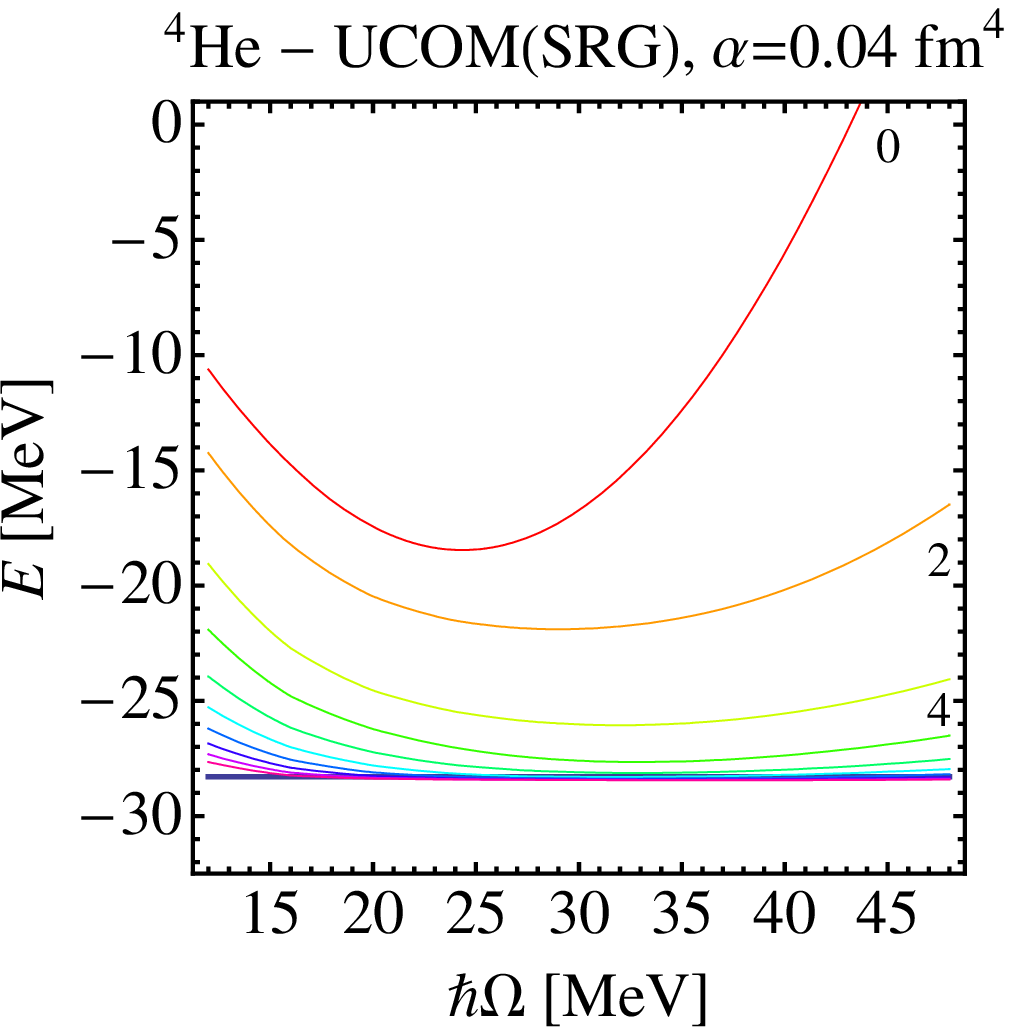}\hfil
  \includegraphics[width=0.31\textwidth]{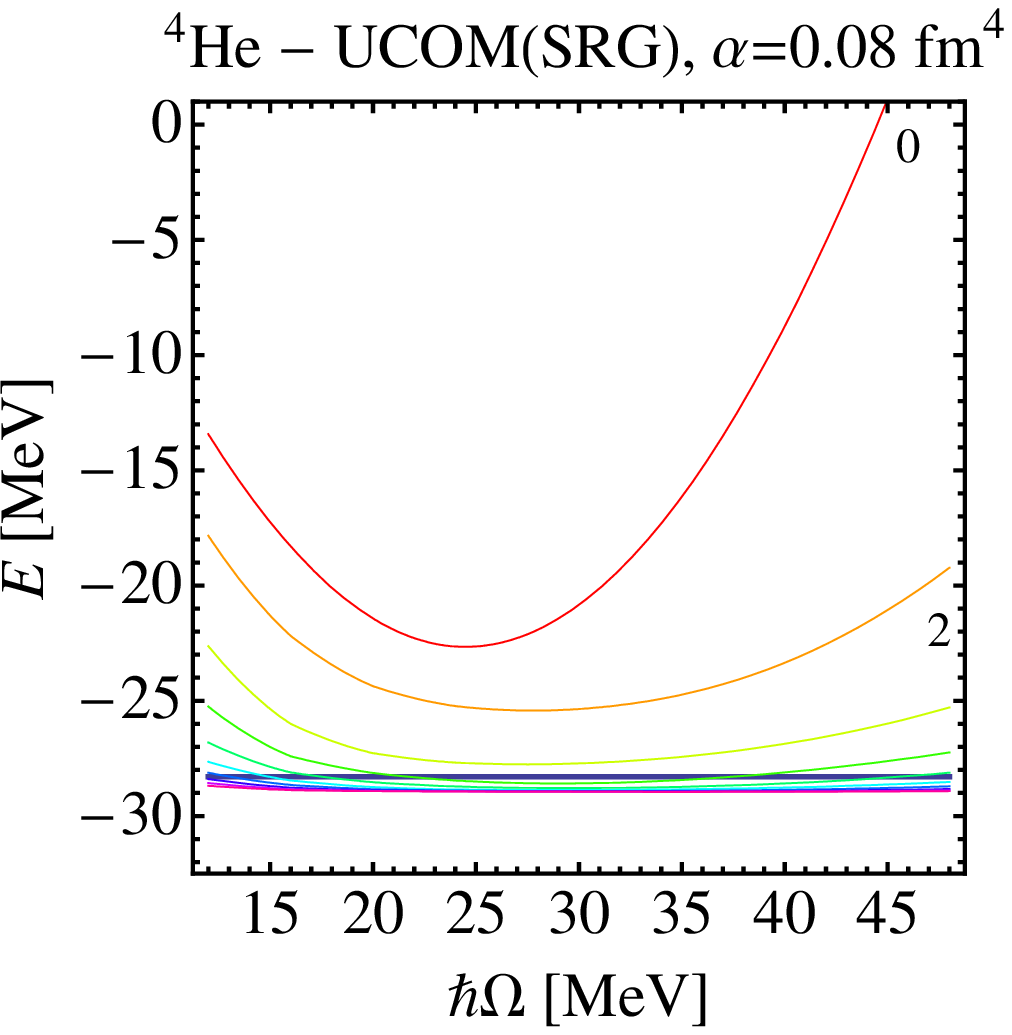}
  \caption{NCSM calculations for the ground state of \elem{He}{4} with UCOM(SRG)
  interactions at different flow parameters ($\alpha=0.01, 0.04, 0.08 \fm^4$) in model spaces with $\Nmax = 0,2,\ldots,18$. The
    horizontal line indicates the experimental binding energy.}
  \label{fig:ncsm-he4-srgD-alpha}
\end{figure}

\begin{figure}
  \includegraphics[width=0.31\textwidth]{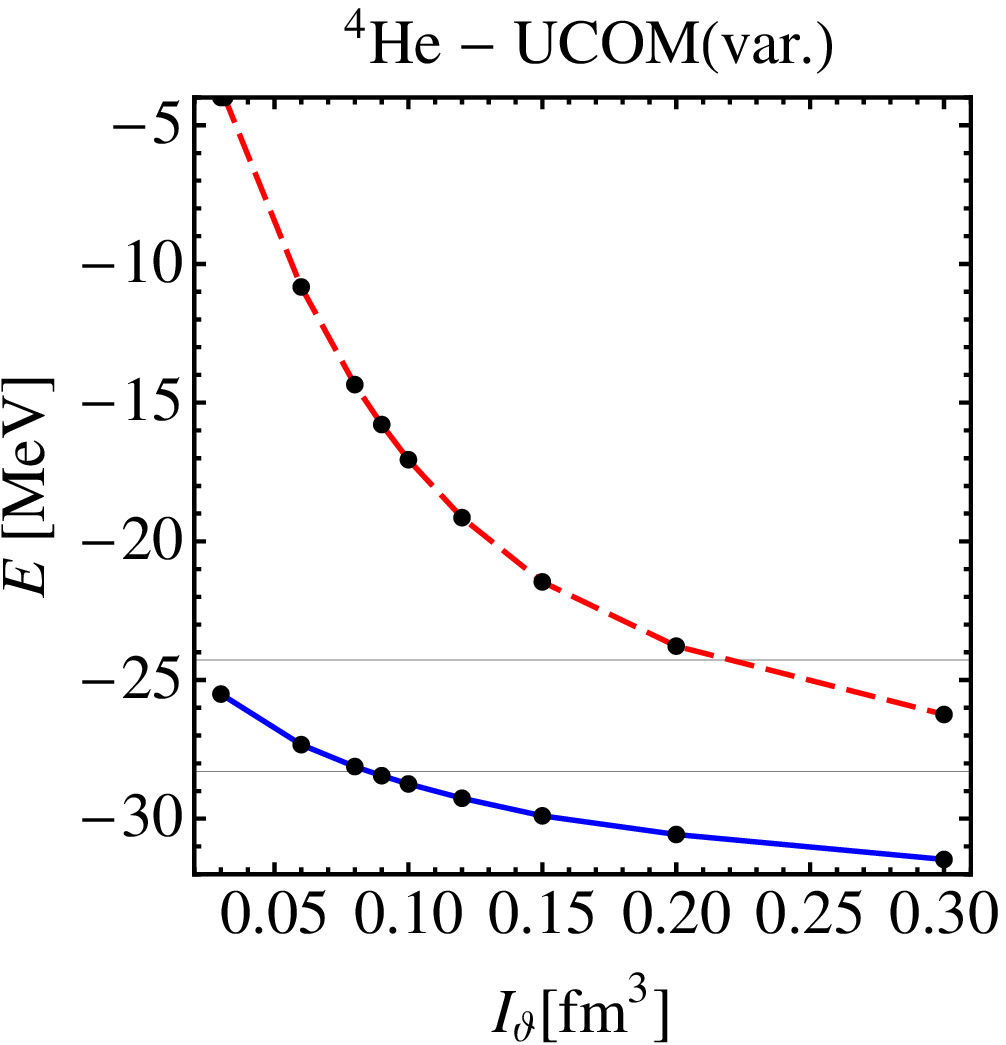}\hfil
  \includegraphics[width=0.31\textwidth]{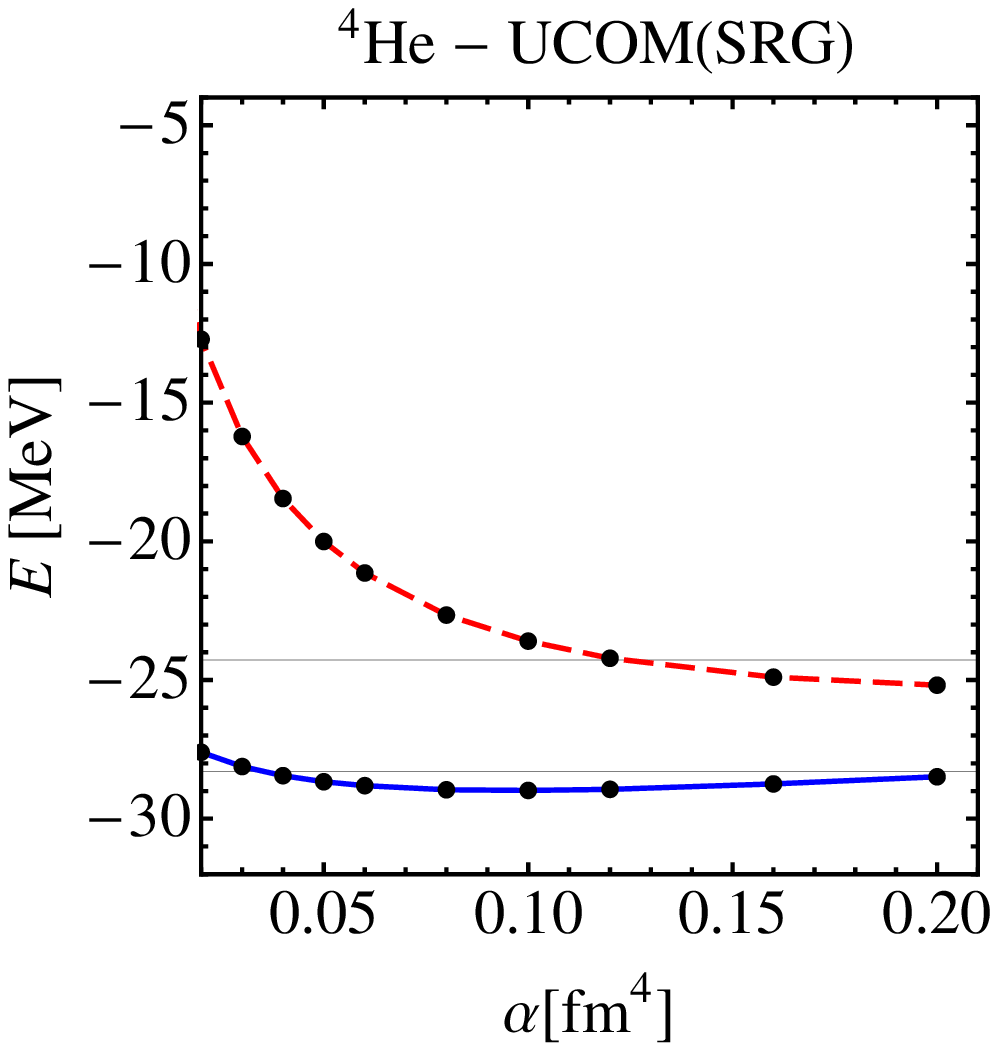}\hfil
  \includegraphics[width=0.31\textwidth]{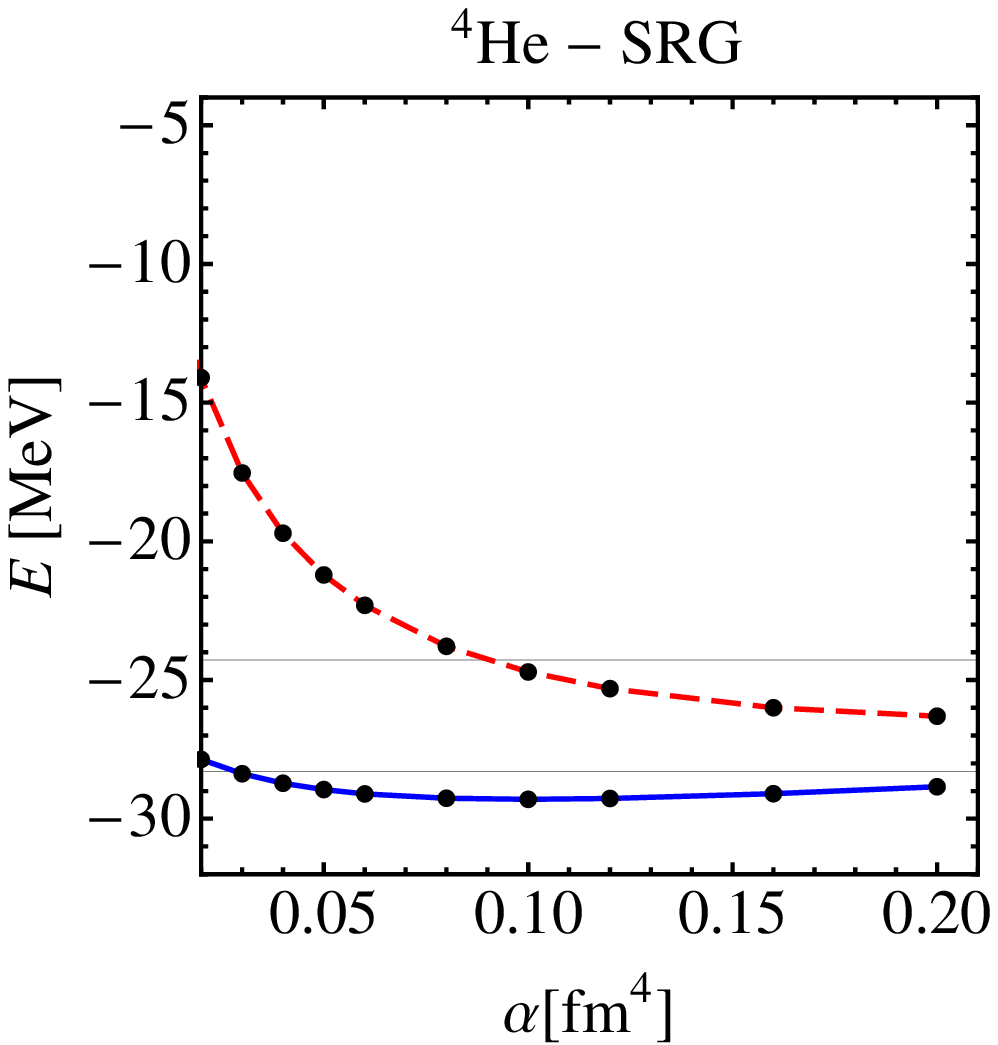}
  \caption{\elem{He}{4} ground-state energy calculated in $0 \hbar\Omega$ space
    (dotted lines) and converged NCSM results (solid lines) as a
    function of correlation range or flow parameter. The exact result for the bare
    Argonne~V18 interaction and the experimental binding energy are
    indicated by horizontal lines.}
  \label{fig:ncsm-he4-convergence}
\end{figure}

The convergence pattern and the converged energy depends on the
parameters of the tensor correlation range or the flow parameter
respectively. This is illustrated in
Fig.~\ref{fig:ncsm-he4-srgD-alpha} for the UCOM(SRG) interaction using
three different flow parameters. With increasing flow parameter the
calculations converge faster and to a lower energy. This is analyzed
in detail in Fig.~\ref{fig:ncsm-he4-convergence}, where we compare the
energy minima in the $0\hbo$ model spaces and the converged energies
as a function of $I_\vartheta$ or $\alpha$. The $0\hbo$
results are getting closer to the converged results with larger
correlation ranges or flow parameters, i.e., the interactions become
``softer''. Note that the converged UCOM(var.) results decrease
monotonically with increasing tensor correlation range, whereas the
UCOM(SRG) and SRG interactions both show a minimum in the converged
energy for flow parameters $\alpha \approx 0.10 \fm^4$. This indicates
that the tensor
correlation range and the
flow parameter in UCOM(SRG) and SRG interactions play a somewhat different role. In the case of the UCOM(var.) interactions $I_\vartheta$ only affects the range of the tensor correlator in the deuteron channel. In the case of UCOM(SRG) and SRG interactions $\alpha$ affects also the central correlations in all channels.
Beyond a certain point, stronger central correlations
actually result in less binding from the central part of the
interaction and the interaction becomes less attractive.

%%%%%%%%%%%%%%%%%%%%%%%%%%%%%%%%%%%%%%%%%%%%%%%%%%%%%%%%%%%%%%%%%%%%%%
%%%%%%%%%%%%%%%%%%%%%%%%%%%%%%%%%%%%%%%%%%%%%%%%%%%%%%%%%%%%%%%%%%%%%%
\subsection{Tjon-line and the role of three-body interactions}
\label{sec:ncsm-tjon-line}

\begin{figure}
  \centering
  \includegraphics[width=0.6\textwidth]{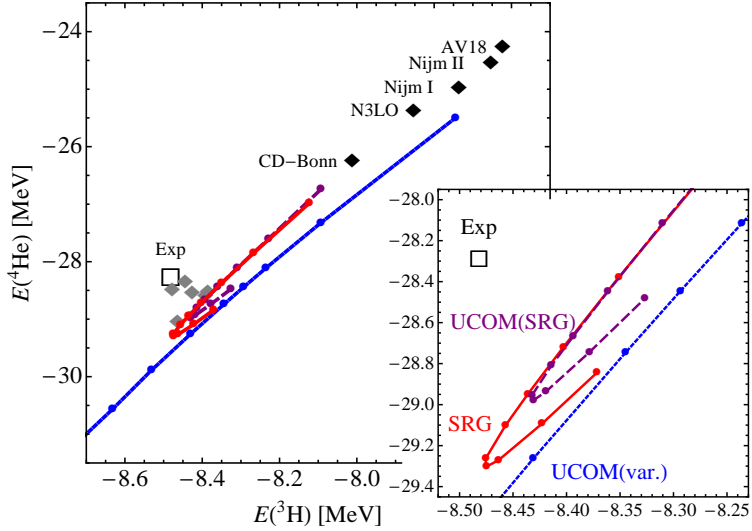}
  \caption{Binding energy of \elem{He}{4} versus binding energy of
    \elem{H}{3} calculated with UCOM(var.), UCOM(SRG) and SRG
    interactions. Also included are results with bare two-body forces (black diamonds) and combinations of two- and three-body forces (grey diamonds) taken from \cite{nogga00} and \cite{deltuva07}.}
  \label{fig:tjon-line}
\end{figure}

As has already been observed by Tjon for local interactions, a
correlation exists between the \elem{H}{3} and \elem{He}{4} binding
energies \cite{tjon75}. When the binding energy of \elem{He}{4} is
plotted against the binding energy of \elem{H}{3} the results for
different interactions fall essentially onto a single line, the so-called
Tjon-line. This has been confirmed also for modern
interactions. Typically the binding energies for bare two-body forces
are too small compared to experiment. This can be corrected by adding
an appropriate three-body force.  In Fig.~\ref{fig:tjon-line} results
from Refs.~\cite{nogga00,deltuva07} for realistic two-body interactions and
combinations of two- plus three-body forces are shown.  In addition we
show the results with UCOM and SRG interactions for which the binding
energies vary as a function of tensor correlation range $I_\vartheta$
or flow parameter $\alpha$ and continue more or less on the Tjon-line
obtained from bare realistic two-body forces. The results obtained
with the UCOM and SRG interactions are very similar and the
trajectories of all interactions pass very close (within
$200\;\mathrm{keV}$) to the experimental binding energies of
\elem{H}{3} and \elem{He}{4}. The results closest to the experimental
binding energies are obtained for the parameters: 
\eq{
  \label{eq:optpar}
  \begin{aligned}
  &\text{UCOM(var.):} \quad &I_\vartheta= 0.09\,\text{fm}^3 \\
  &\text{UCOM(SRG):}  &\alpha = 0.04\,\text{fm}^4 \\
  &\text{SRG:}        &\alpha = 0.03\,\text{fm}^4  \\
  \end{aligned} \;.
}

This is a remarkable result. By choosing a particular tensor
correlation range or particular flow parameters the two-body UCOM or
SRG interactions not only reproduce the experimental nucleon-nucleon
scattering data but also give the correct binding energies in the
three- and four-body systems. In this calculation we neither evaluate
three- and four-body contributions from the correlated two-body
interaction nor do we include genuine three- and four-body forces. At
this point all three- and four-body contributions have to cancel at
least on the level of the expectation value. As explained by Polyzou
and Gl{\"o}ckle \cite{polyzou90} a unitary transformation exists
between different combinations of on-shell equivalent two- and
three-body interactions. UCOM and SRG interactions provide a
particular realization of such a transformation. The three-body
contributions of SRG interactions have been studied explicitly in
\cite{JuNa09} by evolving the three-body matrix elements in the
harmonic oscillator basis.

We will use UCOM and SRG interactions with tensor correlation range
and flow parameters optimized for three- and four-body systems as
described above for NCSM calculations in $p$-shell
nuclei in Sec.~\ref{sec:ncsm-p-shell} and for exploratory studies
using Hartree-Fock and many-body perturbation theory for doubly magic
nuclei up to \elem{Pb}{208} in Sec.~\ref{sec:hf}.

\afterpage{\clearpage}

%%%%%%%%%%%%%%%%%%%%%%%%%%%%%%%%%%%%%%%%%%%%%%%%%%%%%%%%%%%%%%%%%%%%%%
%%%%%%%%%%%%%%%%%%%%%%%%%%%%%%%%%%%%%%%%%%%%%%%%%%%%%%%%%%%%%%%%%%%%%%
\subsection{Properties of $A=6,7$ nuclei}
\label{sec:ncsm-p-shell}

In this section we study properties of the light $p$-shell nuclei
\elem{He}{6}, \elem{Li}{6} and \elem{Li}{7} using the NCSM code
\textsc{Antoine} \cite{caurier99,caurier01}. Matrix elements of UCOM
and SRG interactions are provided as $jj$-coupled matrix elements
generated from the relative matrix elements using the Talmi-Moshinsky
transformation. Again we do not employ the Lee-Suzuki transformation.

Like for the three- and four-body systems the results are studied as a
function of oscillator frequency $\hbo$ and model space size
$\Nmax$. We do not reach full
convergence in the accessible model spaces for these nuclei, even with the UCOM or SRG interactions. For
ground-state energies we can use extrapolations to estimate the
converged binding energy. An error estimate is provided by comparing
results from different oscillator parameters. For observables like the
radii and the quadrupole moment of \elem{Li}{7} the results show a
strong dependence on the model space size and extrapolated results are
not reliable. Spectra on the other hand appear to be much better
converged and can be compared with experiment.
 
The convergence problems are to a large extend not caused by
properties of the interactions but by the cluster or halo nature of
these nuclei. To properly describe the asymptotics of wave functions with
a neutron halo like in \elem{He}{6}, or an underlying cluster
structures in case of the lithium isotopes, large model spaces are
needed in the oscillator basis.

\subsubsection*{Ground-state energies}

\begin{figure}
  \includegraphics[width=0.31\textwidth]{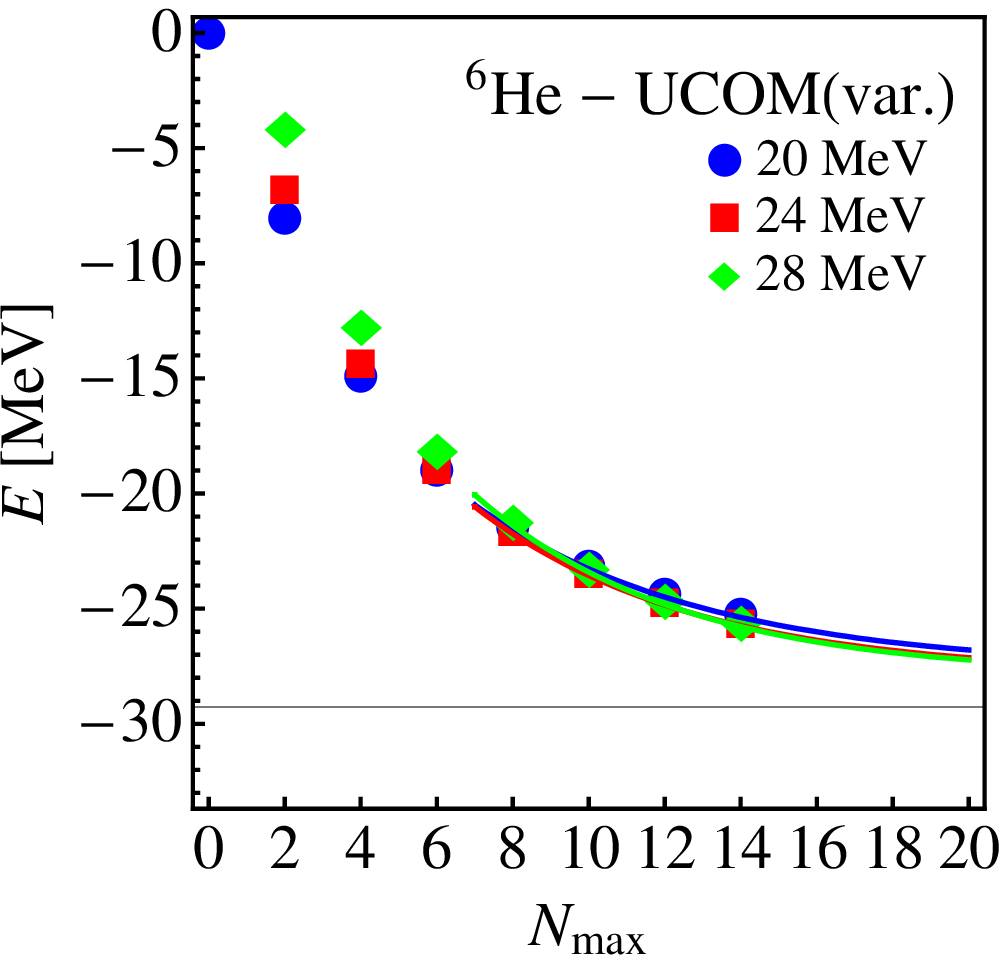}\hfil
  \includegraphics[width=0.31\textwidth]{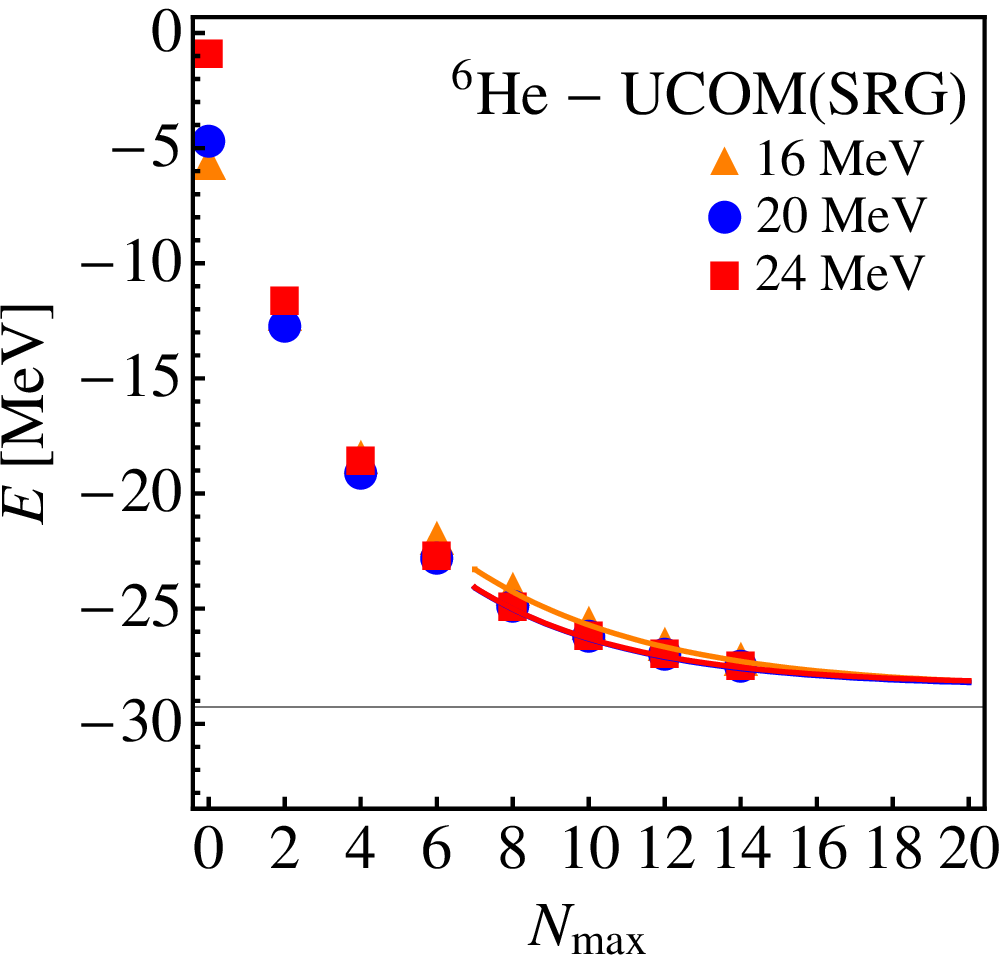}\hfil
  \includegraphics[width=0.31\textwidth]{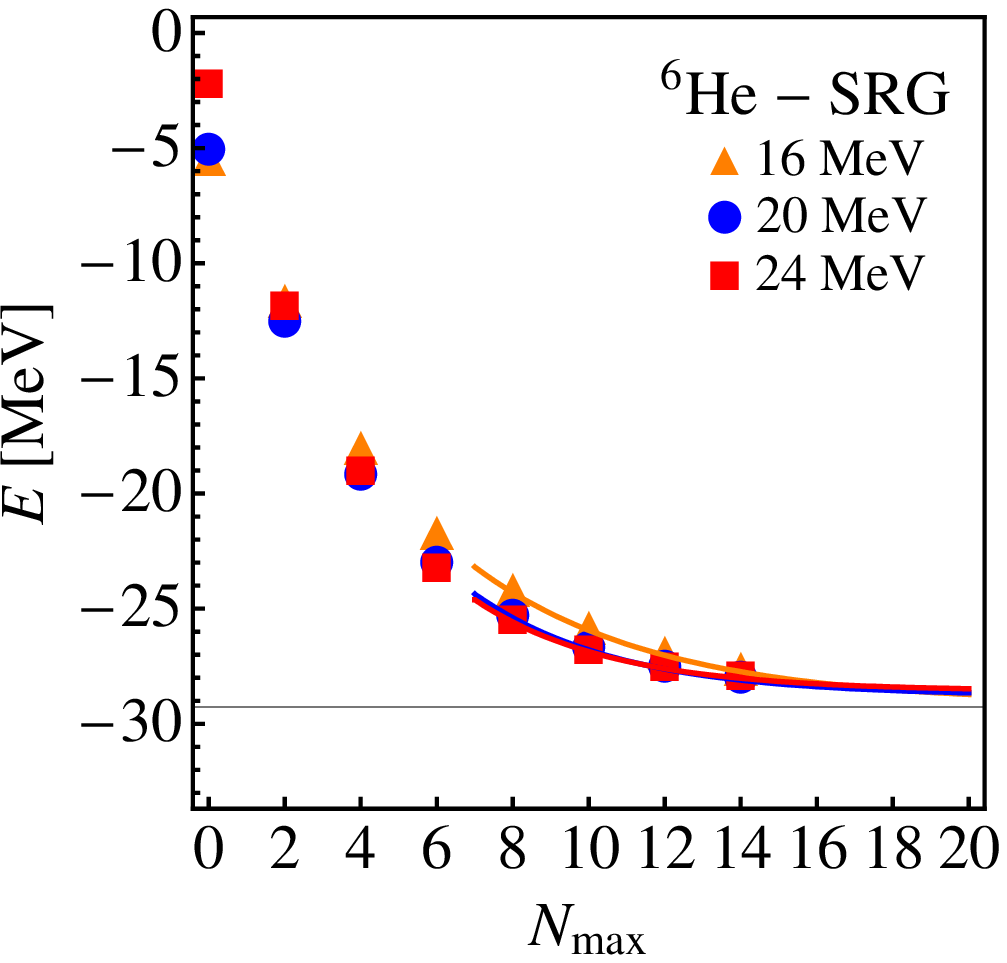}
  \caption{Energy of the \elem{He}{6} $0^+$ state as a function of
    model space size for different oscillator frequencies obtained with UCOM(var.), UCOM(SRG) and SRG interactions. Exponential extrapolations are fitted to the results from the four largest model spaces.}
  \label{fig:ncsm-he6-gs}
\end{figure}

\begin{figure}
  \includegraphics[width=0.31\textwidth]{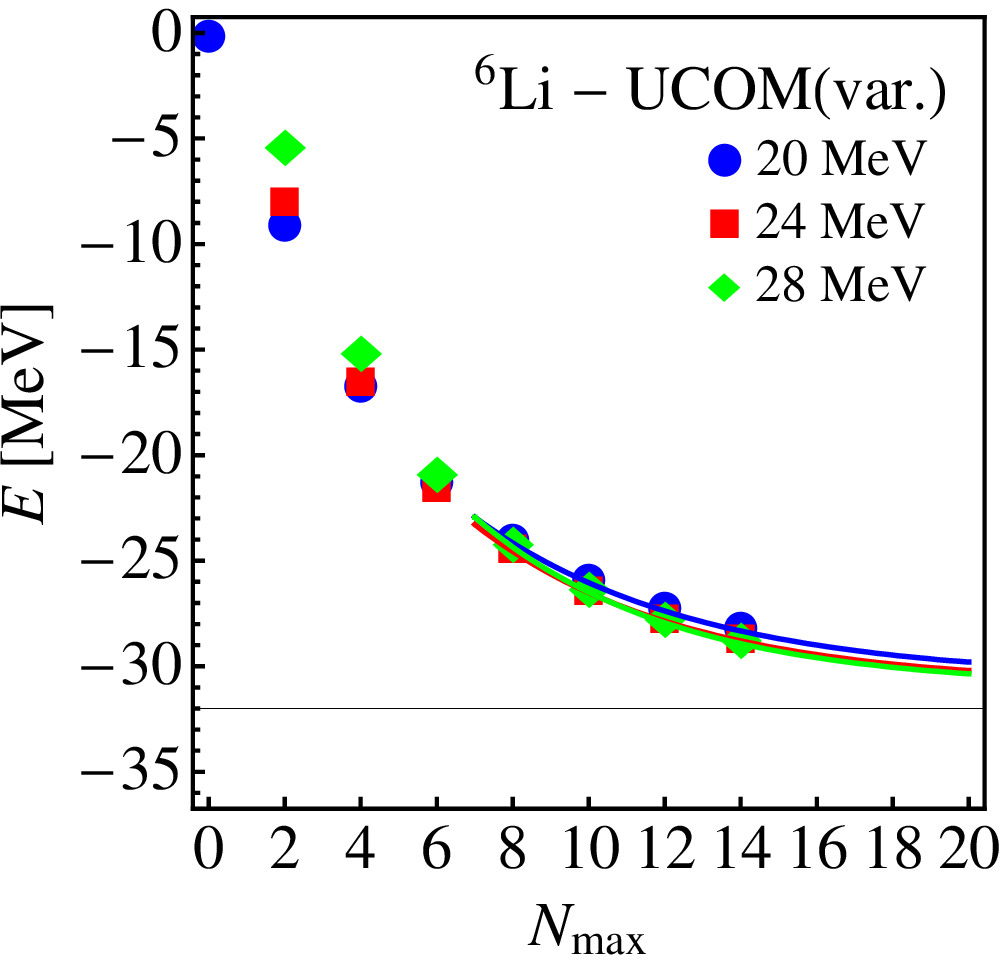}\hfil
  \includegraphics[width=0.31\textwidth]{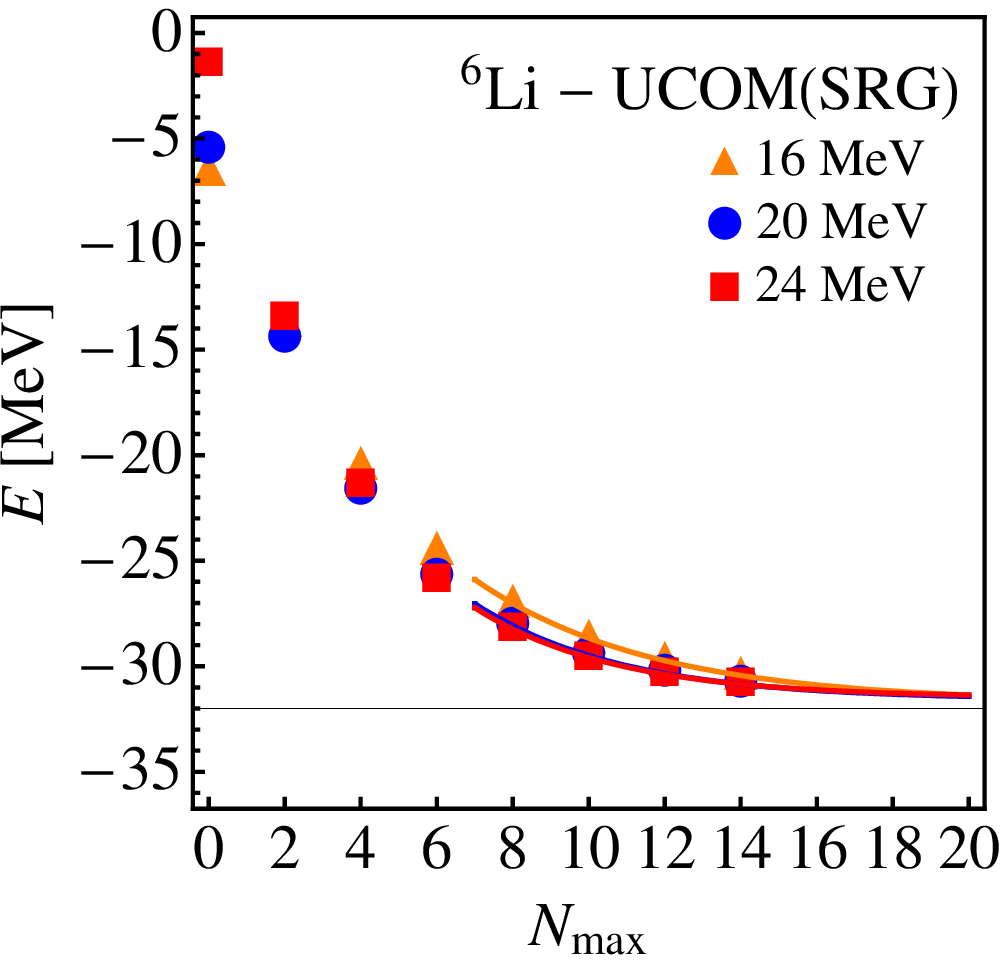}\hfil
  \includegraphics[width=0.31\textwidth]{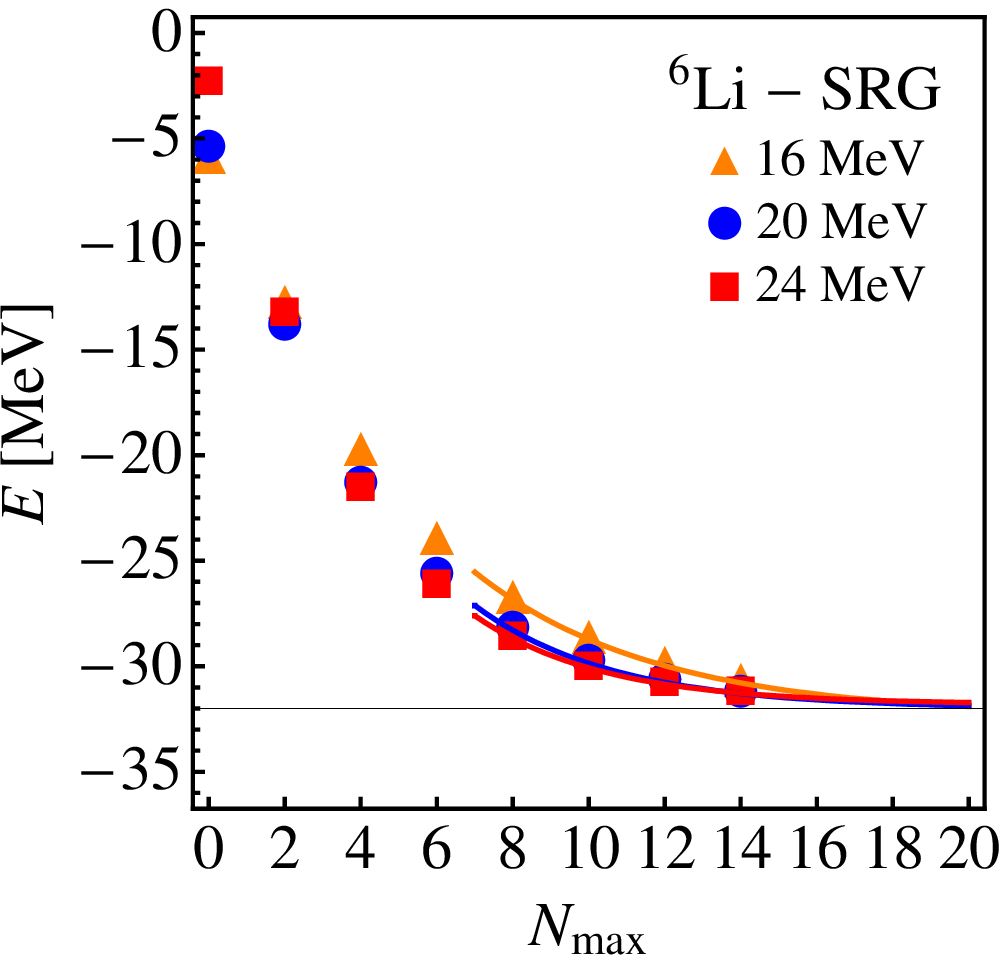}
  \caption{Energy of the \elem{Li}{6} $1^+$ state as a function of
    model space size for different oscillator frequencies obtained with UCOM(var.), UCOM(SRG) and SRG interactions. Exponential extrapolations are fitted to the results from the four largest model spaces.}
  \label{fig:ncsm-li6-gs}
\end{figure}

\begin{figure}
  \includegraphics[width=0.31\textwidth]{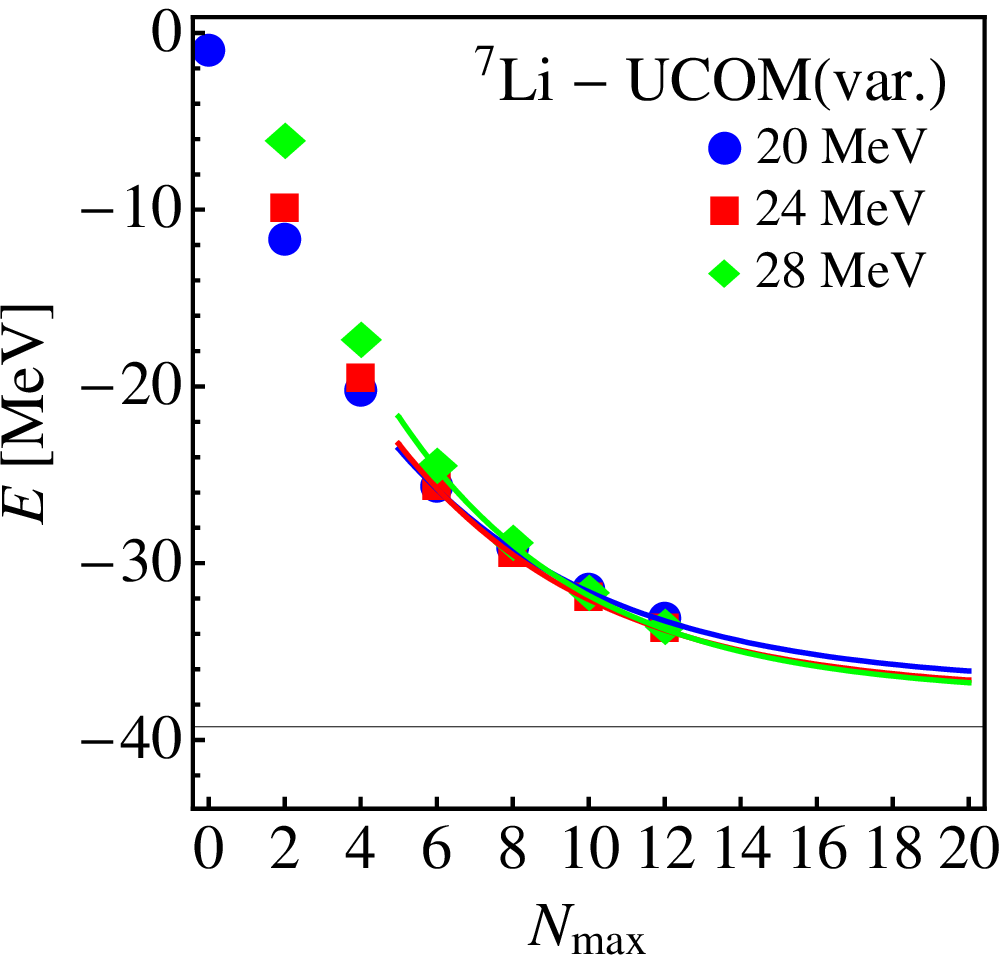}\hfil
  \includegraphics[width=0.31\textwidth]{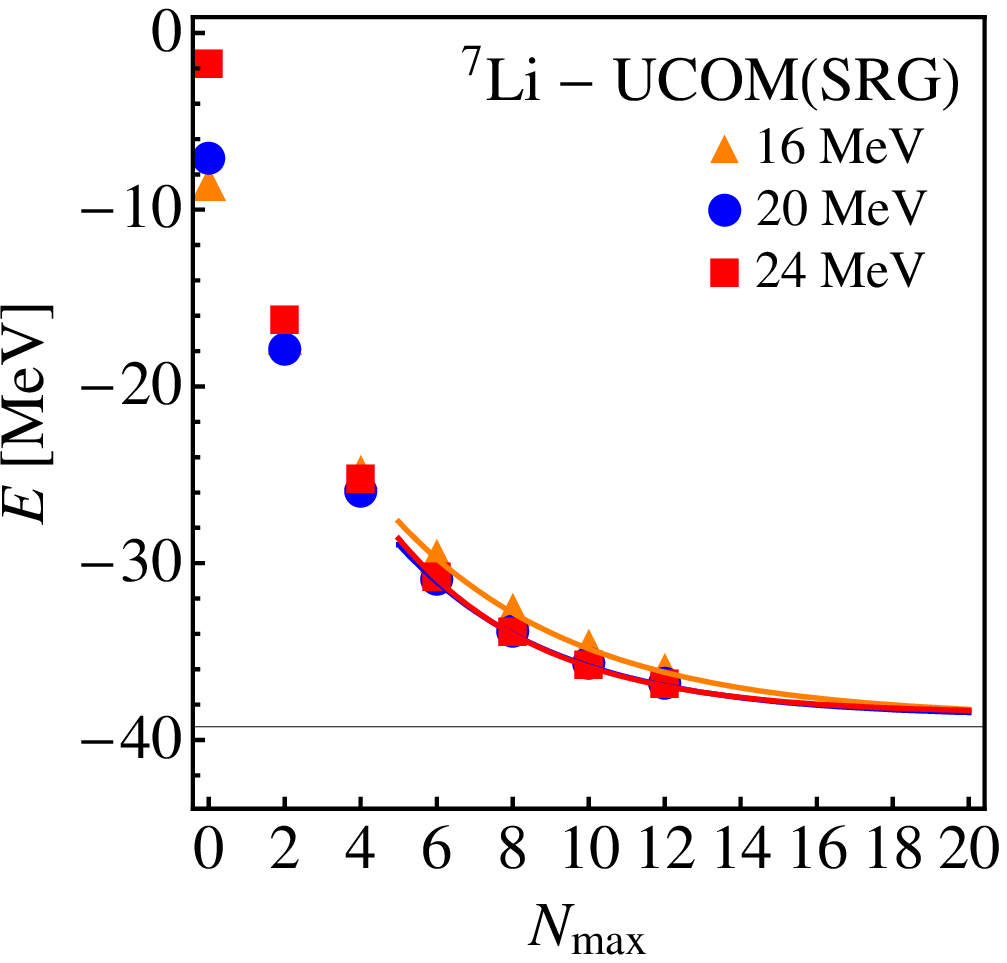}\hfil
  \includegraphics[width=0.31\textwidth]{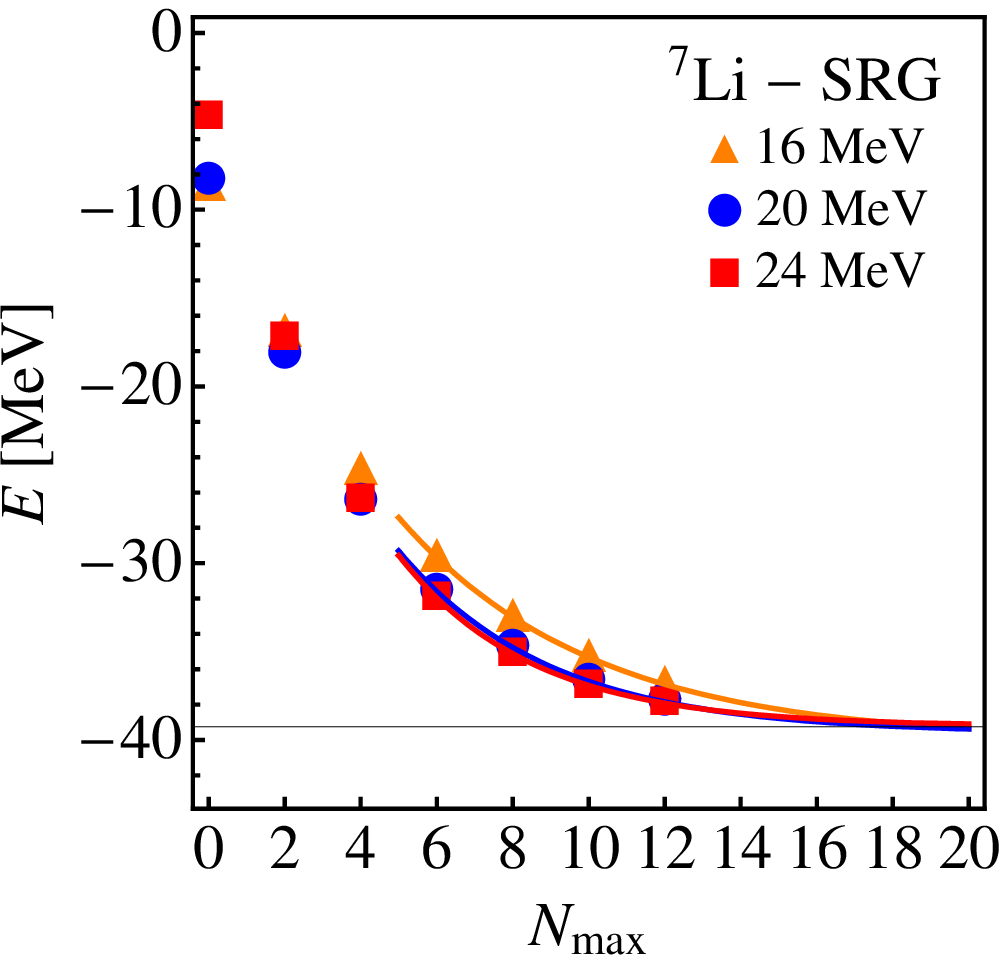}
  \caption{Energy of the \elem{Li}{7} $3/2^-$ state as a function of
    model space size for different oscillator frequencies obtained with UCOM(var.), UCOM(SRG) and SRG interactions. Exponential extrapolations are fitted to the results from the four largest model spaces.}
  \label{fig:ncsm-li7-gs}
\end{figure}

We calculate the ground-state energies in model spaces up to $\Nmax=14$ for \elem{He}{6} and \elem{Li}{6} and up to $\Nmax=12$ for
\elem{Li}{7} for oscillator frequences $\hbo = 12,16, \ldots, 28\MeV$. For all
interactions the lowest energy in the $0\hbo$ space is obtained for an
oscillator frequency of $16\MeV$. In the largest model spaces lowest
energies are found for oscillator constants between $24\MeV$ and
$28\MeV$ for the UCOM(var.) interaction and between $20\MeV$ and
$24\MeV$ for UCOM(SRG) and SRG interactions. To estimate the converged
ground-state energy an exponential extrapolation is performed using
the results obtained in the four largest model spaces. An estimate of
the error can be obtained by comparing the extrapolated energies for
different oscillator frequencies. In Figs.~\ref{fig:ncsm-he6-gs},\ref{fig:ncsm-li6-gs},\ref{fig:ncsm-li7-gs} the calculated energies are shown as a function of the model space
size $\Nmax$ together with the fitted exponentials. Both UCOM(SRG) and
SRG interactions provide extrapolated binding energies that are close
to the experimental values, whereas the binding energies with the
UCOM(var.) interaction are somewhat underestimated for all nuclei.
The results are summarized in Table~\ref{tab:ncsm-pshell-energies}.
Note that we use the interaction parameters \eqref{eq:optpar} as obtained from the Tjon-line analysis---no further adjustments are made here or in the following.

\begin{table}
  \centering\small
  \begin{tabular}{c|ccc|c}
 & UCOM(var.) & UCOM(SRG) & SRG & Experiment \\
 \hline
 $\elem{He}{6}$ & 27.9(4) & 28.4(3) & 28.8(4) & 29.269\\
 $\elem{Li}{6}$ & 30.9(4) & 31.6(3) & 32.0(4) & 31.995\\
 $\elem{Li}{7}$ & 37.4(6) & 38.7(4) & 39.6(5) & 39.245
  \end{tabular}
  \caption{Experimental and extrapolated calculated binding energies (in $\MeV$) obtained with UCOM(var.), UCOM(SRG) and SRG interactions. Error estimates are obtained by comparing extrapolated energies for different oscillator frequencies.}
  \label{tab:ncsm-pshell-energies}
\end{table}

\subsubsection*{Radii, magnetic dipole moments and quadrupole moments}

Electromagnetic properties provide important tests of the wave
function beyond the simple binding energy. We calculate the point
proton radii for all nuclei and the magnetic dipole moment as well as
the electric quadrupole moment for \elem{Li}{6} and
\elem{Li}{7}. The results are very similar for the different UCOM and
SRG interactions and we only show results for the UCOM(SRG)
interaction. The results are also very similar to results from NCSM
calculations using the CD Bonn and INOY interactions
\cite{caurier06,forssen09}.

\begin{figure}
  \includegraphics[width=0.31\textwidth]{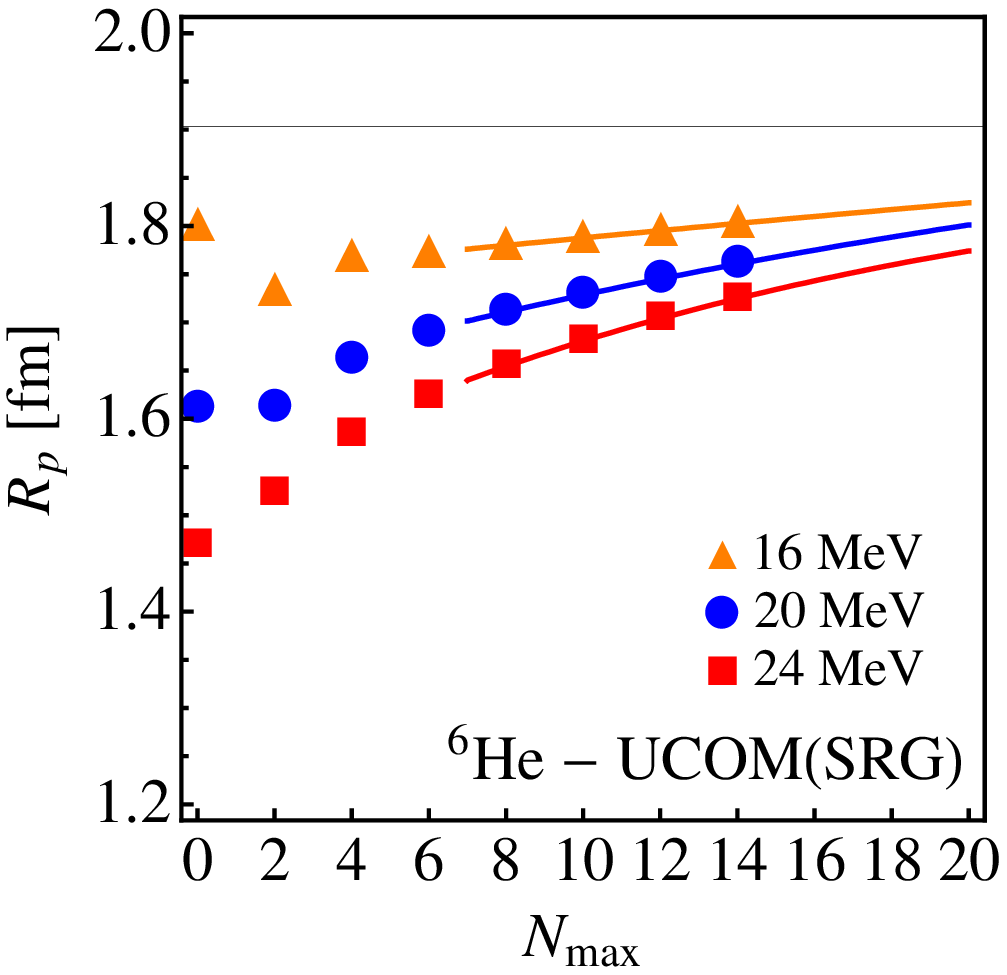}\hfil
  \includegraphics[width=0.31\textwidth]{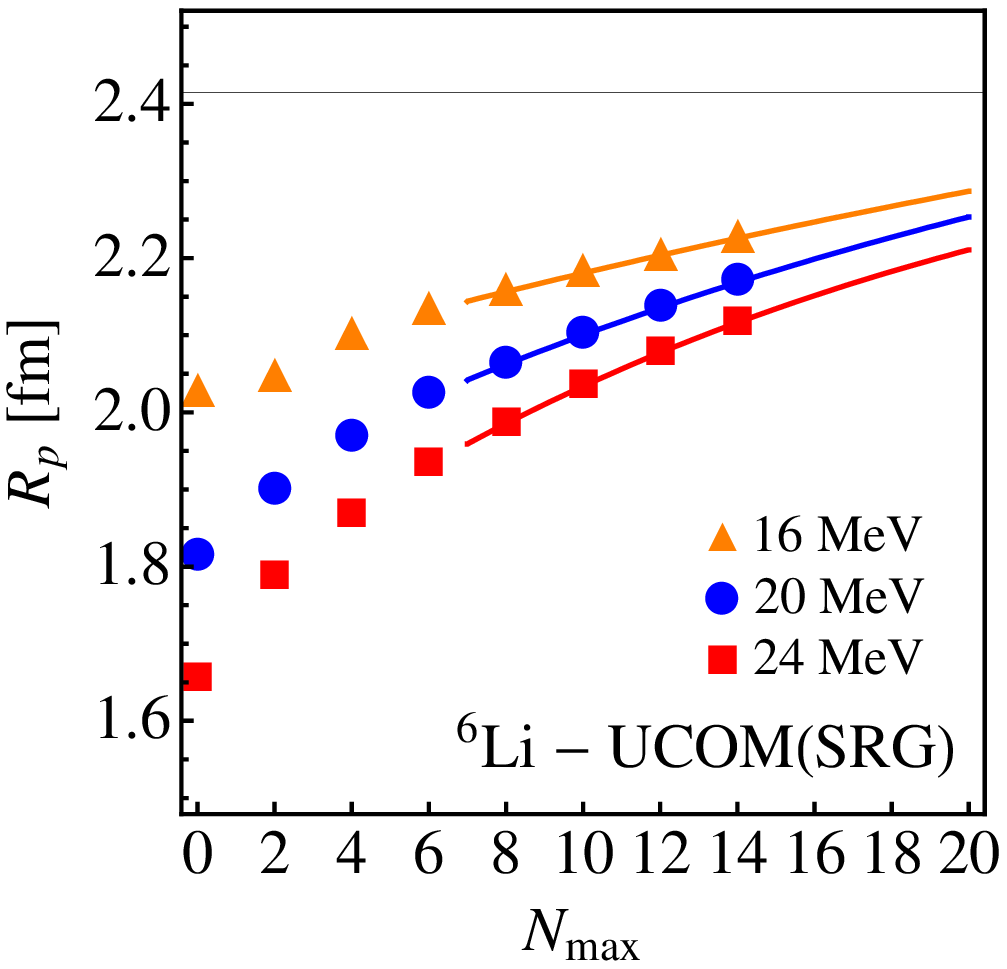}\hfil
  \includegraphics[width=0.31\textwidth]{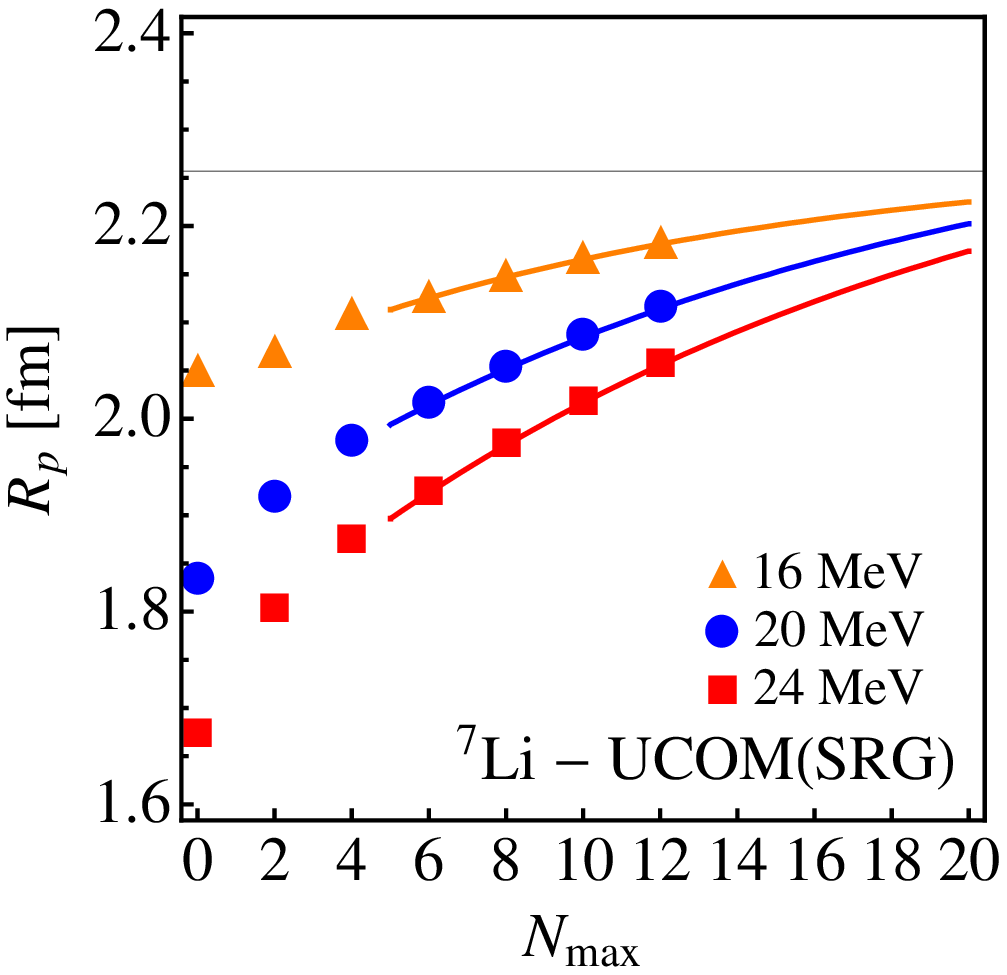}
  \caption{Point proton radii of \elem{He}{6}, \elem{Li}{6} and
    \elem{Li}{7} as a function of model space size for different oscillator frequencies using the UCOM(SRG) interaction. Experimental values from \cite{mueller07} and \cite{sanchez06}.}
  \label{fig:ncsm-radii}
\end{figure}

The point radii shown in Fig.~\ref{fig:ncsm-radii} are calculated for
the oscillator frequencies $\hbo = 16$, $20$, and $24\,\MeV$. There still is a strong
dependence of the calculated radii on the model space size. We do not
believe that the extrapolation of the radii is reliable, but the
results indicate that the radii are slightly too small when compared
to experiment. It is difficult to draw conclusions regarding the
saturation properties of the interaction as the radii for these nuclei
depend strongly on the asymptotic behavior of the wave function due to
their halo or cluster nature.

\begin{figure}
  \centering
  \includegraphics[width=0.31\textwidth]{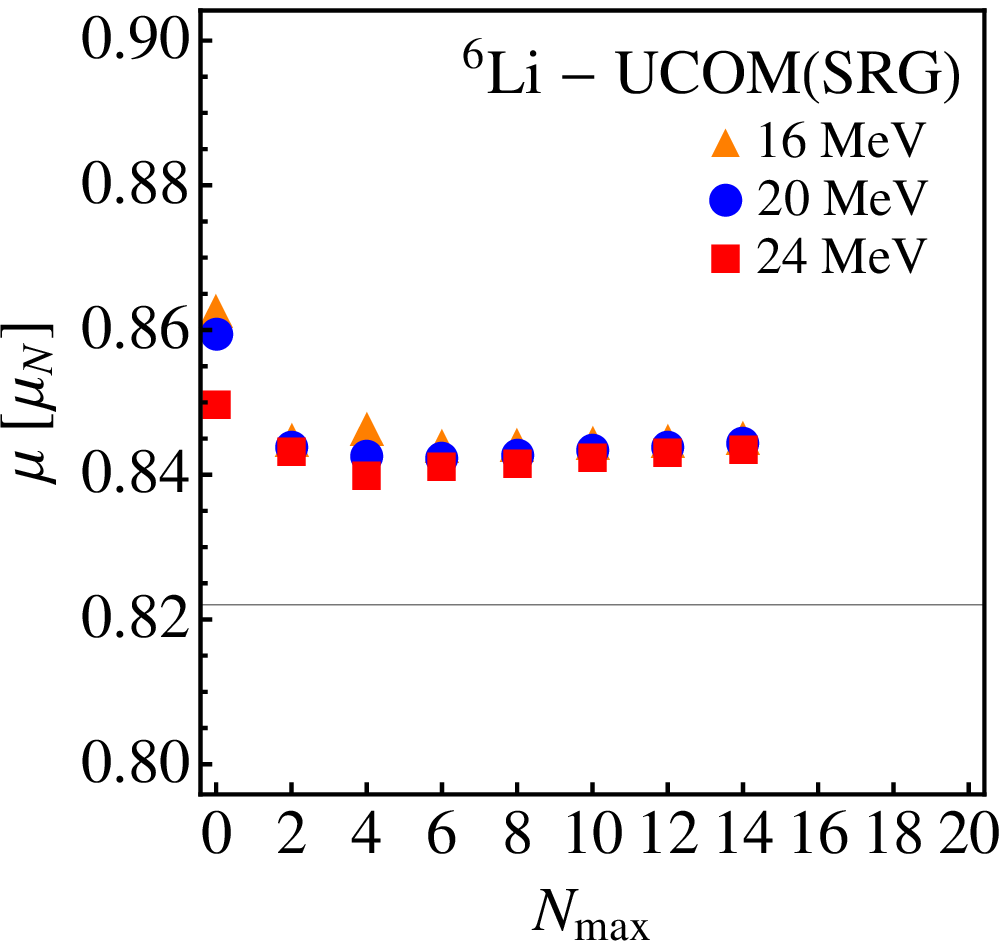}\hspace{0.1\textwidth}
  \includegraphics[width=0.31\textwidth]{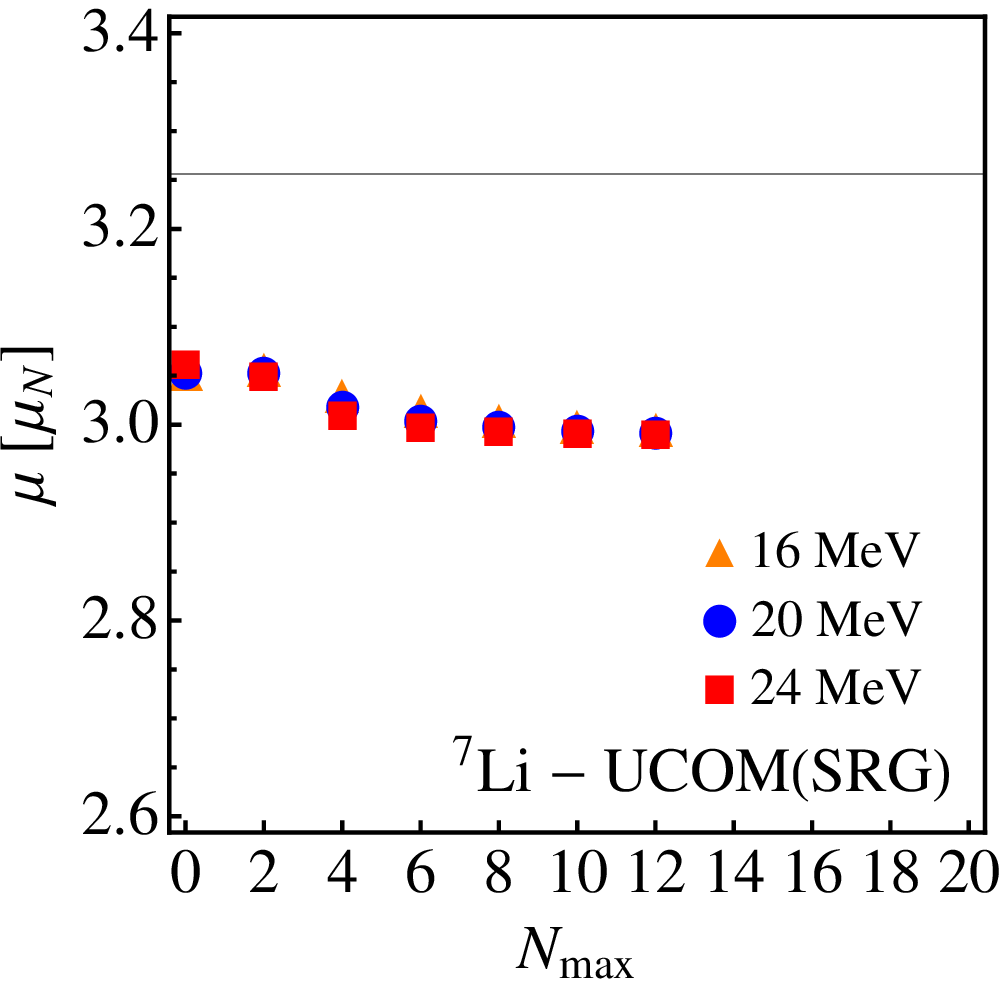}
  \caption{Magnetic dipole moment of \elem{Li}{6} and \elem{Li}{7} as
    a function of model space size calculated with UCOM(SRG). Experimental values from \cite{tilley02}.}
  \label{fig:ncsm-magmom}
\end{figure}

The magnetic dipole moments of \elem{Li}{6} and \elem{Li}{7} shown in
Fig.~\ref{fig:ncsm-magmom} agree reasonably well with the experiment. The
quadrupole moment of \elem{Li}{6} is correctly predicted to be very
small and negative, which confirms the \elem{He}{4} plus deuteron
picture. For the quadrupole moment of \elem{Li}{7} we find a similar
behavior as for the radii. There is still a strong dependence on the
model space size. The extrapolated results underestimate the
experimental value.

\begin{figure}
  \centering
  \includegraphics[width=0.31\textwidth]{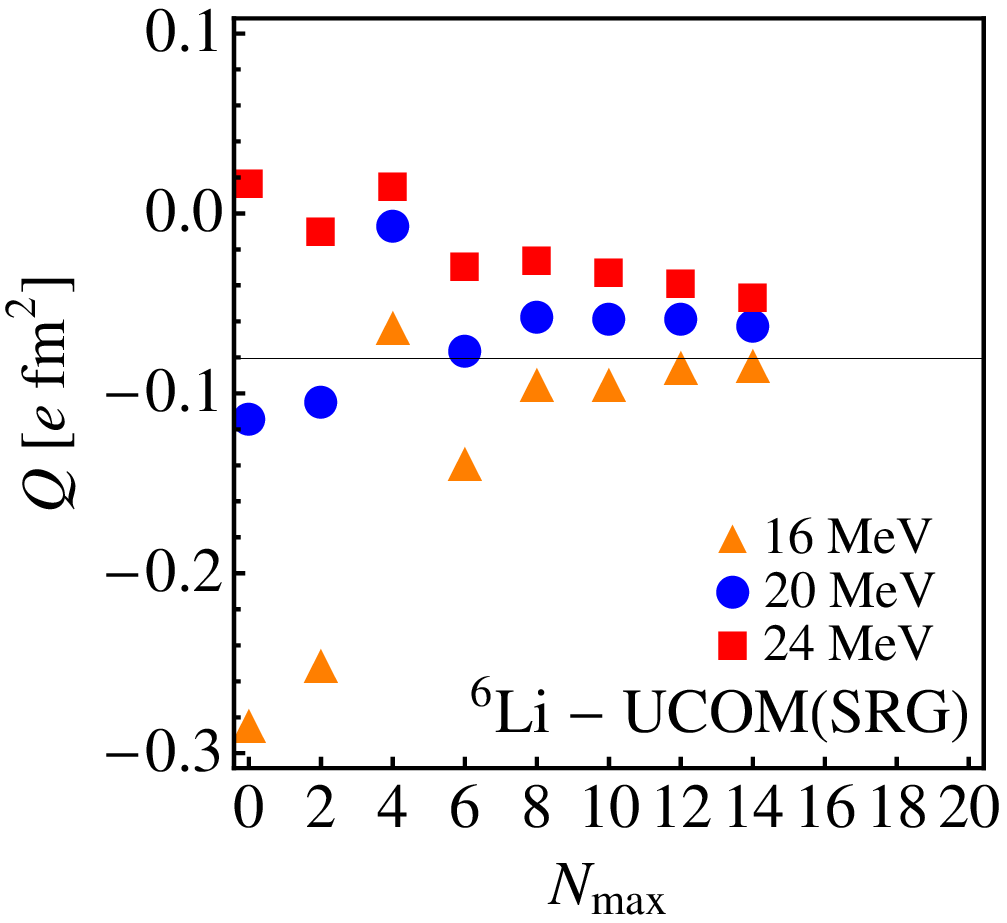}\hspace{0.1\textwidth}
  \includegraphics[width=0.31\textwidth]{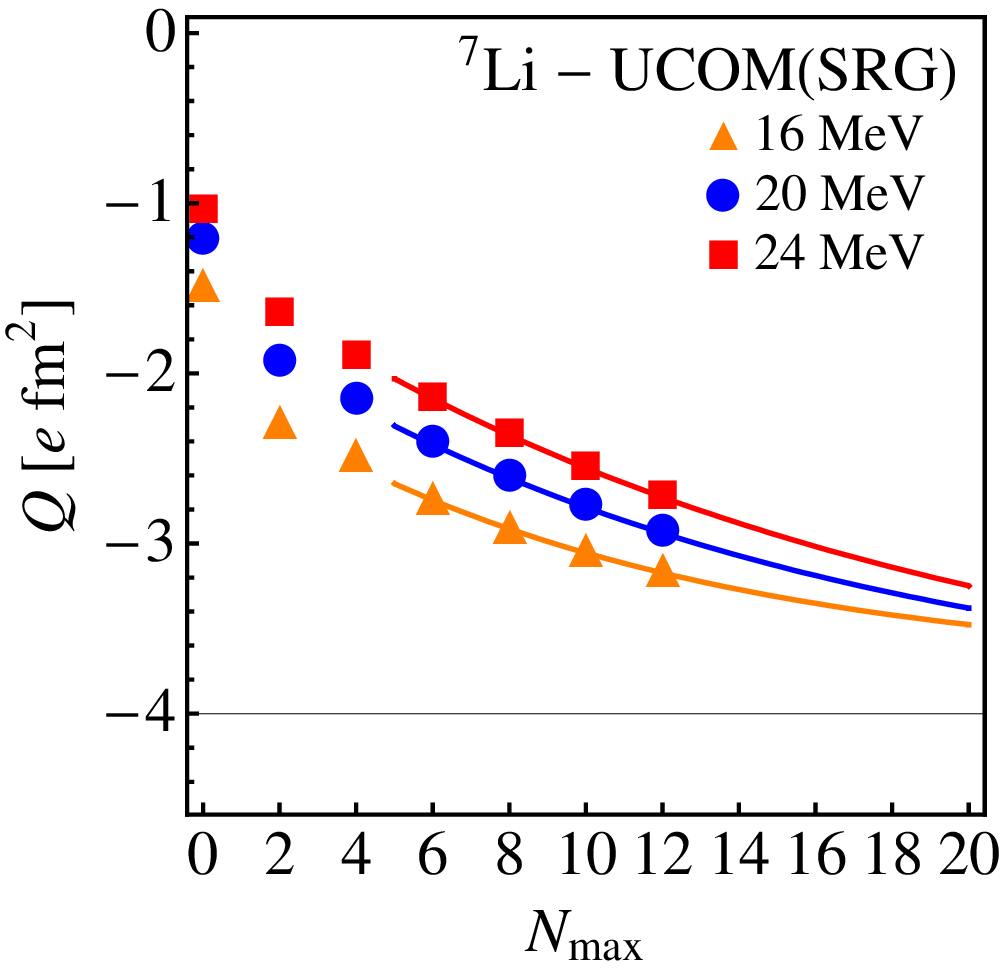}
  \caption{Quadrupole moment of \elem{Li}{6} and \elem{Li}{7} ground
    state as a function of model space size for UCOM(SRG). Experimental values from \cite{tilley02}.}
  \label{fig:ncsm-quadmom}
\end{figure}

\subsubsection*{Spectra}

Further information about the interaction can be obtained from the
spectra of excited states. The spectra are calculated in $12\hbo$ and
$10\hbo$ model spaces for the $A=6$ and $A=7$ nuclei respectively. We
show here the results for the oscillator parameter $16\MeV$ which
corresponds to the ground state minimum in the $0\hbo$ model space and
which shows the fastest convergence for the spectra. For other
oscillator constants the spectra vary more rapidly when enlarging the
model space but the converged results depend only very weakly on the
oscillator parameter.

\begin{figure}
  \centering
  \includegraphics[width=0.72\textwidth]{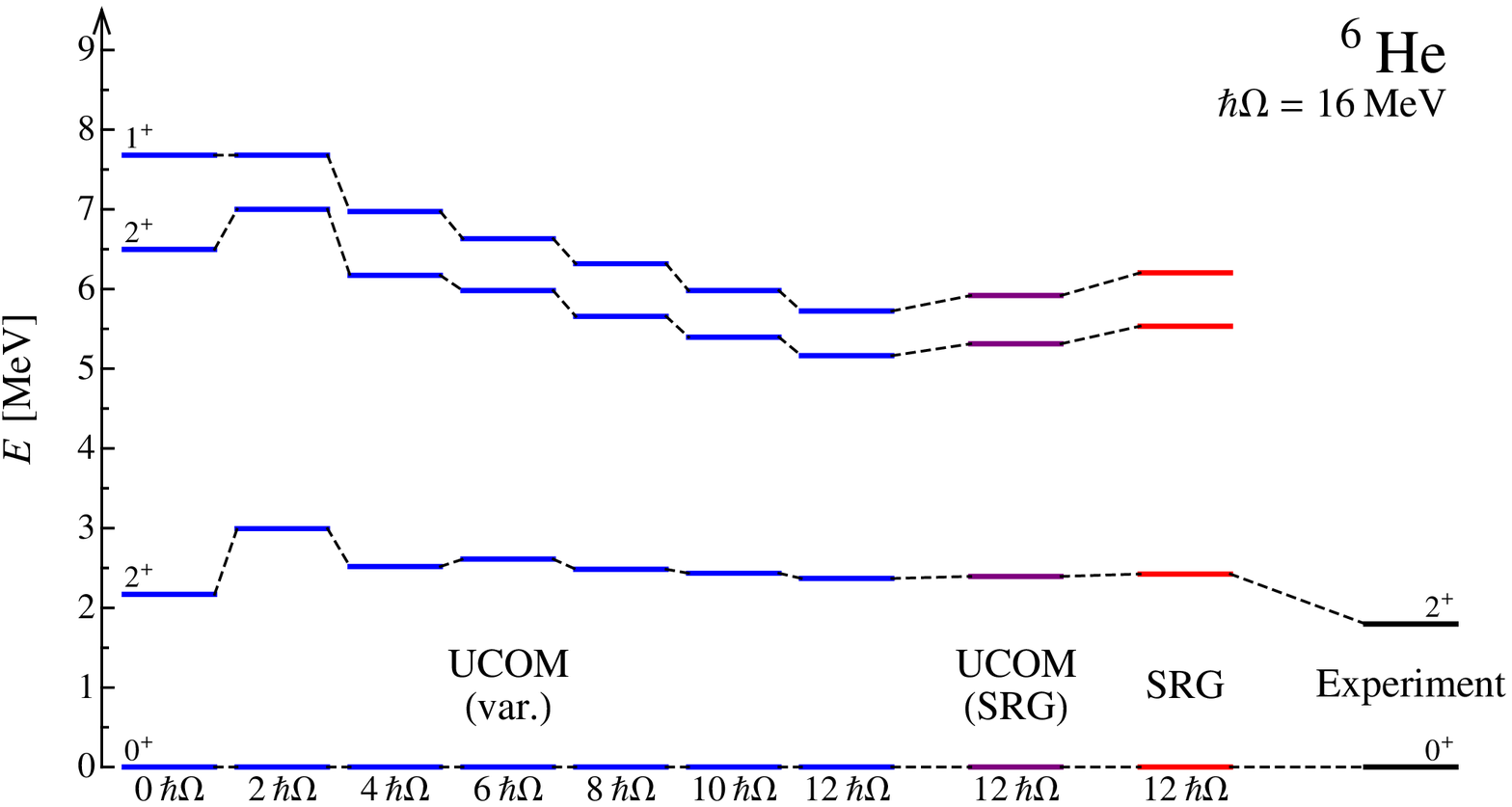}\\[2ex]
  \includegraphics[width=0.72\textwidth]{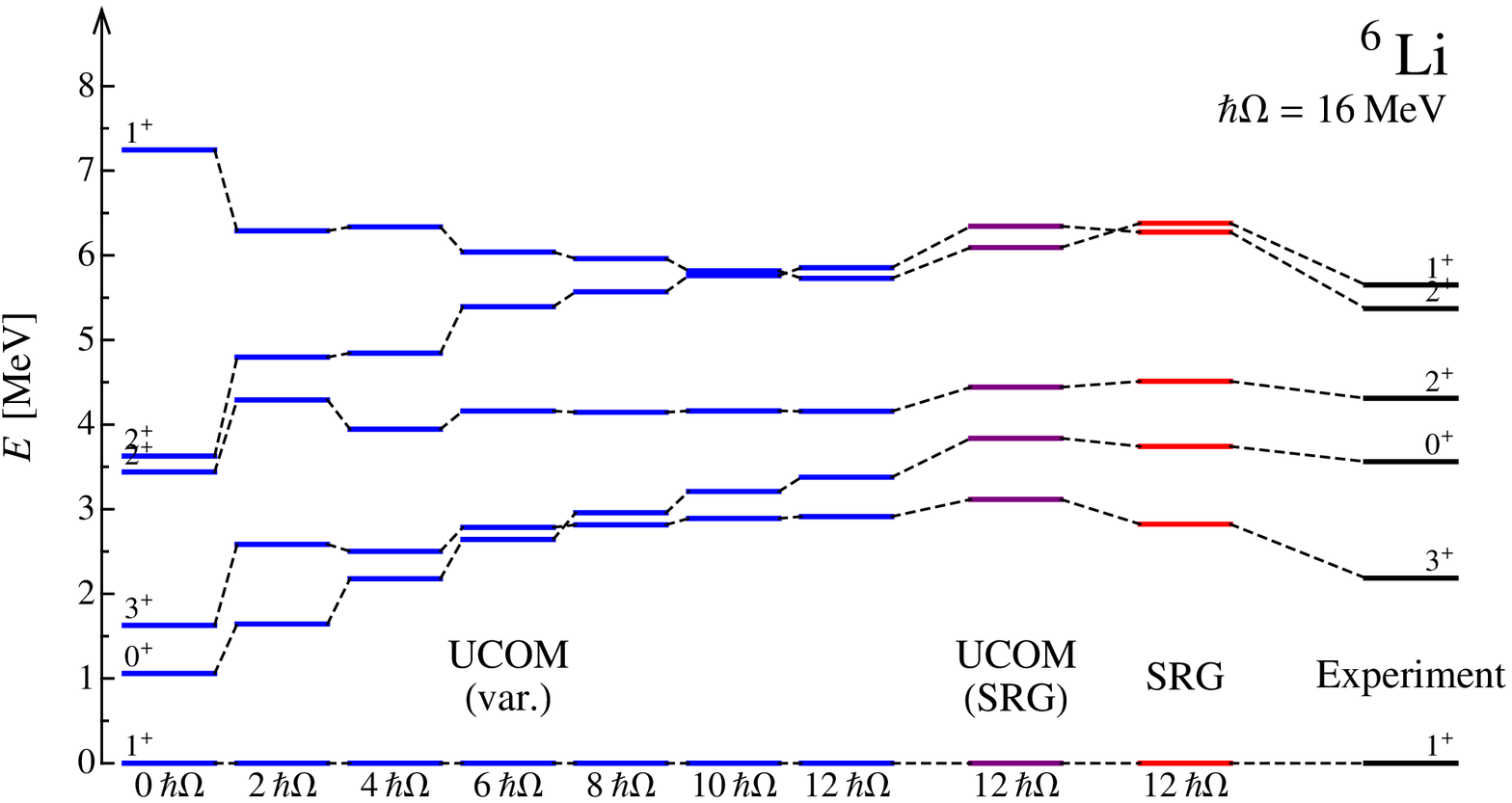}\\[2ex]
  \includegraphics[width=0.72\textwidth]{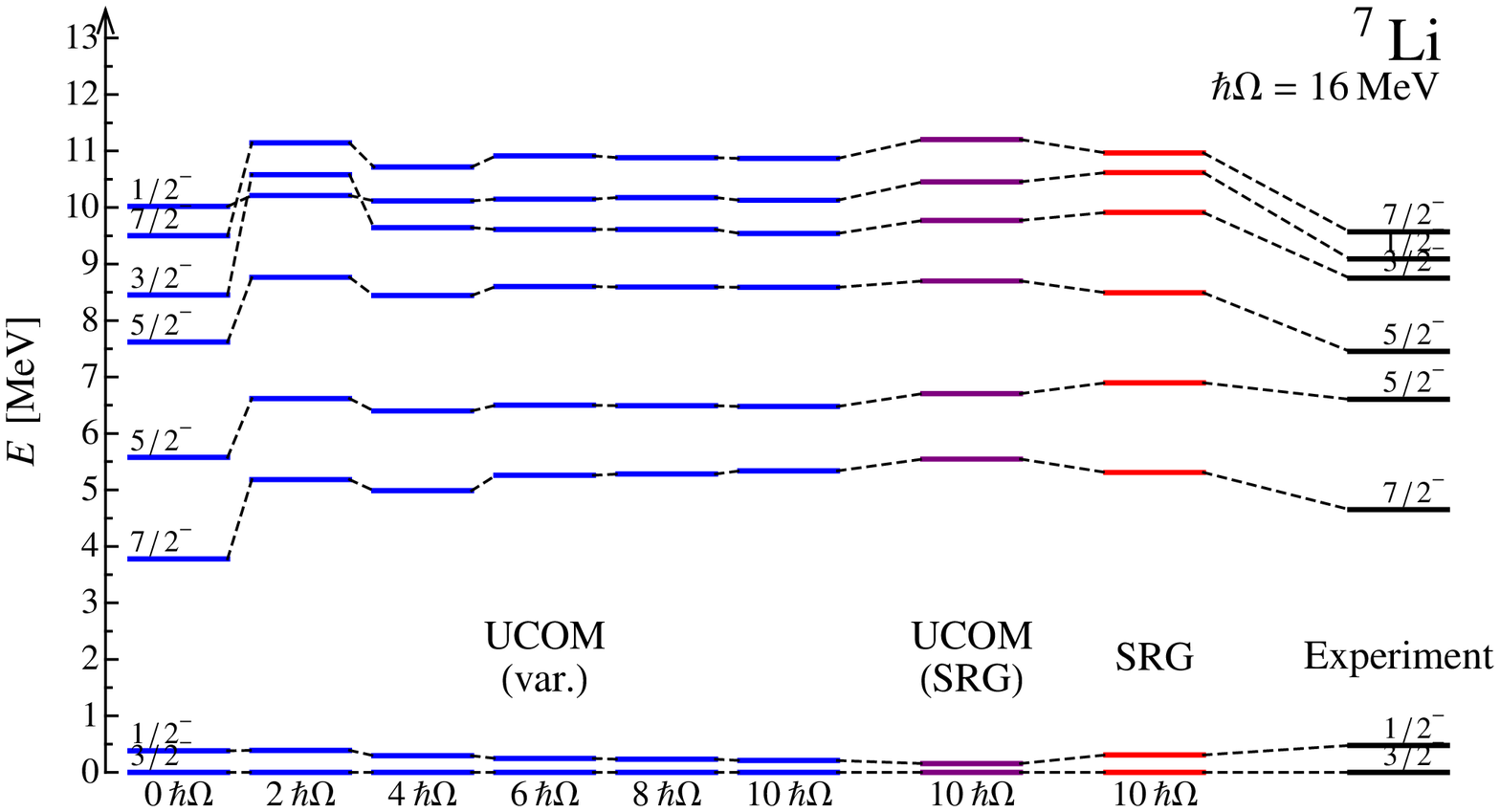}
  \caption{Spectra of low lying states in \elem{He}{6}, \elem{Li}{6} and \elem{Li}{7} calculated with UCOM(var.), UCOM(SRG) and SRG interactions in comparison with experiment \cite{tilley02}.}
  \label{fig:ncsm-spectra}
\end{figure}

The results are very similar for all three interactions. In Fig.~\ref{fig:ncsm-spectra} the results are summarized. For the UCOM(var.) interaction the results are shown starting from the $0\hbo$ model space. For the UCOM(SRG) and SRG interactions we only show the results from the largest model spaces.

In \elem{He}{6} the energy of the $2^+$ state is well converged. This is
not true for the second $2^+$ and the $1^+$ state. These states are
well above the two-neutron separation energy and there is no
experimental confirmation for the existence of these states.

For \elem{Li}{6} the $T=0$
states are well converged in contrast to the $T=1$ states ($2^+_2$ and
$0^+$). We also find the excitation energy of the $3^+$ state to be
too high. This indicates that the effective spin-orbit force in the
UCOM and SRG interactions is too weak. A similar observation can be
made in the spectrum of \elem{Li}{7}. Here the splittings between the
$3/2^-$ and $1/2^-$ states as well as between the $7/2^-$ and $5/2^-$
states are too small compared to the
experimental values, as well. In that respect the UCOM and SRG interactions
perform similarly to other two-body interactions. It has been observed
in GFMC calculations \cite{PiWi01} that three-body forces contribute significantly
to the effective spin-orbit strength. NCSM calculations with chiral
two- and three-body forces \cite{nogga06} also show a strong
dependence of these splittings on the parameters of the three-body
force.

%%%%%%%%%%%%%%%%%%%%%%%%%%%%%%%%%%%%%%%%%%%%%%%%%%%%%%%%%%%%%%%%%%%%%%
%%%%%%%%%%%%%%%%%%%%%%%%%%%%%%%%%%%%%%%%%%%%%%%%%%%%%%%%%%%%%%%%%%%%%%
%%%%%%%%%%%%%%%%%%%%%%%%%%%%%%%%%%%%%%%%%%%%%%%%%%%%%%%%%%%%%%%%%%%%%%
\clearpage
\section{Hartree-Fock and beyond}
\label{sec:hf}

For nuclei beyond the p-shell, many-body calculations in methods like the full NCSM are not feasible anymore, because the dimension of the many-body basis at a given $N_{\max}\hbar\Omega$ truncation level grows factorially with the particle number. One can extend the domain of NCSM calculations to larger particle numbers and model space sizes by using importance truncation methods, as discussed in Refs. \cite{RoNa07,RoNa08,Roth09}. For the description of ground states of closed-shell and neighboring nuclei, coupled cluster methods have been employed quite successfully \cite{WlDe05,KoDe04,HaDe07,HaPa08,HaPa09b,RoGo09,RoGo09b}. These methods are able to cover the majority of correlations in the nuclear many-body system, however, they are computationally demanding as well. 

At the opposite end of the scale regarding the computational cost and the ability the describe correlations is the Hartree-Fock approach \cite{RiSc80}. A simple Hartree-Fock calculation---based on a single Slater determinant for the description of the ground state---can be done easily for any isotope throughout the nuclear chart. Obviously, the Hartree-Fock approach does not allow for the description of any correlations and, therefore, cannot provide a quantitative approximation for nuclear observables when using realistic nuclear interactions. It does, however, provide a variational upper bound for the exact ground-state energy and as such can be used to assess the qualitative systematics, e.g., of the binding energies as function of mass number, throughout the whole nuclear chart. Furthermore, the Hartree-Fock solution can serve as a starting point for improved approximations that take the missing correlations into account. In the simplest case, low-order many-body perturbation theory can be used to estimate the effect of correlations on the energy or other observables. Alternative methods for the inclusion of correlations beyond Hartree-Fock, such as ring-, ladder-, or Pad\'e-resummed perturbation theory \cite{BaPa06,RoLa10} as well as Brueckner-Hartree-Fock schemes and Green's function methods \cite{Barb06,DiBa04} are also feasible. The Hartree-Fock or the corresponding Hartree-Fock-Bogoliubov \cite{HeRo09,HeRo09b} solutions also form the basis for the description of collective excitations in the Random Phase Approximation at different orders \cite{PaPa06,PaRo09,PaRo10}.

In the following we will use the Hartree-Fock scheme as well as second-order many-body perturbation theory to study the systematics of binding energies and charge radii resulting from the UCOM- and SRG-transformed interactions. The aim is not to provide a precise prediction of the ground state energy for heavy nuclei, but to assess the systematic behavior of the transformed two-body interactions with increasing mass number.

%%%%%%%%%%%%%%%%%%%%%%%%%%%%%%%%%%%%%%%%%%%%%%%%%%%%%%%%%%%%%%%%%%%%%%
%%%%%%%%%%%%%%%%%%%%%%%%%%%%%%%%%%%%%%%%%%%%%%%%%%%%%%%%%%%%%%%%%%%%%%
\bigskip
\subsection{Hartree-Fock with UCOM- and SRG-transformed interactions}

In the simplest formulation of a Hartree-Fock (HF) scheme, the many-body state is approximated by a single Slater determinant
\eq{ \label{eq:hf_hfstate}
  \ket{\text{HF}} = \ket{\Phi_{[\nu]}} 
  = \mathcal{A}\; (\ket{\phi_{\nu_1}}\otimes\ket{\phi_{\nu_2}}\otimes\cdots\otimes\ket{\phi_{\nu_A}}),
}
where $\mathcal{A}$ is the antisymmetrization operator acting on an $A$-body product state. The single-particle states $\ket{\phi_{\nu}}$ are used as variational degrees of freedom in a minimization of the expectation value of the many-body Hamiltonian. The formal variational solution of the many-body problem using the trial state \eqref{eq:hf_hfstate} leads to the well known HF equations for the single-particle states, which have to be solved self-consistently \cite{RiSc80,RoPa06}.

The Hamiltonian itself is the same as it was used in Sec. \ref{sec:ncsm} for the NCSM calculations. It consists of the intrinsic kinetic energy $\op{T}_{\intr} = \op{T} - \op{T}_{\cm}$ and the transformed two-body interaction $\op{V}_{\text{NN}}$ including Coulomb and charge-dependent terms 
\eq{
  \op{H}_{\intr} 
  = \op{T} - \op{T}_{\cm} + \op{V}_{\text{NN}} 
  = \op{T}_{\intr} + \op{V}_{\text{NN}} \;,
}
Unlike the NCSM, the use of this translational invariant Hamiltonian does not guarantee that the HF ground state is free of spurious center-of-mass contaminations. The Slater determinant form of the many-body state does not allow a separation of intrinsic and center-of-mass motion for general single-particle states. However, for the purpose of the present discussion, the effect of center-of-mass contaminations on the ground-state energy is irrelevant. A more stringent but computationally expensive treatment of the center-of-mass problem would require an explicit center-of-mass projection \cite{Schm02,RoSc04}.

We formulate the HF scheme in a basis representation using harmonic oscillator single-particle states. Thus the matrix elements entering HF equations are the same as in the NCSM calculations of Sec. \ref{sec:ncsm}. The HF single particle states $\ket{\phi_{\nu}}$ are written as 
\eqmulti{ \label{eq:hf_spstates}
  \ket{\phi_{\nu}} 
  = \ket{\alpha l j m m_t} 
  = \sum_n C^{(\alpha l j m m_t)}_{n} \ket{n l j m m_t} \;,
}
where $\ket{n l j m m_t}$ denotes a harmonic oscillator single-particle state with radial quantum number $n$, orbital angular momentum $l$, total angular momentum $j$ with projection $m$, and isospin projection quantum number $m_t$. Assuming spherical symmetry, only oscillator states with the same quantum numbers $l$, $j$, and $m$ can contribute in the expansion. In the following, we will restrict ourselves to constrained or closed-shell calculations, where $C^{(\alpha l j m m_t)}_{n}=C^{(\alpha l j m_t)}_{n}$ is independent of $m$. The details of the resulting HF equations for the expansion coefficients $C^{(\alpha l j m_t)}_{n}$ and of their solution are discussed in Ref. \cite{RoPa06}.

%%%%%%%%%%%%%%%%%%%%%%%%%%%%%%%%%%%%%%%%%%%%%%%%%%%%%%%%%%%%%%%%%%%%%%
\begin{figure}[t]
\centering\includegraphics[width=0.75\textwidth]{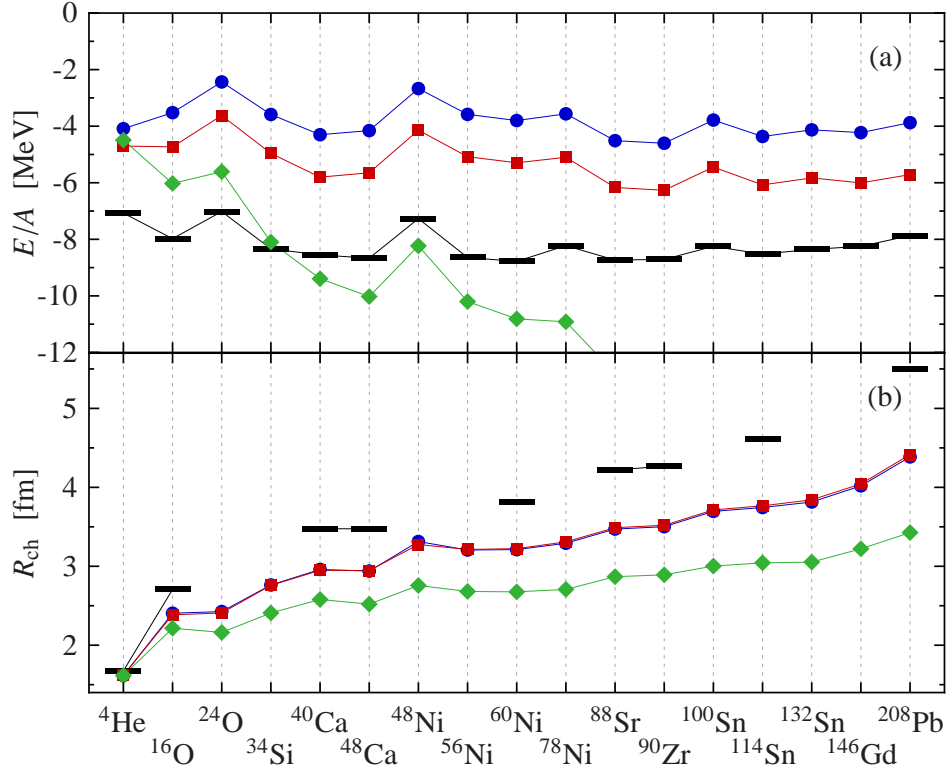}
\caption{Ground-state energies per nucleon (a) and charge radii (b) for a sequence of nuclei with closed $j$-shells obtained at the Hartree-Fock level. The data sets correspond to UCOM(var.) interaction (\symbolcircle[FGBlue]), the UCOM(SRG) interaction (\symbolbox[FGRed]), and the SRG interaction (\symboldiamond[FGGreen]). The range or flow parameter of the different transformations is fixed such that the experimental \elem{He}{4} ground-state energy is reproduced in NCSM calculations (cf. Sec. \ref{sec:ncsm-tjon-line}). The black bars indicate experimental values \cite{AuWa03}.}
\label{fig:hf_compint}
\end{figure}
%%%%%%%%%%%%%%%%%%%%%%%%%%%%%%%%%%%%%%%%%%%%%%%%%%%%%%%%%%%%%%%%%%%%%%

Within the HF approximation, we consider the ground-state energies and the charge radii for a sequence of nuclei with closed $j$-shells from \elem{He}{4} to \elem{Pb}{208} using the different UCOM- and SRG-transformed interactions adopted in Sec. \ref{sec:ncsm}. For these calculations the harmonic oscillator single-particle basis includes 15 major shells, i.e., the sum in \eqref{eq:hf_spstates} is limited to $2n+l\leq14$, which is sufficient to guarantee full convergence of the HF states for all nuclei under consideration. We use a sequence of oscillator lengths $a_{\text{HO}}$ from $1.3\,\text{fm}$ to $2.4\,\text{fm}$ for the underlying oscillator basis. In accord with the variational principle, we adopt the value of the oscillator length that yields the minimal ground-state energy, though, for the basis sizes used here, the energies and radii obtained on the HF level are largely independent of the oscillator length.

A summary of the HF ground-state energies and charge radii obtained with the UCOM and SRG-transformed potentials is given in Fig.~\ref{fig:hf_compint}. Here we adopt the UCOM-correlators obtained variationally and through the SRG-mapping with the range or flow parameters determined by fitting to the experimental \elem{He}{4} ground-state energy in converged NCSM calculations. Similarly, the flow parameter for the SRG-evolved interaction is determined through the \elem{He}{4} binding energy (cf. Sec. \ref{sec:ncsm-tjon-line}). The HF approximation, therefore, yields similar results for the ground-state energy of \elem{He}{4} with all three interactions---UCOM(var.), UCOM(SRG), and SRG. However, the HF energy is above the experimental and the converged NCSM ground-state energy, which is expected---the single determinant describing the HF ground state corresponds to a $0\hbar\Omega$ NCSM eigenstate and cannot describe any of the correlations that the NCSM model space will capture with increasing model space size $N_{\max}\hbar\Omega$. 

With increasing mass number the behavior of the UCOM and the SRG-transformed interactions differs dramatically. The UCOM interactions, both UCOM(var.) and UCOM(SRG), lead to HF ground-state energies per nucleon that are of the order of $4$ or $5$ MeV throughout the whole mass range. Especially for the UCOM(var.) interaction there is a practically constant offset of about $4\,\text{MeV}$ per nucleon between the HF energies and experiment. For the case of \elem{He}{4} we know from the NCSM calculations in Sec. \ref{sec:ncsm} that the missing binding energy resides in correlation energy, because the same effective interaction reproduced the experimental values in large model spaces. We expect that the inclusion of correlations beyond HF will also lower the ground-state energies of the heavier nuclei and bring them closer to experiment. 

The SRG-transformed interaction exhibits a very different trend: The HF binding energy per nucleon increases rapidly with increasing mass number. Already for intermediate masses, the HF energy drops below the experimental ground-state energy. Given that the HF energy gives an upper bound for the exact energy eigenvalue of the Hamiltonian, this discrepancy cannot be remedied through the inclusion of beyond-HF correlations, but hints at the induced many-body forces (see discussion in Secs. \ref{sec:intro} and \ref{sec:ucom_cluster} that are left out in this calculation. In the SRG they have an over-all repulsive effect. 

This intrinsic difference between UCOM and SRG interactions is also reflected in the charge radii depicted in Fig. \ref{fig:hf_compint}(b). The charge radii obtained with the UCOM interaction show a systematic deviation form the experimental trend. The predicted radii are too small and the difference to experiment increases linearly with increasing mass number reaching a deviation of about $1\,\text{fm}$ for \elem{Pb}{208}. For the SRG interaction, the deviation is even more pronounced, the radius of \elem{Pb}{208} is underestimated by about $2\,\text{fm}$.
 
There are two main differences between the UCOM-transformed and the SRG-transformed two-body interactions which cause the different behavior when going to heavier nuclei: (\emph{i}) The UCOM interactions use a unitary transformation that is optimized to account for short-range correlations in the lowest partial wave of each spin-isospin channel only. Thus, higher partial waves, whose impact grows with increasing mass number, are not pre-diagonalized in an optimal way by the UCOM transformation. However, because the wave functions for larger $L$ are suppressed at short distances by the centrifugal barrier, this is a minor effect. The SRG transformation, in contrast, handles each partial wave separately and thus leads to an optimal pre-diagonalization for all.
(\emph{ii}) Even for the lowest partial waves the UCOM transformation does not provide the same perfect pre-diagonalization in the high-momentum sector as the SRG transformation, as discussed in Sec.~\ref{sec:srg_matrixelem}. The residual off-diagonal high-momentum matrix elements together with the less pronounced pre-diagonalization of the higher partial waves stabilize the UCOM interactions against the overbinding observed in the SRG calculations. On the other hand this leads to slower convergence of the UCOM interactions as compared to the SRG. 

One can view the difference between SRG and UCOM from yet another perspective. Both approaches use a unitary transformation, which preserves the eigenvalues of the Hamiltonian in many-body space, provided the transformation is done without any additional truncations. However, here we use the cluster expansion and truncate at two-body level, i.e., we discard all the induced many-body forces. Thus, if we observe a systematic difference in an exact many-body calculation using two-body part of the transformed interactions only, we can conclude that the omitted many-body forces behave systematically different in SRG and UCOM, because the transformed Hamiltonians in their complete form in $A$-body space have to yield the same result. In this view, the simplistic HF calculations already show that the induced three-body and many-body interactions in the SRG framework must have a much larger net effect on binding energies and wave functions---and thus on radii---than in the UCOM framework.

%%%%%%%%%%%%%%%%%%%%%%%%%%%%%%%%%%%%%%%%%%%%%%%%%%%%%%%%%%%%%%%%%%%%%%
%%%%%%%%%%%%%%%%%%%%%%%%%%%%%%%%%%%%%%%%%%%%%%%%%%%%%%%%%%%%%%%%%%%%%%
\bigskip
\subsection{Low-order many-body perturbation theory}

The simplest way to estimate the effect of correlations beyond HF is low-order many-body perturbation theory. Many-body perturbation theory (MBPT) starting from the HF solution is a standard technique in many fields of quantum many-body physics, ranging from quantum chemistry \cite{SzOs96} to nuclear physics \cite{Gold57,FeMa77,StSt01,CoIt03,RoPa06,RoLa10}. It is straightforward to apply and computationally simple, but has inherent limitations. It is well known that the convergence of successive orders of perturbation theory is not guaranteed, on the contrary \cite{DiSc93,RoLa10}. Nevertheless, low-order MBPT provides at least a qualitative measure for the effect of correlations beyond HF.

We will restrict ourselves to second-order calculations for an order-of-magnitude estimate of the correlation energy, i.e., the change in the ground-state energy resulting from beyond-HF correlations. The second order contribution involves antisymmetrized two-body matrix elements of the intrinsic Hamiltonian $\op{H}_{\intr}$ containing two HF single-particle states below the Fermi energy (denoted by $\nu,\nu'$) and two HF single-particle states above the Fermi energy (denoted by $\mu,\mu'$):
\eq{ \label{eq:hf_secondorder}
  E^{(2)} 
  = \frac{1}{4} 
    \sum_{\nu,\nu'}^{<\epsilon_F}  \sum_{\mu,\mu'}^{>\epsilon_F}  
    \frac{|\matrixe{\phi_{\nu}\phi_{\nu'}}{\op{H}_{\intr}}{\phi_{\mu}\phi_{\mu'}}|^2}
      {\epsilon_{\nu} + \epsilon_{\nu'} - \epsilon_{\mu} - \epsilon_{\mu'}} \;.
}
Note that the full two-body part of the many-body Hamiltonian enters, which includes the intrinsic kinetic energy in our case.

%%%%%%%%%%%%%%%%%%%%%%%%%%%%%%%%%%%%%%%%%%%%%%%%%%%%%%%%%%%%%%%%%%%%%%
\begin{figure}[t]
\centering\includegraphics[width=0.75\textwidth]{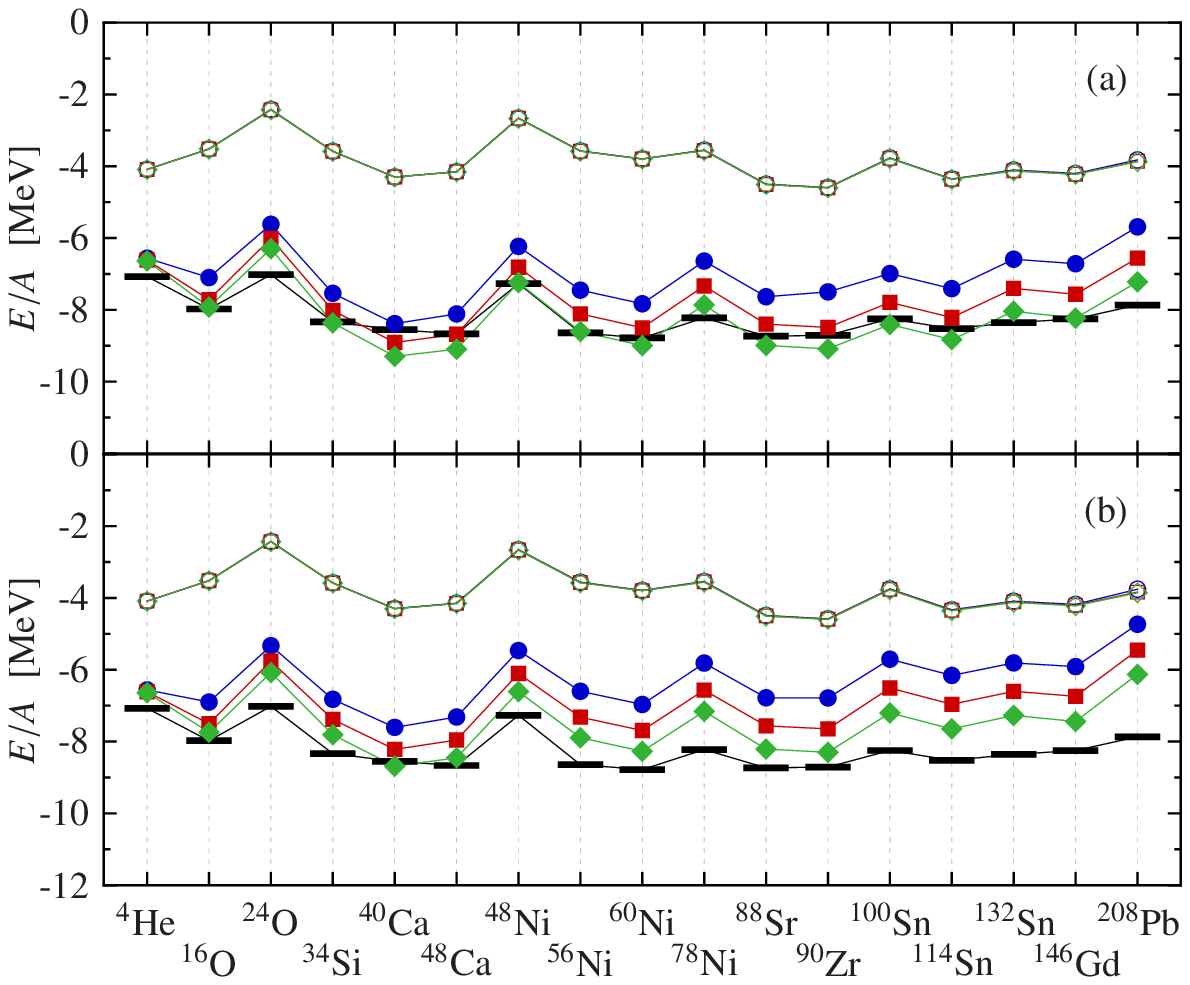}
\caption{Ground-state energies for a sequence of nuclei with closed $j$-shells obtained at the HF level (open symbols) and in second-order MBPT (filled symbols) using the UCOM(var.) interaction. The different curves correspond to different model spaces built from 11 (\symbolcircle[FGBlue]), 13 (\symbolbox[FGRed]), and 15 (\symboldiamond[FGGreen]) major oscillator shells. The oscillator length for each nucleus is chosen according to the HF root-mean-square radius (a) or according to the experimental charge radius (b). The black bars indicate the experimental binding energies \cite{AuWa03}.}
\label{fig:hf_mbpt_ucomvar}
\end{figure}
%%%%%%%%%%%%%%%%%%%%%%%%%%%%%%%%%%%%%%%%%%%%%%%%%%%%%%%%%%%%%%%%%%%%%%
%%%%%%%%%%%%%%%%%%%%%%%%%%%%%%%%%%%%%%%%%%%%%%%%%%%%%%%%%%%%%%%%%%%%%%
\begin{figure}[t]
\centering\includegraphics[width=0.75\textwidth]{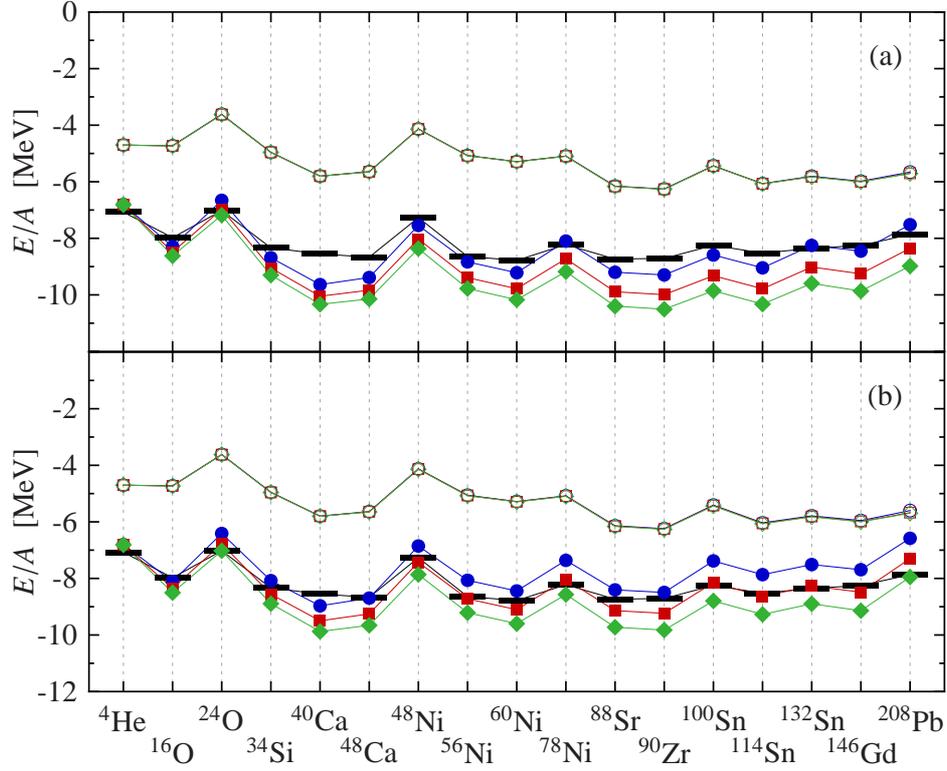}
\caption{Ground-state energies for a sequence of nuclei with closed $j$-shells obtained at the HF level (open symbols) and in second-order MBPT (filled symbols) using the UCOM(SRG) interaction. The different curves correspond to different model spaces built from 11 (\symbolcircle[FGBlue]), 13 (\symbolbox[FGRed]), and 15 (\symboldiamond[FGGreen]) major oscillator shells. The oscillator length for each nucleus is chosen according to the HF root-mean-square radius (a) or according to the experimental charge radius (b). The black bars indicate the experimental binding energies \cite{AuWa03}.}
\label{fig:hf_mbpt_ucomsrg}
\end{figure}
%%%%%%%%%%%%%%%%%%%%%%%%%%%%%%%%%%%%%%%%%%%%%%%%%%%%%%%%%%%%%%%%%%%%%%

The starting point for the evaluation of the correlation energy via perturbation theory is the HF solution yielding a finite set of single-particle states $\ket{\phi_{\nu}}$ and the corresponding single-particle energies $\epsilon_{\nu}$ for the respective nucleus. As discussed before, we use the oscillator basis including a certain number of major shells for a specific oscillator length. Because perturbation theory is not a variational approach, we cannot use variational arguments to find an optimal oscillator length, but have to resort to other prescriptions. We adopt two different schemes for choosing the oscillator length $a_{\text{HO}}$ for each nucleus: Using either the root-mean-square radius predicted by the HF solution or the experimental charge radius we optimize the oscillator length such that a naive shell-model Slater determinant built from harmonic oscillator single-particle states approximately reproduces the respective radius. Because the HF solutions for the UCOM interaction underestimate the charge radii, the oscillator lengths obtained from the HF radius are smaller than the ones obtained from the experimental radius. 

The ground-state energies obtained by including the second-order MBPT contribution on top of the HF energy are presented in Figs. \ref{fig:hf_mbpt_ucomvar} and \ref{fig:hf_mbpt_ucomsrg} for the UCOM-transformed interactions with the UCOM(var.) and the UCOM(SRG) correlators, respectively. The upper and lower panels in each figure are obtained using the HF radii and the experimental radii, respectively, to fix the oscillator length. For each case we show a sequence of calculations in model spaces consisting of 11, 13, and 15 major oscillator shells in order to assess the convergence behavior. The HF energies are fully converged and independent of the model space size. The second-order MBPT estimate of the correlation energy \eqref{eq:hf_secondorder} shows signatures of convergence only for light isotopes, for heavier isotopes there still is a significant change of typically $0.5\,\text{MeV}$ per nucleon when going from 13 to 15 shells. Uncertainties of a similar order of magnitude result from the dependence of the second-order energy on the oscillator length, with decreasing $a_{\text{HO}}$ the second-order energy contribution $|E^{(2)}|$ is increasing. Finally, one should keep in mind that second-order MBPT provides only a crude approximation for the correlation energy. As shown in Ref. \cite{RoLa10} the deviations of the second-order estimate to the exact eigenvalue in the same model space can be sizable.

Despite the uncertainties associated with the second-order MBPT calculation regarding convergence, choice of the oscillator length, and quality of low-order MBPT as such, the results in Figs.~\ref{fig:hf_mbpt_ucomvar} and \ref{fig:hf_mbpt_ucomsrg} prove that the correlations beyond HF cause an essentially constant shift of the ground state energy per nucleon across the whole mass range. For the UCOM interactions, this brings the ground state energies into the same regime as the experimental binding energies. The over-all systematics of binding energies obtained with the UCOM interactions is in agreement with experiment, already at the level of a pure two-body interaction. The role of three-body interactions is reduced to providing corrections, e.g., regarding the charge radii, on top of an already reasonable trend. For the standard SRG-transformed interactions the situation is different: The two-body component alone cannot provide the correct systematics and three-body interaction have to have a strong impact on the binding-energy systematics already.

%%%%%%%%%%%%%%%%%%%%%%%%%%%%%%%%%%%%%%%%%%%%%%%%%%%%%%%%%%%%%%%%%%%%%%
%%%%%%%%%%%%%%%%%%%%%%%%%%%%%%%%%%%%%%%%%%%%%%%%%%%%%%%%%%%%%%%%%%%%%%
%%%%%%%%%%%%%%%%%%%%%%%%%%%%%%%%%%%%%%%%%%%%%%%%%%%%%%%%%%%%%%%%%%%%%%
\clearpage
\section{Conclusion}

The Unitary Correlation Operator Method (UCOM) provides a universal
tool to account for short-range interaction-induced correlations in
the nuclear many-body problem. The correlations can be either
imprinted into many-body states that otherwise would not contain these
correlations, like Slater determinants, or they can be absorbed in
effective operators that are defined through an explicit similarity
transformation. The effective interactions obtained in this way are
well suited for low-momentum Hilbert spaces, because, unlike bare
interactions, they do not scatter strongly to high momenta. UCOM is
very transparent and intuitive as it explicitly introduces correlation
functions for the description of short-range central and tensor
correlations. These correlation functions play the role of variational
degrees of freedom for the many-body states. We propose two methods to
find optimal correlation functions for a given bare Hamiltonian: One is
energy minimization in two-body space; the other employs the close
relationship to the SRG approach, which also aims at separating low-
and high-momentum scales.

UCOM has been developed for both, matrix representation and operator
representation. The matrix representation in the harmonic oscillator
basis is used for NCSM calculations\footnote{The UCOM two-body matrix elements in the harmonic oscillator basis (relative $LS$- or $jj$-coupling for various frequencies and model-space sizes) suitable for no-core calculations are available from the authors upon request.}. We show that binding energies and
spectra converge much more rapidly with increasing size of the Hilbert
space when using effective UCOM-transformed interactions rather than
the bare interaction. It is even possible to do Hartree-Fock
calculations and obtain bound nuclei throughout the nuclear chart. We
show that this \emph{ab initio} Hartree-Fock method can only account for
about half the binding energy, the other half is correlation energy
that cannot be obtained by a single Slater determinant.  

Especially the operator representation of UCOM explains in a very
transparent way, why strong short-range correlations and long mean
free path or the mean-field shell model are not contradicting each
other. There is a separation of length scales, well developed for the
central correlations and less well for the tensor correlations, which
allows to renormalize the bare Hamiltonian to an effective one
appropriate for low-momentum Hilbert spaces. One might be tempted to
believe that short-range correlations are not real, as soft
phase-shift equivalent interactions with a modified off-shell behavior
describe the asymptotic properties of the two-nucleon system and are
successfully used for the description of many-body systems. However,
short-range correlations are revealed by the stiffness of nuclear
matter against compression or the high-momentum tails of nucleon
momentum distributions that are presently gaining renewed interest
\cite{alvioli08,subedi08,HiPi09,FrSa08}. Around the Fermi edge the momentum distributions differ
from the mean-field ones also due to long-range correlations that are
not accounted for by UCOM. These correlations have to be treated by
configuration mixing.

As the short-range correlations are, to a large extend, state
independent and can be treated by a unitary transformation, one can
work in the independent-particle basis using a transformed
Hamiltonian. Also electric and magnetic observables are seeing mainly
the low-momentum nature of the states and change little by the UCOM
transformation. An exception are Gamow-Teller transitions, which are
sensitive to tensor correlations. It is well known that a quenching of
typically 0.8 occurs for Hilbert spaces that do not contain these
correlations.

It turns out that effects from UCOM-induced three-body interactions
cancel to a certain extend those from the original three-body
force. The partial cancellation effect is not yet understood. Heavier
nuclei and the saturation properties of nuclear matter indicate that
three-body forces cannot be substituted completely. The short-range
repulsion of the nuclear interaction is essential for describing the
correct saturation properties. In particular when increasing the
density above twice nuclear saturation density, the short-range
repulsive correlations are expected to become so strong that they
cannot be treated anymore by a two-body approximation for the
effective interaction. From the above arguments it is clear that any
\emph{ab initio} treatment of nuclear matter at higher densities based on
Slater determinants of single-particle plane waves demands a
sophisticated effective interaction with many-body forces.

The Unitary Correlation Operator Method has been formulated in
momentum representation, in harmonic oscillator basis and in operator
representation. The latter can be used in any representation for
example in Fermionic Molecular Dynamics (FMD), which provides
many-body Hilbert spaces especially suited for cluster structures in
nuclei (not discussed in this paper). UCOM is a very versatile
approach that provides effective interactions as well as the
corresponding effective operators.

%%%%%%%%%%%%%%%%%%%%%%%%%%%%%%%%%%%%%%%%%%%%%%%%%%%%%%%%%%%%%%%%%%%%%%
%%%%%%%%%%%%%%%%%%%%%%%%%%%%%%%%%%%%%%%%%%%%%%%%%%%%%%%%%%%%%%%%%%%%%%
%%%%%%%%%%%%%%%%%%%%%%%%%%%%%%%%%%%%%%%%%%%%%%%%%%%%%%%%%%%%%%%%%%%%%%
\section*{Acknowledgments}

This work is supported by the Deutsche Forschungsgemeinschaft through contract SFB 634, by the Helmholtz International Center for FAIR (HIC for FAIR) within the framework of the LOEWE program launched by the State of Hesse, by the BMBF through contract 06DA9040I (NuSTAR.de), and by the GSI Helmholtzzentrum f\"ur Schwerionenforschung.

%%%%%%%%%%%%%%%%%%%%%%%%%%%%%%%%%%%%%%%%%%%%%%%%%%%%%%%%%%%%%%%%%%%%%%
%%%%%%%%%%%%%%%%%%%%%%%%%%%%%%%%%%%%%%%%%%%%%%%%%%%%%%%%%%%%%%%%%%%%%%
%%%%%%%%%%%%%%%%%%%%%%%%%%%%%%%%%%%%%%%%%%%%%%%%%%%%%%%%%%%%%%%%%%%%%%

\end{document}